\documentclass[a4paper,twocolumn,11pt,accepted=2024-02-28]{quantumarticle}
\pdfoutput=1
\usepackage[utf8]{inputenc}
\usepackage[english]{babel}
\usepackage[T1]{fontenc}
\usepackage{amsmath}
\usepackage{hyperref}
\usepackage{amsfonts} 
\usepackage{tikz}
\usepackage{lipsum}

\usepackage{booktabs}

\usepackage{soul}
\usepackage{xcolor}
\usepackage{bm}
\usepackage{dsfont}
\usepackage{hyperref}
\hypersetup{
    colorlinks=true,
    linkcolor=blue,
    filecolor=magenta,      
    urlcolor=cyan,
    pdftitle={Overleaf Example},
    pdfpagemode=FullScreen,
    }

\def\ket#1{\left| #1 \right\rangle}
\def\bra#1{\left\langle #1 \right|}

\usepackage{graphicx}
\usepackage{dcolumn}

\usepackage{cancel}

\usepackage{bm}
\newcommand{\stkout}[1]{\ifmmode\text{\sout{\ensuremath{#1}}}\else\sout{#1}\fi}

\begin{document}

\title{Generation of genuine all-way entanglement in defect-nuclear spin systems through dynamical decoupling sequences}

\author{Evangelia Takou}
\email{etakou@vt.edu}
\affiliation{Department of 
Physics, Virginia Polytechnic Institute and State University, 24061 Blacksburg, VA, USA}
\affiliation{Virginia Tech Center for Quantum Information Science and Engineering, Blacksburg, VA 24061, USA}
\orcid{0000-0002-9963-2205}
 \author{Edwin Barnes}%
 \email{efbarnes@vt.edu}
 \affiliation{Department of 
Physics, Virginia Polytechnic Institute and State University, 24061 Blacksburg, VA, USA}
\affiliation{Virginia Tech Center for Quantum Information Science and Engineering, Blacksburg, VA 24061, USA}
\orcid{0000-0003-1666-9385}
\author{Sophia E. Economou}%
 \email{economou@vt.edu}
\affiliation{Department of 
Physics, Virginia Polytechnic Institute and State University, 24061 Blacksburg, VA, USA}
\affiliation{Virginia Tech Center for Quantum Information Science and Engineering, Blacksburg, VA 24061, USA}
\orcid{0000-0002-1939-5589}

\maketitle

\begin{abstract}
 Multipartite entangled states are an essential resource for sensing, quantum error correction, and cryptography. Color centers in solids are one of the leading platforms for quantum networking due to the availability of a nuclear spin memory that can be entangled with the optically active electronic spin through dynamical decoupling sequences. Creating electron-nuclear entangled states in these systems is a difficult task as the always-on hyperfine interactions prohibit complete isolation of the target dynamics from the unwanted spin bath. While this emergent cross-talk can be alleviated by prolonging the entanglement generation, the gate durations quickly exceed coherence times. Here we show how to prepare high-quality GHZ$_M$-like states with minimal cross-talk. We introduce the $M$-tangling power of an evolution operator, which allows us to verify genuine all-way correlations. Using experimentally measured hyperfine parameters of an NV center spin in diamond coupled to carbon-13 lattice spins, we show how to use sequential or single-shot entangling operations to prepare GHZ$_M$-like states of up to $M=10$ qubits within time constraints that saturate bounds on $M$-way correlations.  
 We study the entanglement of mixed electron-nuclear states and develop a non-unitary $M$-tangling power which additionally captures correlations arising from all unwanted nuclear spins. We further derive a non-unitary $M$-tangling power which incorporates the impact of electronic dephasing errors on the $M$-way correlations. Finally, we inspect the performance of our protocols in the presence of experimentally reported pulse errors, finding that XY decoupling sequences can lead to high-fidelity GHZ state preparation.
\end{abstract}

\section{Introduction}

Generating and distributing entanglement is
one of the most fundamental yet non-trivial requirements for building large-scale quantum networks. Entanglement is the key ingredient of quantum teleportation, robust quantum communication protocols, or alternative quantum computation models such as the one-way quantum computer~\cite{BriegelPRL2001,BriegelNat2009,Raussendorf2012}, or fusion-based quantum computation~\cite{FBQCSparrow2021}. In the context of cryptography, entangled states ensure secure communications via quantum key distribution or quantum secret sharing protocols~\cite{HilleryPRA1999,TittelPRA2001,Chen2005,ChangQIP2013,BellNatCommun2014SecretSharing,BensonNewJPhys2014,MunroPRA2017}. Error detection~\cite{MonroeSciAvd2017,MorenoQIP2018} and correction~\cite{NickersonNatCommun2013,BellNatCommun2014,WratchtrupNature2014,TaminiauNatNano2014,CramerNatCommun2016,TaminiauNature2022} schemes are based on the encoding of information onto entangled states, such that errors can be detected or corrected by utilizing correlations present in entangled states. At the same time, distributed entanglement increases the sensitivity and precision of measurements~\cite{EldredgePRA2018,KoczorNewJPhys2020}. 

Defect platforms offer an electronic qubit featuring a spin-photon interface that enables the generation of distant entanglement~\cite{HansonNature2013,HumphreysNat18,PompiliSci21}, as well as nuclear spins that serve as the long-lived memories needed for information storage and buffering. Several protocols based on remote entanglement or entanglement within the electron-nuclear spin system have already been demonstrated in these platforms, including quantum teleportation~\cite{HermansNature2022}, quantum error correction~\cite{TaminiauNatNano2014,CramerNatCommun2016,TaminiauNature2022}, enhanced sensing~\cite{WrachtrupNatCommun2016,CappellaroPRapplied2019,WrachtrupNpj2021}, and entanglement distillation~\cite{KalbSci2017}. The electron-nuclear spin entanglement is usually realized through dynamical decoupling (DD) sequences; large entangled states can be created by applying consecutive sequences with the appropriate interpulse spacings. By tuning the interpulse spacings, one can select different nuclear spins to participate in entangling gates while decoupling the electron from the remaining always-on coupled nuclear spin bath~\cite{TaminiauPRL2012}. 

{\color{black}A major issue is that entangled states often decohere faster than product states~\cite{CiracPRL1997,CarvalhoPRL2004}}, so entanglement must be preserved for long enough times to complete quantum information tasks. Additionally, in defect platforms, the unwanted spin bath sets a lower bound on the duration of entangling gates, which should be long enough to suppress unwanted interactions that lower the quality of target entangled states. Despite this challenge, generation of electron-nuclear entangled states has been realized experimentally in NV~\cite{BradleyPRX19}, {\color{black}in SiV centers in diamond~\cite{LukinPRL2019,LukinPRB2019}, and in SiC defects~\cite{BourassaNatMater2020}}. In particular, Bradley \textit{et al.} demonstrated {\color{black} experimentally in an NV defect in diamond} electron-nuclear GHZ states involving up to 7 qubits~\cite{BradleyPRX19} by combining DD sequences with direct RF driving of the nuclear spins, accompanied by refocusing pulses to extend the entanglement lifetime. This direct driving was necessary to improve the selective coupling of individual nuclear spins. Although successful, this method introduces experimental overhead and heating of the sample due to the direct RF driving of the nuclei, potentially creating scalability issues. Another problem is that as the number of parties contributing to the entangled state increases, a larger gate count and longer sequences are needed, exacerbating dephasing. Failure to provide optimal isolation from {\color{black} unwanted} nuclei further deteriorates the electron-nuclear entangled states. Therefore, a more efficient approach to generating electron-nuclear spin entanglement within time constraints and verifying its existence is necessary for large-scale applications.

In this paper, we address these challenges by introducing a framework for preparing high-quality GHZ states of up to 10 qubits within time constraints. Using experimental parameters from a 27 nuclear spin register well-characterized by Taminiau \textit{et al.}~\cite{TaminiauNat2019}, we show how to improve sequential entanglement generation methods by minimizing the cross-talk induced by the unwanted nuclear spin bath. We find that it is possible to prepare GHZ-like states with maximal all-way correlations in excess of $95\%$ and gate errors lower than 0.05$\%$ due to residual entanglement with  {\color{black} unwanted} nuclei. We show how to prepare GHZ-like states with single-shot operations, reducing the gate times at least two-fold compared to the sequential scheme while offering decoupling capabilities comparable to the latter. We present a closed-form expression for the  {\color{black}$M$-tangling} power of the evolution operator and use this to develop a method for verifying genuine multipartite entanglement. Remarkably, this metric depends only on two-qubit Makhlin invariants and is closely related to the one-tangles we introduced in Ref.~\cite{EconomouPRX2023}. This simplification allows us to systematically determine the DD sequences that maximize all-way correlations as desired for generating multipartite entangled states. {\color{black} Further, we analyze the entanglement of mixed electron-nuclear states and derive a non-unitary  {\color{black}$M$-tangling} power that captures residual entanglement links arising from}  {\color{black} unwanted nuclei. We incorporate electronic dephasing errors into the $M$-tangling power and derive a simple closed-form expression. Finally, we study the impact of pulse control errors on the $M$-way correlations.}

The paper is organized as follows. In Sec.~\ref{Sec:Summary}, we discuss methods for generating electron-nuclear spin entangled states using DD sequences. In Sec.~\ref{Sec:MwayEP}, we introduce the  {\color{black}$M$-tangling} power of an evolution operator. In Sec.~\ref{Sec:GHZprotocols}, we show how to prepare GHZ$_M$-like states through sequential or single-shot entanglement protocols. In Sec.~\ref{Sec:Mixed_States}, we quantify the entanglement of electron-nuclear mixed states by tracing out unwanted spins. {\color{black}In Sec.~\ref{Sec:Non_Uni_Ep}, we study the non-unitary {\color{black}$M$-tangling} power of the electron-nuclear target subspace, which encodes correlations arising from the entire electron-nuclear system. Finally, in Sec.~\ref{Sec:Additional_Errors} we study the effect of electronic dephasing errors as well as pulse control errors on the $M$-tangling power.}

\section{Establishing electron-nuclear entanglement \label{Sec:Summary}}

\begin{figure*}
    \centering
    \includegraphics[scale=0.75]{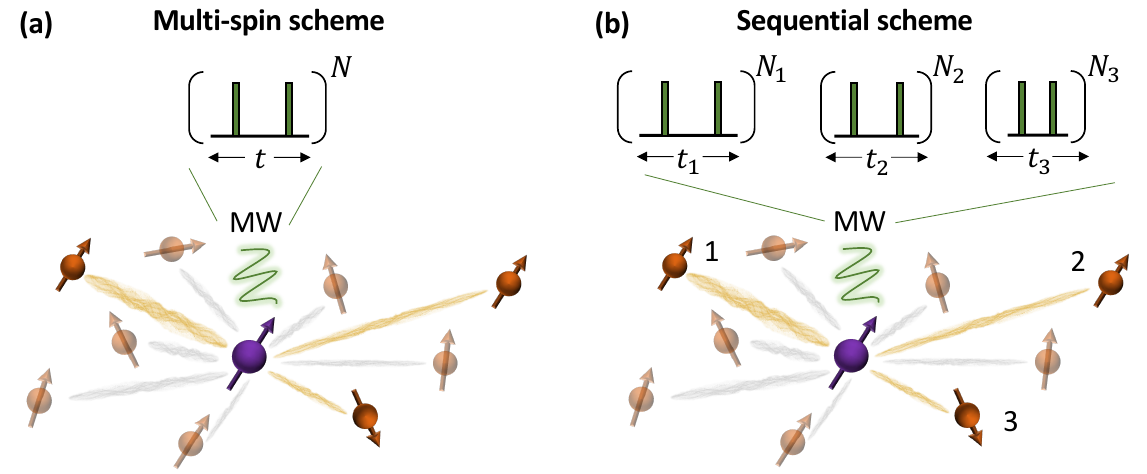}
    \caption{Schematics of two protocols for generating GHZ$_M$-like states (shown for $M=4$ and using the CPMG sequence). (a) The multi-spin scheme is capable of generating direct entanglement between the electron and  a subset of nuclei from the nuclear spin register in a single shot. (b) The sequential scheme requires $M-1$ consecutive entangling gates to prepare a GHZ$_M$-like state. }
    \label{fig:Schematics_of_Protocols}
\end{figure*}

One way to generate electron-nuclear entanglement is through DD sequences, which are trains of $\pi$-pulses applied on the electron spin that are interleaved by free evolution periods. These sequences are constructed by concatenating a basic $\pi$-pulse unit of time $t$ a certain number of iterations, $N$. Well-known examples are the Carr-Purcell-Meiboom-Gill (CPMG)~\cite{CarrPurcellPhysRev54,MeiboomGill58,deLangeSci10,Gullion_Journal_Mag_Res1969} and Uhrig (UDD)~\cite{UhrigNewJPhys08,UhrigPRL07} sequences, which have been used experimentally for example in Refs.~\cite{RBLiuNatNanotechnol2011,TaminiauPRL2012,TaminiauNatNano2014,BradleyPRX19,HansonPRB2012}, or proposed for defects in Ref.~\cite{Dong2020}. In the defect-nuclear spin system, DD sequences serve a two-fold purpose; i) they average out the interactions between the electron and unwanted nuclei and, ii) under so-called resonance conditions, they selectively entangle a target nuclear spin with the electron~\cite{TaminiauPRL2012}. These resonances correspond to specific values of $t$, and they are generally distinct for each nuclear spin, as they depend on the hyperfine (HF) parameters of each nucleus. By composing consecutive sequences with different parameters (i.e., unit times $t$ and iterations $N$), we can thus create entanglement between the electron and multiple nuclei. We refer to this standard experimental approach of entangling one nuclear spin at a time with the electron as ``sequential". In Ref.~\cite{EconomouPRX2023}, we showed that alternatively single-shot operations can be used to entangle the electron with a subset of nuclei from the register, significantly reducing gate times compared to the sequential approach. Figure~\ref{fig:Schematics_of_Protocols} summarizes the two approaches to generating electron-nuclear entanglement.

Figure~\ref{fig:Schematics_of_Protocols}(a) depicts the multi-spin scheme implemented by a single-shot entangling operation, and Fig.~\ref{fig:Schematics_of_Protocols}(b) shows the sequential scheme which requires $M-1$ entangling gates to prepare GHZ$_M$-like states. The $M-1$ entangling gates are realized by composing sequences of different interpulse spacings and iterations of the unit, such that a different nuclear spin is selected from the register based on the resonance condition.

GHZ$_M$ states are created by initializing the electron in the $|+\rangle$ state, polarizing the nuclear spins in the $|0\rangle$ state~\cite{PfaffNatPhys2013,TaminiauNatNano2014}, and then performing electron-nuclear entangling gates via DD sequences. Probabilistic (deterministic) initialization of the nuclear spins in $|0\rangle$ can be achieved through measurement-based (SWAP-based) initialization~\cite{AbobeihThesis2021}. After state initialization, one can perform a DD sequence to generate entanglement. Setting the sequence unit time $t$ to be a nuclear spin resonance time forces that nuclear spin to rotate along an axis conditioned on the electron spin state. If the axes are anti-parallel and along the $\pm \textbf{x}$ direction, and the unit is iterated an appropriate number of times, the nuclear spin can accumulate a rotation angle of $\phi=\pi/2$, realizing a CR$_x(\pm\pi/2)$ electron-nuclear entangling gate~\cite{TaminiauPRL2012}. Repeating this process by changing the unit time $t$ and iterations $N$ of the subsequent sequence (and assuming the $k$-th target spin is rotated only during the $k$-th evolution; trivial non-entangling rotations could happen on the $k$-th nucleus during the remaining evolutions) creates a state equivalent to GHZ$_M=(|0\rangle^{\otimes M}+|1\rangle^{\otimes M})/\sqrt{2}$ up to local operations:
\begin{equation}
    \text{CR}_x\left(m_k\frac{\pi}{2}\right)^{\otimes k}|+\rangle |0\rangle^{\otimes k}=\frac{1}{\sqrt{2}}\sum_{j\in\{0,1\}}|j\rangle |m_{j}^{(k)}y\rangle^{\otimes k},
\end{equation}
where $m_k=\pm 1$, $m_{j}^{(k)}$ can be $+1$ or $-1$ for the $k$-th nucleus, and we defined $|{\pm} y\rangle = (|0\rangle\pm i|1\rangle)/\sqrt{2}$. For example, if all the gates are CR$_{x}(\pi/2)$, then the resulting state is $(|0\rangle |{-}y\rangle^{\otimes k}+|1\rangle |y\rangle^{\otimes k})/\sqrt{2}$. This is the idea behind the sequential entanglement protocol. The multi-spin entanglement protocol produces a more complicated state since the electron-nuclear evolution now becomes a CR$_{xz}$ gate (see Ref.~\cite{EconomouPRX2023}), but the resulting state is still locally equivalent to GHZ$_M$. In this latter setup, multiple nuclei are simultaneously entangled with the electron via a single-shot operation.

Quantifying the quality of GHZ$_M$-like states using a fidelity overlap is computationally costly since we would need to optimize over local gates. This optimization becomes more difficult for larger entangled states. In the following section, we introduce an entanglement metric that allows us to quantify genuine all-way correlations in an arbitrarily large electron-nuclear spin system, using only the information of the evolution operator generated by $\pi$-sequences. {\color{black} With this analysis we are able to translate the $M$-tangling capability of such an evolution operator irrespective of the total system size, into simple spin-spin correlation metrics between the central spin and each of the $M-1$ nuclear spins.} Our analysis throughout the paper is completely general and is not restricted by the choice of DD sequence or by the choice of the defect system. It is further generically valid for any type of nuclear spins present in the register (provided they have spin $I=1/2$). {\color{black}The code used to simulate all our following results can be found in}~\cite{GitHubCode}.

\section{\texorpdfstring{$M$}{M}-tangling power \label{Sec:MwayEP}}

Genuine multipartite entanglement of GHZ$_M$-like states can be verified with metrics such as entanglement witnesses~\cite{Chruscinski2014,CarvachoSciRep2017,ZhaoPRA2019}, concentratable entanglement~\cite{CerezoPRL2021}, or the so-called $M$-tangles~\cite{Coffman2000,WongPRA2001,DafaQInf2012}. We choose to work with the latter, which are usually defined in terms of a state vector (or, more generally, a density operator). The $M$-tangles, $\tau_M(|\psi\rangle)\in[0,1]$, distinguish the GHZ entanglement class from other entanglement classes. For example, the three-tangle saturates to 1 for the GHZ$_3$ state, whereas it vanishes for the $|\text{W}\rangle = 1/\sqrt{3}(|100\rangle+|010\rangle+|001\rangle)$ state. The $M$-tangles are  invariant under permutations of the qubits and SLOCC, and are entanglement monotones\footnote{More precisely, the $M$-tangles monotonically decrease under SLOCC, due to single qubit measurements.} ~\cite{Horodecki2009}. Since they are defined based on a state vector, their calculation requires that we make an assumption about the initial state of the register, which is then evolved under the $\pi$-pulse sequence. We circumvent this issue by focusing instead on the capability of a gate to saturate all-way correlations and thus prepare states locally equivalent to GHZ$_M$.

We introduce the  {\color{black}$M$-tangling} power of a unitary, which  {\color{black} we define to be} the average of the $M$-tangle $\epsilon_{p,M}(U):=\langle \tau_M(U\rho_0 U^\dagger) \rangle$, over all initial product states, $\rho_0$. The formulas of $M$-tangles expressed in terms of the state vector can be found in Appendix~\ref{App:MTangles} and in Refs.~\cite{Coffman2000,DafaQInf2012,WongPRA2001}. 

In our system, the total evolution operator (of the electron and $M-1$ nuclear spins) produced by DD sequences has the form:
\begin{equation}\label{Eq:U}
    U = \sum_{j\in{0,1}}\sigma_{jj}\otimes_{l=1}^{M-1}R_{\textbf{n}_j}^{(l)},
\end{equation}
where $\sigma_{jj}\equiv|j\rangle \langle j|$ are projectors onto two of the levels in
the electron spin multiplet, and $R_{\textbf{n}_j}^{(l)}:=e^{-i\phi_j^{(l)}/2(\boldsymbol{\sigma}\cdot \textbf{n}_j^{(l)})}$ is the rotation that acts on the $l$-th nuclear spin and, in general, depends on the electron spin state. Due to the fact that the evolution is controlled only on the electron, we find that we can compactly express the $M$-tangle of a given state as [see Appendix~\ref{App:MwayEP}]:
\begin{equation}\label{Eq:MTangle}
    \tau_M(\rho)=\text{tr}[\rho^{\otimes 2}\tilde{P}], 
\end{equation}
where $\rho^{\otimes 2}$ is the density matrix of two copies of the $M$-qubit system living in the Hilbert space $\mathcal{H}_{M^2}$. $\tilde{P}$ is the product of projectors onto the antisymmetric (or symmetric) space of the $i$-th subspace and its copy, i.e.
\begin{equation}
    \tilde{P}=
    \begin{cases}
    \prod_{j=1}^MP^{(-)}_{j,j+M}, \text{ even } M \\
    P_{1,1+M}^{(+)}\prod_{j=2}^MP^{(-)}_{j,j+M} \text{ odd } M,
    \end{cases}
\end{equation}
where $P^{(\pm)}_{j,j+M}=1/2(\mathds{1}\pm\text{SWAP}_{j,j+M})$, and with $\text{SWAP}_{j,j+M}=\sum_{\alpha,\beta\in\{0,1\}}|\alpha\rangle_j |\beta\rangle_{j+M}\langle\beta|_{j}\langle \alpha|_{j+M}$. Note that we have fixed system ``1'' to be the electron's subspace. (If the control qubit corresponds to a different subspace, then for odd $M$, the product of antisymmetric projectors needs to exclude the control system from the copies, and instead we apply $P^{(+)}$ on the sectors of the control qubit). For even $M\geq 4$, Eq.~(\ref{Eq:MTangle}), holds for density matrices $\rho$ obtained by any arbitrary $U$, whereas for odd $M$, Eq.~(\ref{Eq:MTangle}) holds strictly for states obtained by CR-evolutions. (We verified this numerically, comparing with the exact expressions of $\tau_M$ that hold for states obtained by arbitrary evolutions.) By averaging Eq.~(\ref{Eq:MTangle}) over all product initial states, we prove that the  {\color{black}$M$-tangling} power of an operator $U$ has the simple expression [Appendix~\ref{App:MwayEP}]:
\begin{equation}\label{Eq:MwayEp_of_any_U}
    \epsilon_{p,M}(U):=2^M \text{Tr}[U^{\otimes 2}\Omega_{p0}(U^\dagger)^{\otimes 2}\tilde{P}],
\end{equation}
with $\Omega_{p0}=(d+1)^{-M}\prod_{j=1}^M P^{(+)}_{j,j+M}$, where $d=2$ (dimension of qubit subspace). Equation~(\ref{Eq:MwayEp_of_any_U}) is exact for any arbitrary $U(M)$, for {\color{black}$M$ even and $M\geq 4$}, whereas for odd $M$ it holds only for CR-type evolution operators of the form of Eq.~(\ref{Eq:U}). The simple structure of CR-type evolution operators generated by $\pi$-sequences allows us to derive analytically a closed-form expression for the  {\color{black}$M$-tangling} power for an arbitrarily large nuclear spin register coupled to the electron. In Appendix~\ref{App:MwayEP_CRevol}, we prove that the  {\color{black}$M$-tangling} power of the operator $U$ of Eq.~(\ref{Eq:U}), is given by:
\begin{equation}\label{Eq:MwayEP}
    \epsilon_{p,M}(U)=\left(\frac{d}{d+1}\right)^M \prod_{k=1}^{M-1}(1-G_1^{(k)}),
\end{equation}
where $G_1$ is the so-called Makhlin invariant~\cite{Makhlin2002}, which we have shown that for a $\pi$-pulse sequence is given by~\cite{EconomouPRX2023}: 
\begin{equation}\label{Eq:G1}
\resizebox{0.97\hsize}{!}{
    $G_1^{(k)}=\Big(\cos\frac{\phi^{(k)}_0}{2}\cos\frac{\phi^{(k)}_1}{2}+(\textbf{n}_0\cdot \textbf{n}_1)^{(k)}\sin\frac{\phi^{(k)}_0}{2}\sin\frac{\phi^{(k)}_1}{2}\Big)^2,$}
\end{equation}
with $G_1\in[0,1]$. Here $\phi_{0}^{(k)}$ ($\phi_1^{(k)}$) is the rotation angle that the $k$th nuclear spin undergoes when the electron is in the $|0\rangle$ ($|1\rangle$) state, and $\textbf{n}_0^{(k)}$ ($\textbf{n}_1^{(k)}$) is the corresponding nuclear rotation axis. $G_1^{(k)}$ characterizes the evolution of the $k$th nuclear spin under the DD sequence and is related to the bipartition entangling power of the nuclear spin with the remaining register, i.e., the nuclear one-tangle given by $\epsilon_p^{\text{nuclear}}=2/9(1-G_1)$~\cite{EconomouPRX2023}. The nuclear one-tangle is a faithful metric of selectivity; its minimization ensures that the spin is decoupled from the register, while its maximization implies that the particular spin attains maximal correlations with the register.  

The remarkable simplicity of Eq.~(\ref{Eq:MwayEP}) allows us to check whether a controlled unitary of the form of Eq.~(\ref{Eq:U}) is capable of preparing GHZ$_M$-like states. Saturating the bounds of all-way correlations between the electronic defect and the nuclear register translates into minimizing $G_1$ (or, equivalently maximizing $\epsilon_p^{\text{nuclear}}$) of the nuclei that will be part of the GHZ state, and implies that the maximum
 {\color{black}$M$-tangling} power is $[d/(d+1)]^M$.

At this point, we should mention a caveat of the $M$-tangle metric. In Ref.~\cite{WongPRA2001} it was mentioned that the four-tangle cannot discriminate the entanglement of a GHZ$_4$-like state from that of bi-separable maximally entangled states, e.g., $|\Phi^+\rangle^{\otimes 2}$, where $|\Phi^+\rangle$ is one of the four Bell states; in both cases the four-tangle is 1. Similarly, $\tau_6(|\text{GHZ}_6\rangle)=1=\tau_6(|\Phi^+\rangle^{\otimes 3})$, but $\tau_6(|\text{GHZ}_3\rangle^{\otimes 2})=0$. Analogously, the even $M$-tangle cannot discriminate the entanglement of a GHZ$_M$ state from the entanglement of $M/2$ copies of two-qubit maximally entangled states $|\text{Bell}\rangle^{\otimes M/2}$ (see also Ref.~\cite{Liarxiv2010}).
Nevertheless, in our system we have the extra condition that $G_1$ of all $M-1$ nuclei should be minimized to saturate all-way correlations. Hence, if we fail to minimize at least one out of the $M-1$ $\{G_1\}$ quantities, we know that the evolution operator will never be able to prepare genuine multipartite entanglement between all $M$ parties.
Another way to understand this is to consider the example of 4 qubits, and the entangling gate $\text{CR}_x(\pi/2)^{\otimes 3}=\sigma_{00}\otimes_{i=1}^3 R_x(\pi/2)+\sigma_{11}\otimes_{i=1}^3 R_x(-\pi/2)$. If we act with $\text{CR}_x(\pi/2)^{\otimes 3}$ on arbtirary initial states, we can at most prepare a GHZ$_4$-like state, but we will never be able to prepare two individual Bell pairs, since correlations are constrained to be distributed among all parties connected by the electron (since we are dealing with a central spin type system). Similar statements hold for the multi-spin method that utilizes CR$_{xz}$ gates. Therefore, we can safely use the $M$-tangle metrics to detect genuine multipartite electron-nuclear (and nuclear-nuclear) spin entanglement.

Let us further comment that the $M$-tangling power is derived by averaging over all product (and pure) states, so it only tells us how well the dynamics of $U$ can saturate the all-way correlations of pure product states. No conclusion can be made, however, if one has a statistical mixture of such states (i.e., a mixed state).

In the following section, we show how to optimize the sequential or multi-spin schemes and guide the selection of nuclei from the register to generate genuine all-way correlations within time constraints.

\section{Creation of GHZ\texorpdfstring{$_M$}{M}-like states \label{Sec:GHZprotocols}}

\begin{table}[ht]
\begin{center}
\resizebox{\columnwidth}{!}{
\begin{tabular}{ |c|c|c|c|c| } 
\hline
 Symbol  & Meaning \\ \hline
 $|\text{GHZ}|$  & GHZ size \\ \hline
 $L_{nuc}$ & Total $\#$ of nuclei \\ \hline
 $j$ & Spin index $\in[1,L_{nuc}]$ \\
 \hline
 $N$ & Sequence iterations \\
 \hline
 $k$ & Resonance $\#$ \\
 \hline
 $(\textbf{n}_0\cdot \textbf{n}_1)^{(j)}$ & Dot product of rotation axes for $j$-th spin \\
 \hline
 $\epsilon_{p}^{\text{nuc} (j)}$ & Nuclear one-tangle of $j$-th spin \\
 \hline
 $\{N_k^*(j)\}$ & Iterations for $k$-th resonance where  \\
 & $\epsilon_p^{\text{nuc} (j)}$ is maximal \\ \hline
 $\delta_{t/u}$ & Tolerance of target/unwanted one-tangles \\
 \hline
 $\delta_E$ & Gate error tolerance \\
 \hline
 $t$ & Unit time \\
 \hline
 $t_k^{(j)}$ & Unit time at $k$-th resonance of $j$-th spin \\
 \hline
$T$ & Gate time of total sequence \\
\hline
$T_{max}$ & Gate time tolerance \\
\hline
$G_1^{(j)}$ & Makhlin-invariant of $j$-th spin \\
 \hline
\end{tabular}
}
\end{center}
\caption{Meaning of the symbols that we use in the flowchart of Fig.~\ref{fig:FlowChart} to describe the optimization steps for the sequential and multi-spin protocols. }
\label{Tab:0}
\end{table}

\begin{figure*}[!htbp]
    \centering
    \includegraphics[scale=0.6]{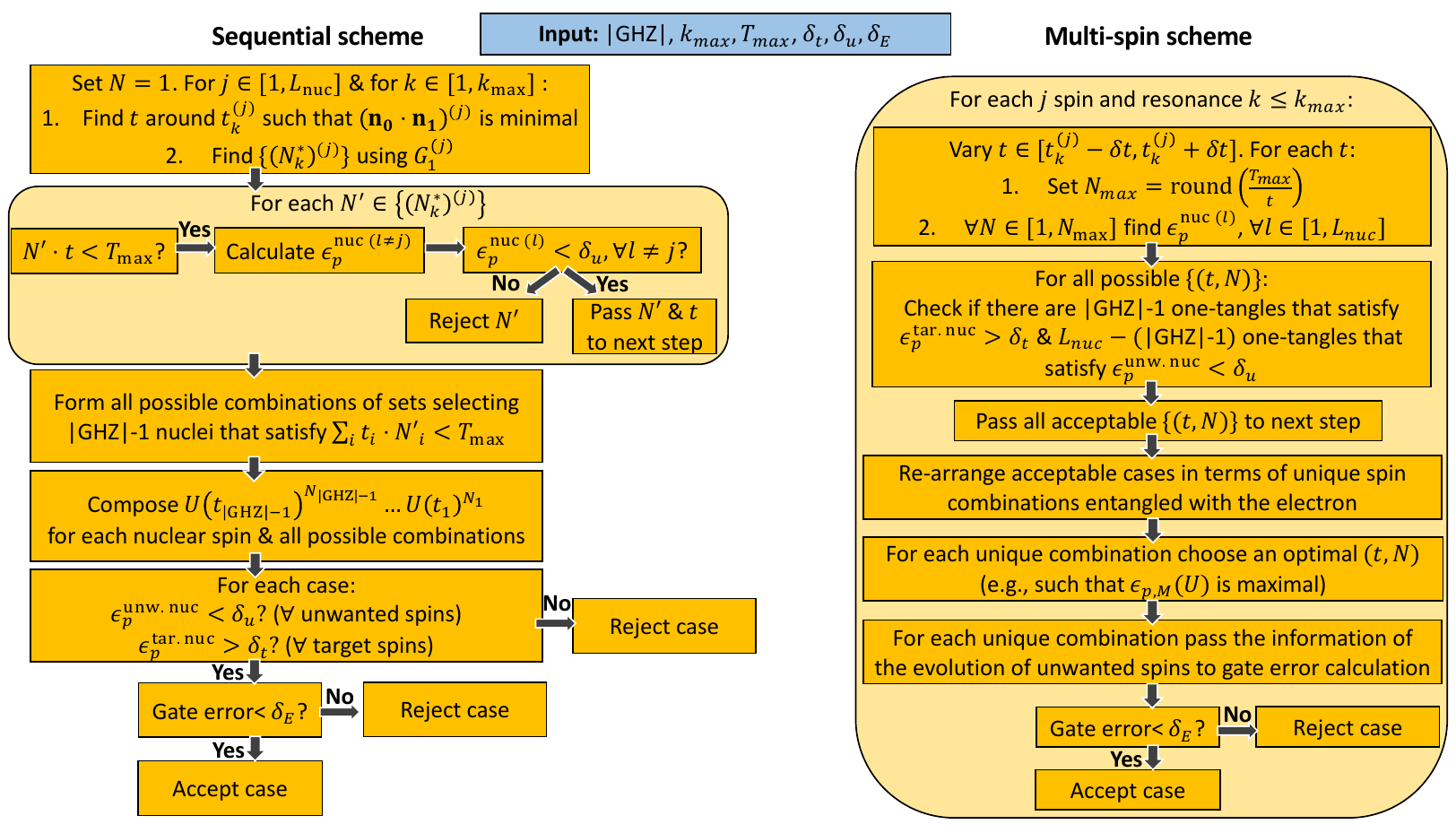}
    \caption{Flowcharts that summarize the steps we follow  to find optimal sequences that prepare GHZ$_M$-like states, utilizing the sequential or multi-spin scheme. The various symbols are defined in Table~\ref{Tab:0}.}
    \label{fig:FlowChart}
\end{figure*}

\begin{figure*}[!htbp]
    \centering
    \includegraphics[scale=0.7]{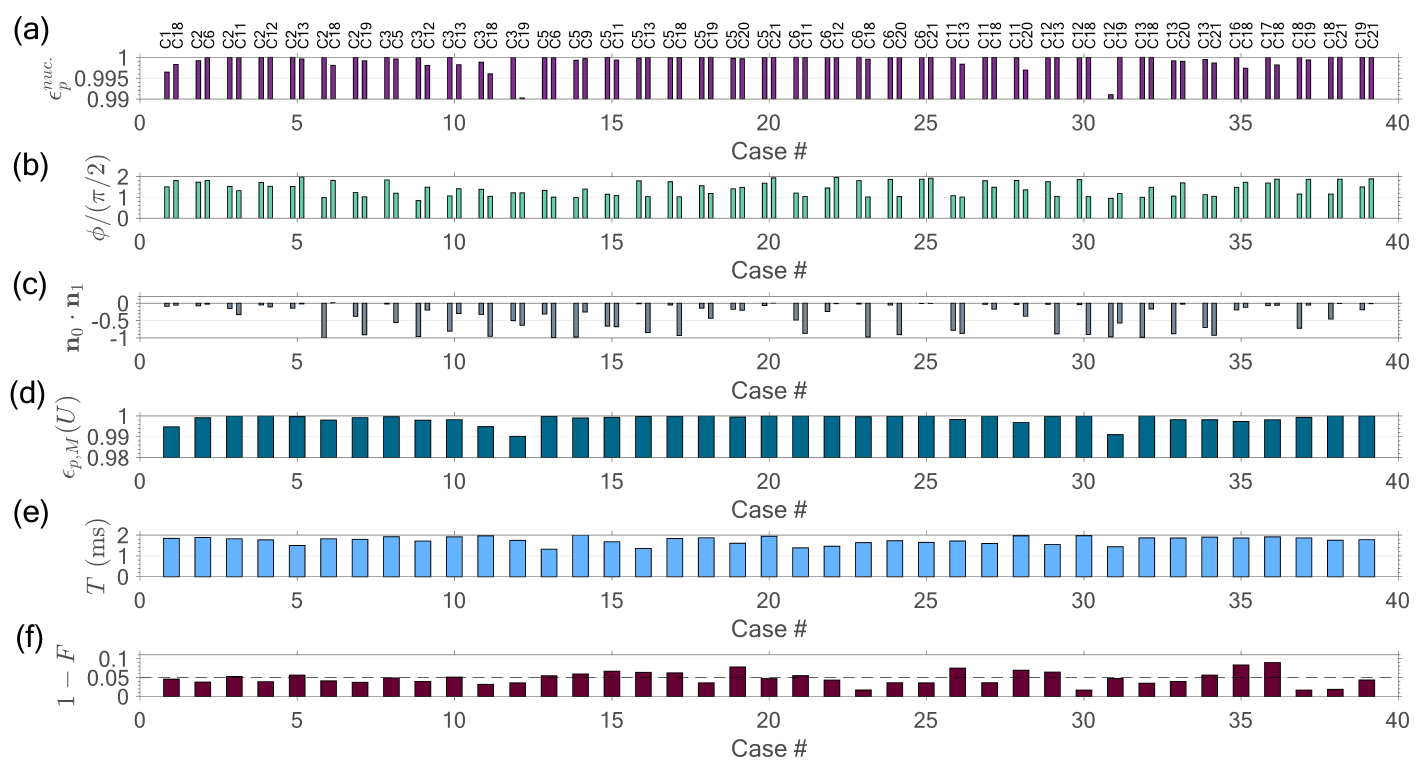}
    \caption{Generation of GHZ$_3$-like states via the sequential protocol. Each case $\#$ corresponds to a different composite DD sequence that sequentially entangles a pair of nuclei with the electron to generate GHZ$_3$-like states. (a) Nuclear one-tangles (scaled by 2/9) after the two entangling gates. The first (second) bar of each case corresponds to a particular nuclear spin, labeled as C$j$. (b) Nuclear rotation angles ($\phi :=\phi_0$) and (c) dot products of nuclear rotation axes at the end of the composite evolution of the two entangling gates.  (d)  {\color{black}$M$-tangling} power for each case of composite evolution. (e) Total gate time of the sequence and (f) gate error of the composite evolution due to residual entanglement with the remaining nuclear spins that are not part of the GHZ$_3$-like state. }
    \label{fig:GHZ3_Sequential}
\end{figure*}

\subsection{Sequential scheme for GHZ\texorpdfstring{$_3$}{3} states}

We first begin with the task of generating GHZ$_M$-like states using the sequential scheme, which requires $M-1$ consecutive gates. Hereafter, we refer to the collection of $M-1$ entangling gates as the composite evolution.

Under this scheme, to entangle the nuclei with the electron, we need to ensure that during each evolution we rotate conditionally only one particular nuclear spin, meaning we maximize its one-tangle. At the same time, all other nuclear one-tangles should be minimal in each evolution, so that we suppress cross-talk arising from the nuclear spin bath. 
The procedure of identifying optimal nuclear spin candidates and sequence parameters is shown in Fig.~\ref{fig:FlowChart} and
explained in more detail in Appendix~\ref{App:Optimization_Procedure}. Depending on the size of the GHZ state, we set different tolerances for unwanted/target one-tangles, and gate time restrictions within $T_2^*$ of the nuclei (see Table~\ref{Tab:1} of Appendix~\ref{App:Optimization_Procedure}). At the end of this procedure, we are left with different options for selectively entangling a single nuclear spin with the electron.

After identifying the optimal sequence parameters and nuclear spin candidates, we compose the $M-1$ entangling gates and create a ``case'' of a composite evolution operator, from which we can extract the Makhlin invariants $G_1$ associated with the evolution of each nuclear spin. This process allows us to calculate the  {\color{black}$M$-tangling} power of the composite gate (using Eq.~(\ref{Eq:MwayEP})) and verify if it can prepare genuine multipartite entanglement. Further, as we showed in Ref.~\cite{EconomouPRX2023}, we can analytically calculate the induced gate error due to residual entanglement links with the unwanted nuclei. This unresolved entanglement could make the target gate deviate from the ideal evolution, which we take to be the composite $M-1$ entangling operations in the absence of unwanted spins; minimization of this gate error thus ensures a better quality of the GHZ-like states that we produce.

We start with the simplest case of generating GHZ$_3$-like states using the sequential scheme. Throughout the rest of the paper, we focus on the CPMG sequence $(t/4-\pi-t/2-\pi-t/4)^N$, where $t$ is the unit time, $\pi$ represents a $\pi$-pulse, and $N$ is the number of iterations of the basic unit. For concreteness, we consider an NV center in diamond and define the qubit states of the electron to be $|0\rangle\equiv |m_s=0\rangle$ and $|1\rangle\equiv |m_{s}=-1\rangle$. We further use the hyperfine parameters of the $^{13}\text{C}$ nuclear spins from the 27 nuclear spin register characterized by the Taminiau group~\cite{TaminiauNat2019}, and following their conventions we  label the nuclear spins as $\text{C}j$ with $j\in[1,27]$. Despite these choices, the analysis that follows is general and valid for any $\pi$-sequence or electronic defect in diamond or SiC, and for nuclei with spin $I=1/2$ (e.g., $^{13}\text{C}$ in diamond, $^{13}\text{C}$ and $^{29}\text{Si}$ in silicon carbide).

\begin{figure*}[!htbp]
    \centering
    \includegraphics[scale=0.66]{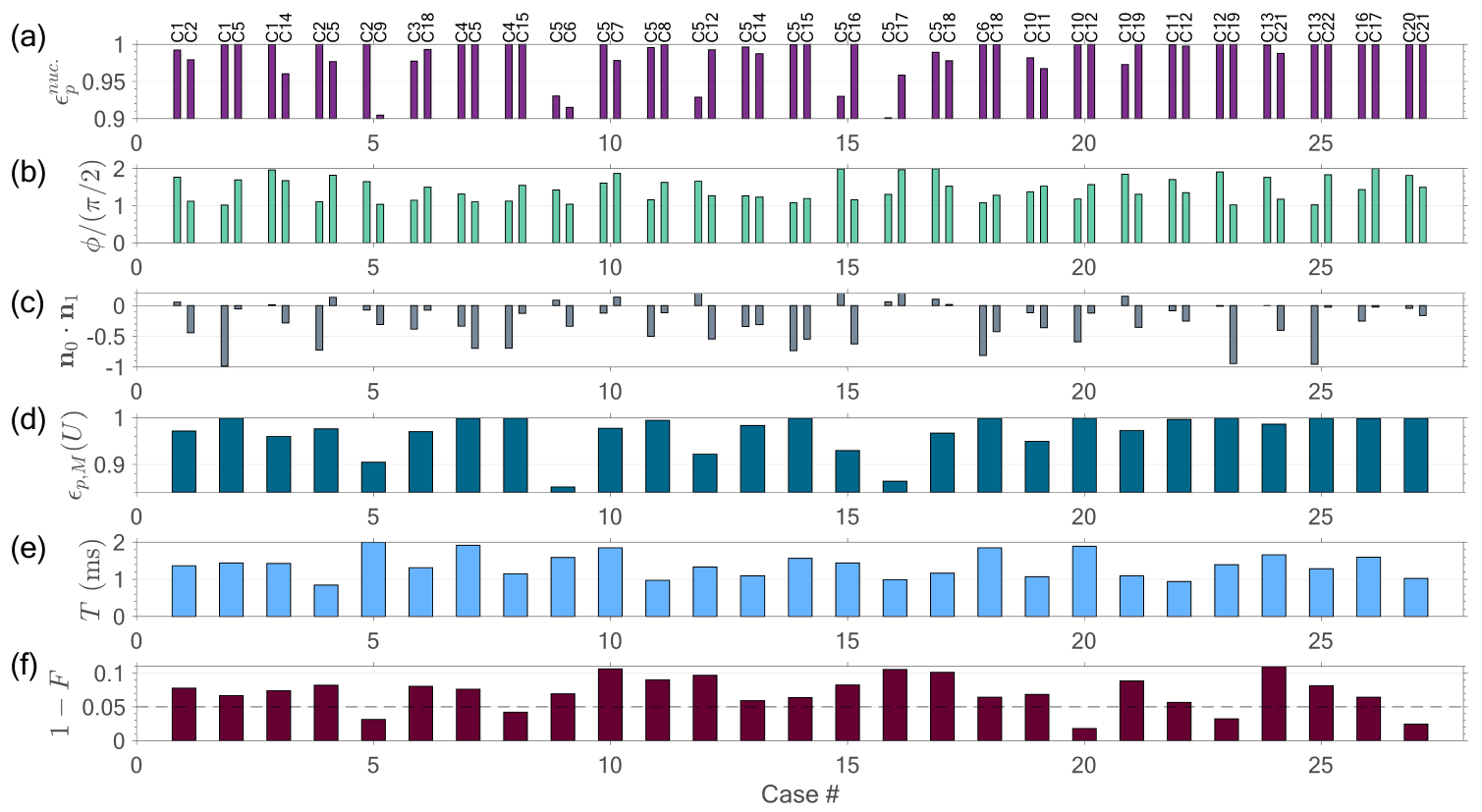}
    \caption{Generation of GHZ$_3$-like states using the multi-spin protocol. Each case $\#$ corresponds to a different DD sequence that simultaneously entangles a pair of nuclei with the electron to generate GHZ$_3$-like states. (a) Nuclear one-tangles (scaled by 2/9) after the single-shot operation. (b) Nuclear rotation angles and (c) dot products of nuclear rotation axes at the end of the single-shot entangling operation.  (d)  {\color{black}$M$-tangling} power for each case. (e) Total gate time of the sequence and (f) gate error of the evolution due to residual entanglement with the remaining nuclear spins that are not part of the GHZ$_3$-like state.}
    \label{fig:GHZ3_MultiSpin}
\end{figure*}

In Fig.~\ref{fig:GHZ3_Sequential}(a), we show the nuclear one-tangles (scaled by the maximum value of 2/9) for  {\color{black}39} different composite evolutions (labeled as ``cases'') that can prepare GHZ$_3$-like states. Each realization corresponds to a unique nuclear spin combination that is entangled with the electron (the text above the bars indicates which $^{13}$C nuclear spins are involved). The first (second) bar of each case corresponds to the one-tangle of the first (second) target nuclear spin at the end of the composite evolution of two entangling gates. Each entangling gate is implemened using an optimal unit time $t$ of the sequence and an optimal number $N$ of iterations of the unit. Since we optimize $t$ close to the resonance time of the respective nuclear spin, the dot product of the nuclear axes is $\textbf{n}_0\cdot \textbf{n}_1\approx -1$. Thus, the maximization of the first (second) one-tangle happens when the first (second) nuclear spin rotates by $\phi=\pi/2$ [see Eq.~(\ref{Eq:G1})], during the first (second) evolution. (Note that for each CPMG evolution it holds that $\phi_0=\phi_1\equiv \phi$, see also Ref.~\cite{EconomouPRX2023}).

The composition of the two entangling operations gives rise to total rotation angles $\phi_0\neq \phi_1$ for each nuclear spin. The nuclear rotation angles $\phi:=\phi_0$ are shown in Fig.~\ref{fig:GHZ3_Sequential}(b), and the dot products of the rotation axes are depicted in Fig.~\ref{fig:GHZ3_Sequential}(c). Due to the composite evolution, the total nuclear rotation angles deviate from $\pi/2$, and the rotation axes are no longer anti-parallel for all cases. However, as we noted in Ref.~\cite{EconomouPRX2023}, $G_1$ can be minimized as long as $\textbf{n}_0\cdot \textbf{n}_1\leq 0$. Due to the unwanted one-tangle tolerances we impose, we ensure that only one out of the $M-1$ gates will conditionally rotate the target spin we intend to entangle with the electron. The remaining evolutions will approximately leave the entanglement of that specific spin with the electron intact. For example, the first (second) target spin will evolve approximately trivially (i.e., in an unconditional manner) during the second (first) evolution, hence preserving the high entanglement between all parties at the end of all the gates. This behavior is confirmed in Fig.~\ref{fig:GHZ3_Sequential}(d) where we show the  {\color{black}$M$-tangling} power (scaled by $[d/(d+1)]^M$, which is higher than 0.99 across all different cases. In Fig.~\ref{fig:GHZ3_Sequential}(e), we further show the total gate time, which we have restricted to less than 2~ms so that we are within the coherence times of the nuclei ($T_2^*\in[3,17]$ ms; see Ref.~\cite{BradleyThesis2021}). Further, in Fig.~\ref{fig:GHZ3_Sequential}(f), we show the gate error due to residual entanglement of the target subspace with unwanted nuclei. We note that for various realizations, the gate fidelity can exceed 95$\%$ (e.g., cases ``2'', ``7'', {\color{black}``19''}, {\color{black}``23''},``30'', {\color{black}``37''}), implying small levels of cross-talk with the remaining 25 unwanted nuclear spins. 

\begin{figure*}
    \centering
    \includegraphics[scale=0.73]{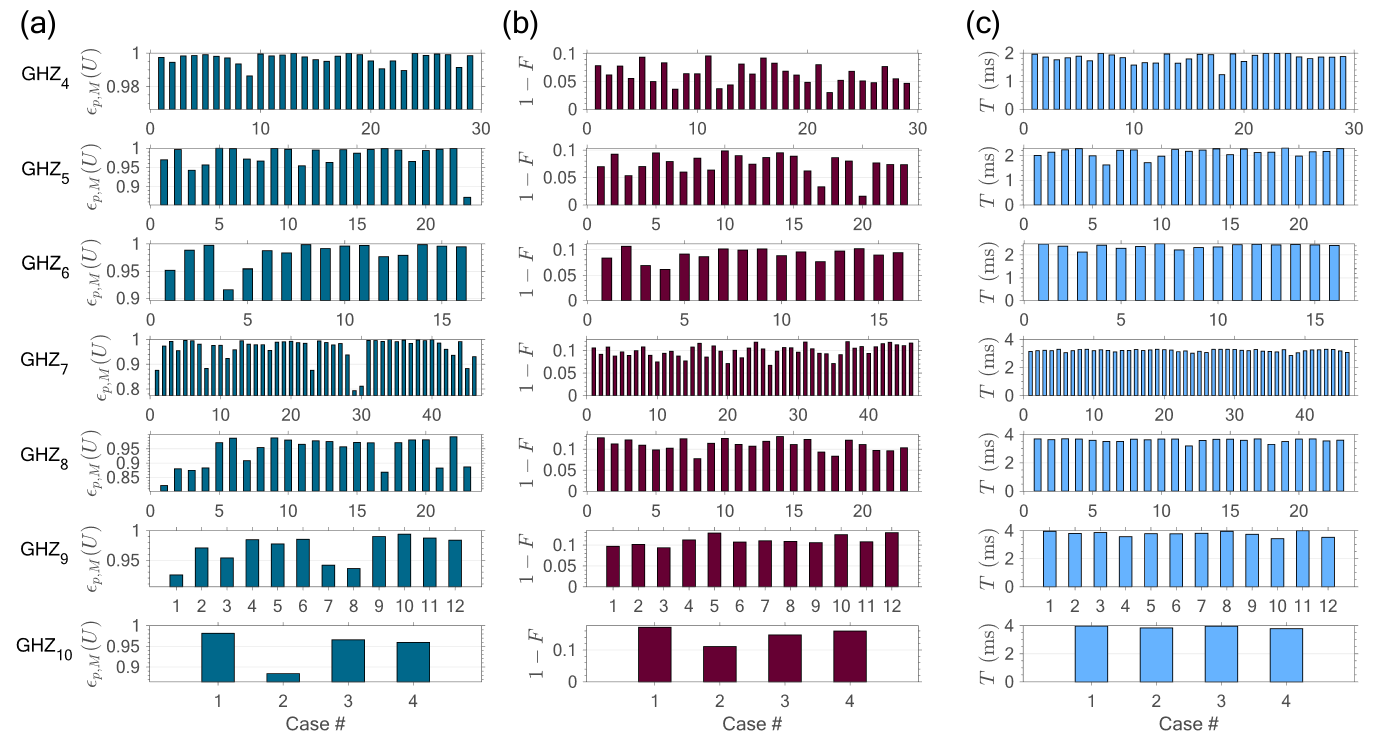}
    \caption{Preparing states locally equivalent to GHZ$_M$ via the sequential scheme. Each case $\#$ corresponds to a unique composite DD sequence that selects $M-1$ nuclei from the register. (a)  {\color{black}$M$-tangling} power of the composite $M-1$ entangling gates for various cases. (b) Gate error of the composite evolution due to the presence of unwanted nuclei, and (c) gate time of the total sequence for each case and each different size of GHZ-like states. The panels from top to bottom correspond to GHZ$_4$, GHZ$_5$, GHZ$_6$, GHZ$_7$, GHZ$_8$, GHZ$_9$ and GHZ$_{10}$ states. The nuclear one-tangles for each case and value of $M$ can be found in Appendix~\ref{App:OneTangles_Seq_GHZ}. }
    \label{fig:GHZM_Sequential}
\end{figure*}

\subsection{Multi-spin scheme for GHZ\texorpdfstring{$_3$}{3} states}

Next, we proceed with the multi-spin scheme, which can prepare GHZ$_3$-like states with a single-shot entangling operation. To identify optimal nuclear spin candidates and sequence parameters, we follow a slightly different procedure, shown in Fig.~\ref{fig:FlowChart} and explained more in Appendix~\ref{App:Optimization_Procedure}. In this case, we require that two nuclear one-tangles are maximized for the same sequence parameters and under the gate time restriction of 2 ms. At the same time, we require small values of unwanted one-tangles to minimize the cross-talk with the remaining nuclear spin bath. We summarize the results of this method in Fig.~\ref{fig:GHZ3_MultiSpin}. In Fig.~\ref{fig:GHZ3_MultiSpin}(a), we show the one-tangles of two nuclei that are maximized by a single-shot gate for 15 different DD sequences. To accept these cases as ``optimal'', we require that the target one-tangles are greater than a threshold, here set to 0.9. Figure~\ref{fig:GHZ3_MultiSpin}(b) shows the nuclear rotation angles, and Fig.~\ref{fig:GHZ3_MultiSpin}(c) shows the dot products of the nuclear rotation axes for each case. Since now the sequence unit time will not in principle coincide with the resonance of both target spins (since their HF parameters are distinct), the nuclear rotation axes will not be anti-parallel for both nuclei. However, when $\textbf{n}_0\cdot \textbf{n}_1\leq 0$ for a particular nuclear spin, then if it is also rotated by the appropriate angle, the one-tangle can be maximal (see for example cases  {\color{black}``2''},  {\color{black}``7''}, {\color{black}``8''}, ``11'' and Ref.~\cite{EconomouPRX2023} for a more detailed explanation). Thus, the feature of non-antiparallel axes can be compensated by rotating the nuclear spins by angles that deviate from $\pi/2$. On the other hand, if $\textbf{n}_0\cdot \textbf{n}_1>0$, $G_1$ cannot go to 0, and the one-tangle can never be 1 (see case  {\color{black}``4'', for nuclear spin C5}). In Fig.~\ref{fig:GHZ3_MultiSpin}(d) we show the  {\color{black}$M$-tangling} power (scaled by $[d/(d+1)]^M$) of the multi-spin gate. As expected, the  {\color{black}$M$-tangling} power saturates to 1 when the nuclear one-tangles are also maximal. In Fig.~\ref{fig:GHZ3_MultiSpin}(e), we show the gate time of the multi-spin operation, and in Figure~\ref{fig:GHZ3_MultiSpin}(f), the gate error. We note that for various cases, the multi-spin gate duration is at least two times shorter (see for example, cases {\color{black} ``4'',} ``11'', {\color{black}``22''} ) compared to the sequential one. Thus, the multi-spin gates can be equally reliable as the sequential gates, but with the additional advantage of being significantly faster compared to the latter. 

\subsection{Generating GHZ\texorpdfstring{$_M$}{M} states}
We can use similar ideas to extend the size of the GHZ$_M$-like states. In Fig.~\ref{fig:GHZM_Sequential}(a), we show the  {\color{black}$M$-tangling} power of composite evolutions that can prepare GHZ$_M$-like states up to $M=10$ qubits. The nuclear one-tangles for all different realizations are provided in Appendix~\ref{App:OneTangles_Seq_GHZ}. Figure~\ref{fig:GHZM_Sequential}(b) shows the gate error of the composite evolution due to the presence of the remaining $27-(M-1)$ nuclear spins, whereas Fig.~\ref{fig:GHZM_Sequential}(c) shows the total gate time of the composite sequences. As the size of the GHZ-like states increases, it becomes more difficult to eliminate the cross-talk between the target nuclei since we need to ensure that only one out of the $M-1$ evolutions rotates conditionally one of the $M-1$ target nuclei. However, we still find acceptable cases that can prepare up to GHZ$_{10}$-like states with  {\color{black}$M$-tangling} power over 0.95. The infidelity tends to increase as we perform more entangling gates because it becomes more likely that at least one of the $M-1$ entangling operators will induce cross-talk between the target subspace and the remaining nuclear spin bath. However, it is still possible to create GHZ$_6$-like states within $\sim 2$~ms, and even GHZ$_{9}$- or GHZ$_{10}$-like states within $4$~ms, a significant improvement compared to gate times reported in Ref.~\cite{BradleyPRX19}, which could be as high as 2.371~ms for controlling only a single nuclear spin.
\begin{figure*}
    \centering
    \includegraphics[scale=0.7]{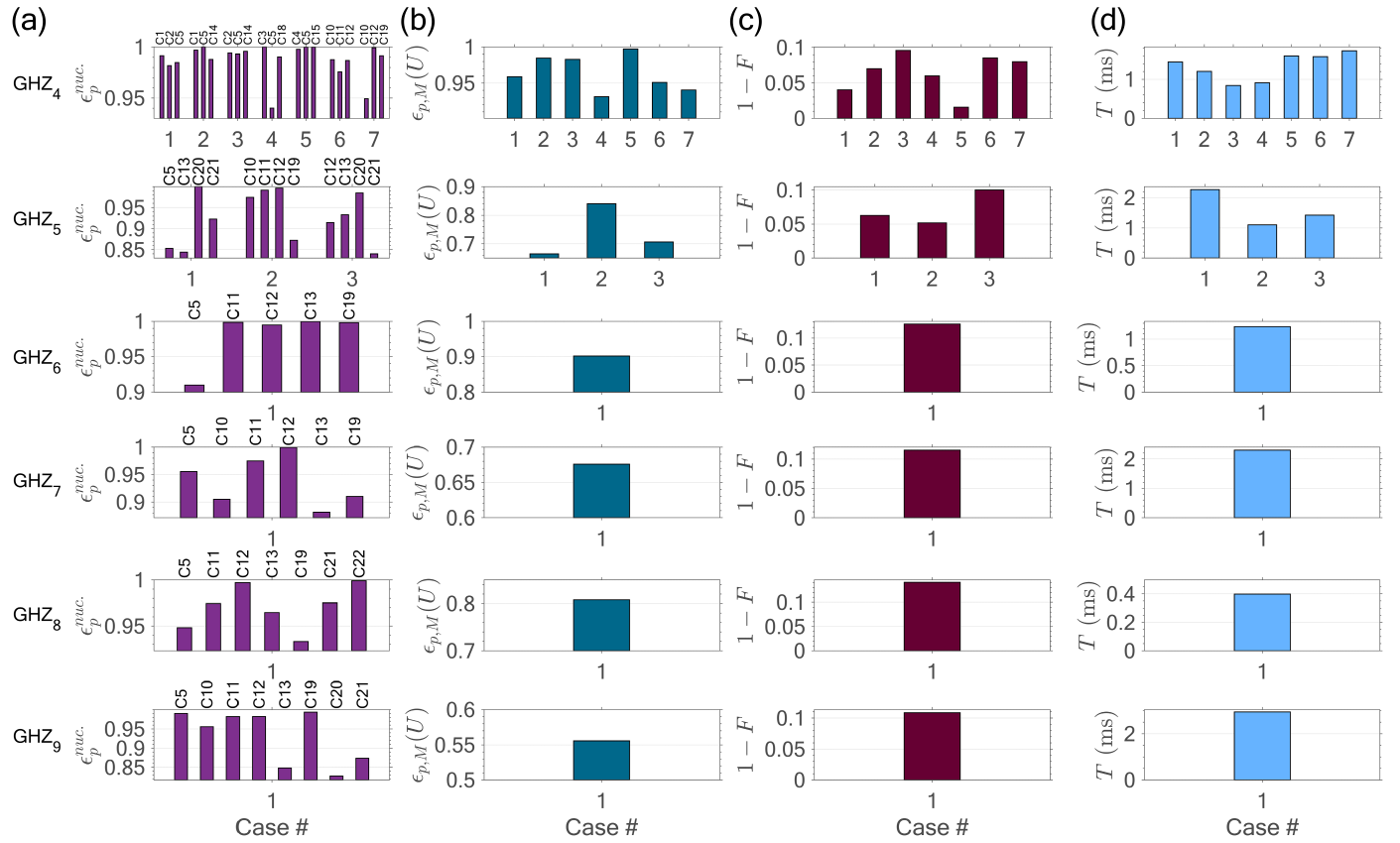}
    \caption{Creating GHZ$_M$-like states using the multi-spin protocol. (a) One-tangles of nuclei that participate in the GHZ-like state. (b)  {\color{black}$M$-tangling} power of the single-shot operation. (c) Gate error due to residual entanglement with unwanted nuclei from the register. (d) Gate time of the single-shot operation.}
    \label{fig:GHZM_Multispin}
\end{figure*}

We also study the possibility of creating GHZ$_M$-like states for $M> 3$ via the multi-spin scheme. In principle, due to the more constrained optimization that requires the simultaneous maximization of $M-1$ nuclear one-tangles, we expect that we will find fewer acceptable cases that respect the bounds we set for target/unwanted one-tangles. In Fig.~\ref{fig:GHZM_Multispin}(a), we show the nuclear one-tangles for various cases of single-shot entangling operations, which can create up to GHZ$_9$-like states. Figure~\ref{fig:GHZM_Multispin}(b) shows the  {\color{black}$M$-tangling} power of the single-shot gate, whereas Fig.~\ref{fig:GHZM_Multispin}(c) shows the gate error. We note that the infidelity can in principle be less than 0.1, but the  {\color{black}$M$-tangling} power is not high enough in all cases, due to imperfect entanglement between target nuclei and the electron.

Figure~\ref{fig:GHZM_Multispin}(d) shows the gate time of the multi-spin operation. Once again, the gates are significantly faster compared to the sequential scheme. While for the sequential protocol increasing the number of parties of the GHZ$_M$-like states implies durations of the total gate that only increase, the multi-spin operations follow a different trend. To understand this, note that the sequential scheme requires higher-order resonances (i.e., longer unit times) to suppress unintended couplings with unwanted nuclei. Higher-order resonances are also needed to ensure that only one target nuclear spin is rotated conditionally by each of the $M-1$ sequential gates. Thus, increasing the GHZ size makes the requirement to resort to higher-order sequences, in principle, more stringent. On the other hand, increasing the size of the GHZ$_M$-like state in the multi-spin protocol means that we need to gradually \textit{reduce} how well we decouple the electron from the spin bath and allow interactions of the electron with increasingly more nuclear spins. In contrast to the sequential scheme, we thus need to shorten the unit time, which gives us acceptable cases of realizing GHZ$_M$-like states. Of course, the duration of the single-shot gate also depends on the number of times, $N$, we repeat the  unit and on the constraints we impose on the target all-way correlations and unwanted nuclear one-tangles. In principle, a shorter basic unit can lead to faster operations. For example, it is remarkable that multi-partite entangled states can be prepared within 1 ms or even faster with only one entangling gate (see for example, the case of the GHZ$_8$-like state). Experimentally, depending on the nuclear HF parameters, this method could be highly beneficial for preparing entangled states involving nuclear spin clusters fast, whereas the entanglement could be boosted using distillation protocols~\cite{KalbSci2017}.

Overall, both the sequential and multi-spin protocols can generate high-quality GHZ$_M$-like states with reasonable gate times. Our formalism based on the  {\color{black}$M$-tangling} power allows us to identify optimal scenarios of preparing entangled states with minimal cross-talk. Interestingly, a hybrid entanglement generation scheme involving both single-shot and sequential gates could offer a more realistic path to scalability by drastically reducing the number of entangling operations and gate times.

\section{Entanglement of mixed states \label{Sec:Mixed_States}}

So far we have made no assumption about the initial state of the system and focused instead on the capability of the gates to produce entangled states. We have used the gate error due to residual entanglement links and the unwanted nuclear one-tangles as a metric of mixedness in the GHZ$_M$-like states prepared by DD sequences. We now work with a density matrix for the total system and inspect the entanglement of a target subspace after we trace out the unwanted nuclear spins. This partial trace operation will result, in general, in a mixed state for the target subspace, for which we cannot  use directly the $M$-tangles that are defined for pure states. 

In the case of mixed states, the $M$-tangles are calculated via so-called convex roof constructions~\cite{Coffman2000,OsterlohPRA2008}:
\begin{equation}\label{Eq:tangle_mixed}
    \tau_M(\rho)= \text{min} \sum_i p_i \tau(|\psi_i\rangle),
\end{equation}
where the minimum is taken over all ensembles with probabilities $p_i$ and pure states $|\psi_i\rangle$ that reconstruct the mixed state density operator. The ensemble that gives the minimum value of the tangle is known as optimal. Such convex roof extensions are necessary to ensure an entanglement monotone, but since the minimum of a sum of convex functions is not always convex, we need to take the convex hull of Eq.~(\ref{Eq:tangle_mixed}) (see also Refs.~\cite{LohmayerPRL2006,OsterlohPRA2008}).

Finding the optimal ensemble is a non-trivial task as it entails searching over all possible decompositions of the mixed state. This task becomes even harder when the rank of the reduced density matrix is large. However, for our problem, we find that starting from an arbitrary pure state of the total system and tracing out any number of nuclei, the maximum rank of the reduced density matrix is 2. This is due to the form of the total evolution operator, which can produce at most GHZ$_M$-like states (if extra single-qubit gates are not allowed) given the appropriate initial state. Further, the maximum rank proves that the creation of $|W\rangle$-like states is impossible if one allows only entangling gates generated by DD sequences. In Appendix~\ref{App:MixedState}
we derive the eigenvectors and eigenvalues of the density matrix, of which two are nonzero, provided the electron starts from a superposition of $|0\rangle$ and $|1\rangle$. 

The analytical expression of the eigenvalues and eigenvectors of the reduced density matrix of the target subspace allows us to find the optimal ensemble that minimizes the $M$-tangle of the mixed state. This can be done using the methods developed in Ref.~\cite{OsterlohPRA2008}. The first step is to diagonalize the reduced density matrix of the system. Based on the eigendecomposition we can then construct the trial state:
\begin{equation}\label{Eq:Trial_State}
    |\psi_{\text{trial}}\rangle=\sqrt{\lambda_+} |v_+\rangle -e^{i\chi}\sqrt{1-\lambda_+}|v_-\rangle,
\end{equation}
where $|v_\pm\rangle$ are the eigenvectors of the rank-2 mixed state and $\lambda_{\pm}$ the two nonzero eigenvalues. Since any ensemble can be obtained by acting with unitaries on the diagonalized reduced density matrix, minimizing the entanglement of the trial state by varying the angle $\chi$ implies that we obtain the entanglement of the optimal ensemble that reconstructs the mixed state. The angle $\chi$ controls the relative phase of the eigenvectors. 

Our system consists of 28 qubits, so simulating the total density matrix is computationally hard. The fact that we obtain the reduced density matrix analytically [see Appendix~\ref{App:MixedState}] allows us to find the impact of the unwanted spins on the target subspace. Working with a density matrix necessitates that we specify the system's  initial state. In the literature, it is often assumed that the nuclear spin bath starts from the maximally mixed state. This assumption, however, would result in maximal entanglement in the target subspace (provided we perform close to perfect gates and once the electron-nuclear target register is purified) free from cross-talk, since the maximally mixed state of the unwanted spins would remain invariant under any unitary evolution. This is far from true since the target subspace suffers from cross-talk from unwanted nuclei introduced by the entangling gates. On the other hand, considering a pure state for the entire electron-nuclear register is again unrealistic, but we choose this convention to find how classical correlations due to the partial trace operation manifest in the target subspace. 

For the following analysis, we will use the  {\color{black} 39} cases we found for creating GHZ$_3$-like states (generalization to larger dimension of the target subspace is straightforward) from Sec.~\ref{Sec:GHZprotocols} via the sequential scheme. We further use the analytical expression of the three-tangle, first introduced in Ref.~\cite{Coffman2000}. To prepare an entangled state with genuine all-way correlations, we need to find the appropriate initial state for the target subspace, upon which we act with the entangling operations. In most cases, this state needs to be $|+\rangle |0\rangle^{\otimes M-1}$. (Recall that the $\text{CR}_x(\pi/2)$ gates are locally equivalent to CNOT.) If this initial state does not give rise to a three-tangle of at least 0.95, we then optimize over the initial three-qubit state (we assume a sampling of $0.05\pi$ for $\theta\in[0,\pi]$ and $0.1\pi$ for $\gamma\in[0,2\pi]$ for each qubit state $|\xi\rangle=\cos(\theta/2)|0\rangle+e^{i\gamma}\sin(\theta/2)|1\rangle$). It is thus possible that other initial states might give slightly larger three-tangles than the ones we will show later on. We further assume that the initial state of the bath is $|0\rangle^{\otimes 27-(M-1)}$.

We begin with the following setup. In Sec.~\ref{Sec:GHZprotocols}, we found {\color{black} 39} cases of preparing GHZ$_3$-like states using the sequential scheme, via the composition of two consecutive DD sequences. The composite evolution rotates both target and  {\color{black}unwanted} nuclei. Some  {\color{black}unwanted} spins might evolve slightly conditionally on the electron. Thus, we need to update the initial state of all spins under the composite evolution. Using the analytical expression of the reduced density matrix, we obtain its eigenvalues and eigenvectors and arrange the {\color{black} 39} cases in increasing value of the largest eigenvalue, $p=\lambda_+$.

\begin{figure}[!htbp]
    \centering
    \includegraphics[scale=0.7]{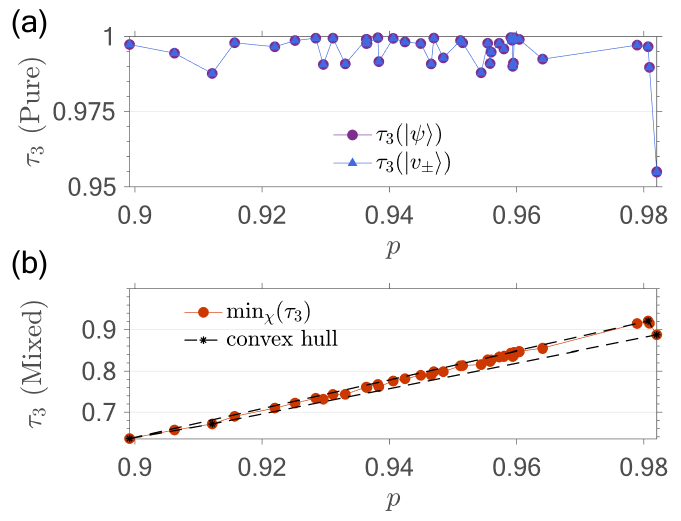}
    \caption{(a) Three-tangle of the pure state of the target subspace ignoring unwanted nuclei (purple), and three-tangle of either eigenvector (blue) of the mixed state for each of the {\color{black} 39} cases of Fig.~\ref{fig:GHZ3_Sequential}. The  {\color{black} 39} cases are sorted in terms of largest eigenvalue, $\lambda_+=p$, shown along the $x$-axis. (b) Three-tangle of the mixed electron-nuclear state (red), obtained by minimizing the entanglement of the trial state
    of Eq.~(\ref{Eq:Trial_State}) for each value of $p$. The dashed line shows the convex hull.}
    \label{fig:Three_Tangle_Mixed}
\end{figure}

In Fig.~\ref{fig:Three_Tangle_Mixed}(a), we show the three-tangle of the pure state prepared using the composite evolutions of Fig.~\ref{fig:GHZ3_Sequential} by ignoring the presence of unwanted nuclei. Using the three-tangle, we verify that all-way correlations are maximal across all cases, as expected based on the fact that we approximately saturated the  {\color{black}$M$-tangling} power (and we also use the appropriate three-qubit initial state). We further show the three-tangle of the two eigenvectors, $|v_\pm\rangle$, as a function of the largest eigenvalue $p$, obtained by diagonalizing the mixed state (after tracing out unwanted nuclei). The three-tangle of $|v_+\rangle$ coincides with the three-tangle of $|v_-\rangle$, which means that the unwanted bath creates a mixture of two terms, which both have the same $M$-way entanglement and are GHZ$_3$-like states. Another feature we observe is that the all-way correlations of the eigenvectors $|v_\pm\rangle$ coincide with those of the pure GHZ$_3$-like state of the target electron-nuclear register. Hence, tracing out the unwanted spins makes the reduced state classically correlated, but the amount of entanglement in each term of the mixed state is unaffected.

In Fig.~\ref{fig:Three_Tangle_Mixed}(b), we then use the two eigenvectors to build the trial state and find the minimum of the three-tangle as we vary $\chi$ for each value of $p$. Since we are tracing out 25  {\color{black}unwanted} nuclei, we see that the entanglement of the mixed state reduces substantially. One simple way to understand this feature is to think of a trial (pure) two-qubit state which is a superposition of two Bell states. When we have the equal superposition, the two-qubit concurrence goes to 0, whereas for different weights between the two terms, the concurrence is $0<C(|\psi\rangle)\leq 1$. A similar thing happens here, with the difference that the superposition terms are two GHZ$_3$-like states. In Appendix~\ref{App:C_VS_Tau3}, we show that the three-tangle of the superposition of two orthogonal GHZ$_3$ states is much more sensitive compared to the concurrence of the superposition of two orthogonal Bell states for the two-qubit case. Even for large values of $p$ (i.e. $p=0.9$) the minimum three-tangle is around $\sim 0.64$ for the GHZ$_3$ mixture compared to the minimum concurrence which is $\sim 0.8$ for the Bell mixture. In Fig.~\ref{fig:Three_Tangle_Mixed}(b), we further show the convex hull of the minimum value of the three-tangle. In the scenario we are studying, the minimum value of the three-tangle is not convex since the pure state of the target subspace does not have perfect correlations due to slightly imperfect gates (i.e., gates that cause over- or under-rotation of the nuclei). 
If the gates create perfect all-way entanglement then both eigenvectors are perfect GHZ$_3$-like states, and the minimum of the three-tangle is convex in terms of $p$, and is given by $\tau_{\text{min}}(p)=(1-2p)^2$ (see Appendix~\ref{App:C_VS_Tau3}). 

We should further comment that our results depend highly on the initial state of the system. Since the evolution is conditional on the electron, if the initial state of the unwanted spins is not the appropriate one to generate the maximum possible entanglement, cross-talk will be suppressed. Also, as the system's initial state is generically mixed, the rank of the reduced density matrix after tracing out  {\color{black}unwanted} nuclei will be larger than 2. However, the simple model we assumed in this section captures qualitatively how the $M$-way entanglement changes as the target subspace becomes more mixed, as a result of residual entanglement with unwanted nuclei. 

In the next section, we show another way of understanding the unwanted residual entanglement in terms of the non-unitary entangling power of the entire quantum channel, where now the partial trace operation is translated into the operator-sum representation~\cite{nielsen_chuang_2010}.

\section{Non-unitary entangling power \label{Sec:Non_Uni_Ep}}

An alternative way to include the impact of unwanted nuclei on the target subspace is to use the Kraus operators associated with the partial trace channel. In Ref.~\cite{EconomouPRX2023}, we derived the Kraus operators for an arbitrary number of nuclei coupled to a single electron spin. 
Here, we use this result to derive the  {\color{black}$M$-tangling} power of the evolution that the target system undergoes due to the presence of the unwanted nuclei. Because this is a generalization for non-unitary dynamics, we will refer to this metric for brevity as the non-unitary $M$-tangling power.

The density matrix of the total system evolves under the unitary $U$ given by Eq.~(\ref{Eq:U}). If we trace out the unwanted nuclei, the target subspace evolves under the quantum channel $\mathcal{E}(\rho)=\sum_{k=0}^{2^{L-K}-1} E_k\rho_0 E_k^\dagger$, where $E_k$ are the Kraus operators corresponding to {\color{black}unwanted} nuclei, $L$ is the number of total nuclei in the bath, and $K$ is the number of target nuclear spins. In this case, the {\color{black}$M$-tangling} power (where $M=K+1$) of the quantum channel reads~\cite{KongPRA2015}:
\begin{equation}
\resizebox{0.98\hsize}{!}{$
    \epsilon_{p,M}(\mathcal{E}):=2^M \sum_{r,s=0}^{2^{L-K}-1}\text{Tr}[(E_r\otimes E_s)\Omega_{p0}(E_r\otimes E_s)^\dagger \tilde{P}],$}
\end{equation}
with the same definitions for $\Omega_{p0}$ and $\tilde{P}$ we introduced in Sec.~\ref{Sec:MwayEP}. Since the total evolution is controlled on the electron, it is possible to derive a closed-form expression for the  {\color{black}$M$-tangling} power of the channel, which reads (see Appendix~\ref{App:Non_Uni_Ep}):
{\color{black}
\begin{equation}\label{Eq:Non_uni_Mway_Ep}
    \epsilon_{p,M}(\mathcal{E})=\frac{\epsilon_{p,M}(U)}{2} \Bigg(1+
    \prod_{\substack{j\in  \\ \text{unw. nuc.}} }\Bigg\{G_1^{(j)}+W^{(j)}\Bigg\}\Bigg)
\end{equation}
with $W^{(j)}$ given by:
\begin{equation}
\begin{split}
  W^{(j)}&=\Big(n_{z,0}^{(j)}\cos\frac{\phi_1^{(j)}}{2}\sin\frac{\phi_0^{(j)}}{2}-n_{z,1}^{(j)}\cos\frac{\phi_0^{(j)}}{2}\sin\frac{\phi_1^{(j)}}{2}
    \\&
    -(n_{x,1}^{(j)}n_{y,0}^{(j)}-n_{x,0}^{(j)}n_{y,1}^{(j)})\sin\frac{\phi_0^{(j)}}{2}\sin\frac{\phi_1^{(j)}}{2}\Big)^2  
\end{split}    
\end{equation}}
We find that the  {\color{black}$M$-tangling} power of the channel is upper-bounded by the unitary  {\color{black}$M$-tangling} power of the target space, i.e.  $\epsilon_{p_M}(\mathcal{E})\leq \epsilon_{p,M}(U)$. The equality is satisfied when the unwanted subspace evolves trivially (i.e., irrespective of the electron's state), in which case it holds that {\color{black} $G_1^{(j)}=1$ and $W^{(j)}=0$, $\forall j$}. Therefore, whenever $\epsilon_{p,M}(\mathcal{E})<\epsilon_{p,M}(U)$, we know that the unwanted spin bath possesses nonzero correlations with the target subspace. This is an alternative way to see that the entanglement of the mixed state is lower compared to the pure case, a feature we already observed in Sec.~\ref{Sec:Mixed_States}.

\begin{figure}[!htbp]
    \centering
    \includegraphics[scale=0.74]{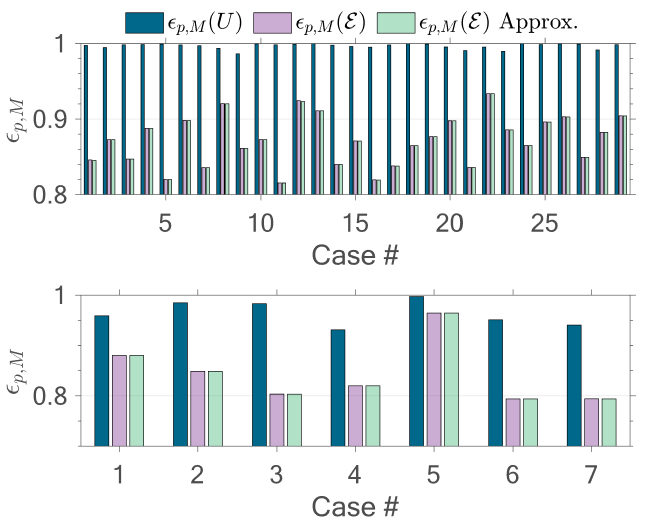}
    \caption{(a) Comparison of the unitary, $\epsilon_{p,M}(U)$, with the non-unitary, $\epsilon_{p,M}(\mathcal{E})$, {\color{black}$M$-tangling} power for the  {\color{black}29} cases of Fig.~\ref{fig:GHZM_Sequential} that prepare GHZ$_4$-like states in the target subspace, via the sequential protocol. (b) Comparison of $\epsilon_{p,M}(U)$ with $\epsilon_{p,M}(\mathcal{E})$ for the  {\color{black} 7} cases of Fig.~\ref{fig:GHZM_Multispin} for preparing GHZ$_4$-like states via the multi-spin protocol. {\color{black}In both panels, the dark blue bars show the unitary  {\color{black}$M$-tangling} power, the pink bars the non-unitary  {\color{black}$M$-tangling} power, whereas the green bars show an approximation of the latter.} }
    \label{fig:Non_Uni_Ep}
\end{figure}

The remarkable simplicity of Eq.~(\ref{Eq:Non_uni_Mway_Ep}) allows us
to obtain full information of the entanglement distribution within the electron-nuclear register, irrespective of the total number of qubits in the system, or number of nuclear spins we are tracing out. {\color{black} Using  $\epsilon_{p,M}(\mathcal{E})$, we can find how well the gate we design saturates the $M$-way correlations of the target subspace, given the fact that there might be residual unwanted correlations linking the target subspace with unwanted spins; those unwanted correlations are encoded in the Kraus operators. An extra advantage of this approach is that we never need to assume an initial state of the electron-nuclear register, as we did for example in Sec.~\ref{Sec:Mixed_States}.} To compare the non-unitary  {\color{black}$M$-tangling} power against $\epsilon_{p,M}(U)$, we consider as an example the optimal cases we identified for preparing GHZ$_4$-like states from Sec.~\ref{Sec:GHZprotocols} for either the sequential or the multi-spin schemes. The results for the sequential scheme are shown in Fig.~\ref{fig:Non_Uni_Ep}(a). The  {\color{black} dark blue} bars show the unitary  {\color{black}$M$-tangling} power across  {\color{black}29} realizations, and the  {\color{black}pink} bars show the non-unitary  {\color{black}$M$-tangling} power. As expected, {\color{black}$\epsilon_{p,M}(\mathcal{E})$} is lower in all cases {\color{black}due to unresolved crosstalk}, and this deviation is enhanced when the gate error is larger [see Fig.~\ref{fig:GHZM_Sequential}(b)]. A similar feature is observed in Fig.~\ref{fig:Non_Uni_Ep}(b) for the  {\color{black}7} cases of preparing GHZ$_4$-like states using the multi-spin scheme. Specifically, for case  {\color{black}``5''}, for which we found the lowest gate error emerging from residual entanglement across the  {\color{black}7} cases [see Fig.~\ref{fig:GHZM_Multispin}(b)], the non-unitary  {\color{black}$M$-tangling} power is closer to the unitary one. Hence, this verifies our previous observation that the mixed electron-nuclear states we create in the target subspace remain GHZ$_M$-like, but their entanglement is reduced whenever correlations between the target subspace and the unwanted spin bath are present. 

{\color{black} In Fig.~\ref{fig:Non_Uni_Ep}, we also display with green bars an approximate expression for $\epsilon_{p,M}(\mathcal{E})$, where we set $W^{(j)}=0, \forall j$.  We note that the approximate expression agrees very well with the exact expression of $\epsilon_{p,M}(\mathcal{E})$.
}

Finally, let us discuss the computational complexity related to the various entanglement metrics we introduced. The calculation of the $G_1$ Makhlin invariants and hence of target/unwanted one-tangles scales linearly with the number of nuclear spins, since each $G_1$ quantity describes the evolution of a single nuclear spin. Thus, the  {\color{black}$M$-tangling} power, $\epsilon_{p,M}(U)$, can be calculated without computational difficulty. {\color{black} Similarly, $\epsilon_{p,M}(\mathcal{E})$ scales linearly with the size of the total size of the register (including target and unwanted nuclei), and can also be calculated without difficulty.} The infidelity of the target gate due to the presence of unwanted spins requires a summation over $2^{L-K}$ Kraus operators. Therefore, the starting point to identify optimal cases for preparing GHZ$_M$-like states is to calculate the target/unwanted one-tangles in order to maximize (minimize) target (unwanted) correlations. To further inspect the optimal cases in terms of gate error, one can then proceed with  {\color{black}this metric} to obtain full information about the dynamics of the system. 

{\color{black}

\section{Additional error analysis \label{Sec:Additional_Errors}}

In this section, we study the effect of pulse control errors and dephasing errors that can occur on the electronic qubit. To find the impact of those errors, we consider only their contribution to the $M$-tangling power of the target subspace.    

\subsection{Target subspace \texorpdfstring{$M$}{M}-way entanglement under electronic dephasing}

\begin{figure*}[!htbp]
    \centering
    \includegraphics[scale=0.9]{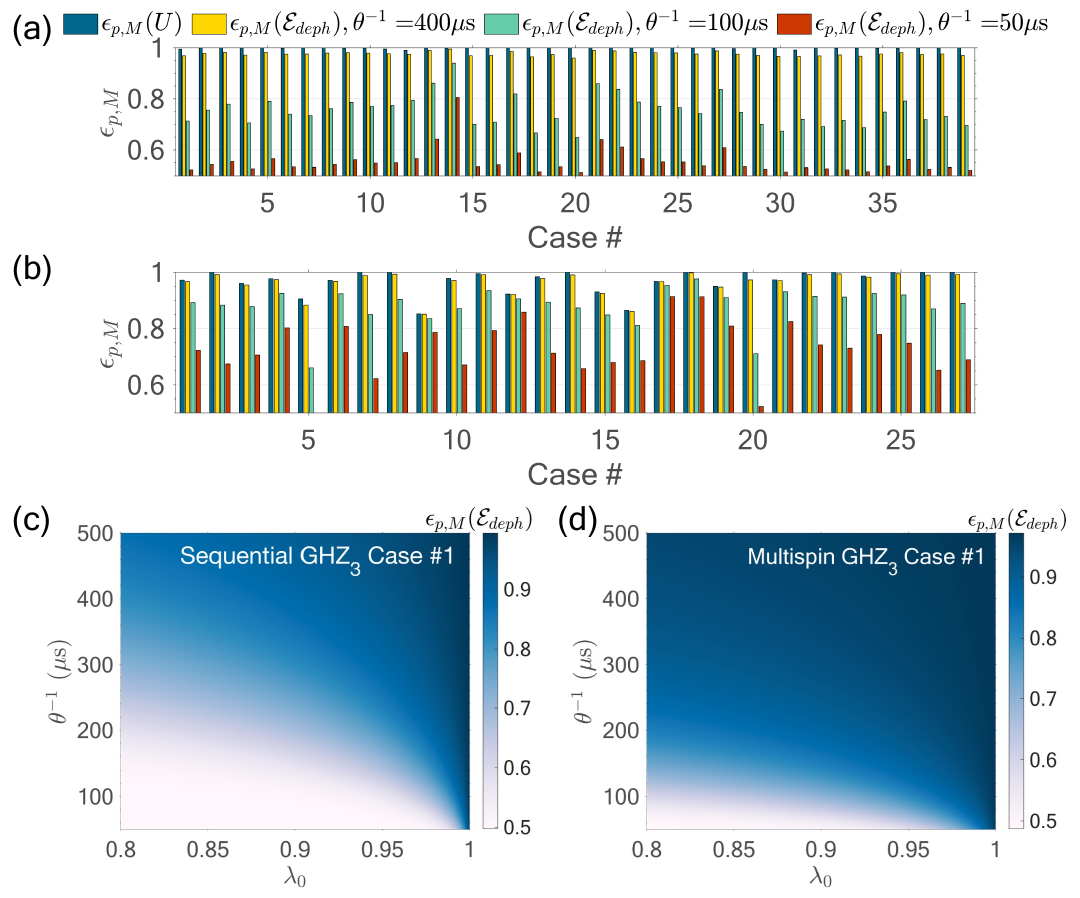}
    \caption{{\color{black}$M$-tangling power of the target subspace when the electronic qubit experiences dephasing. (a) $M-$tangling power for the 39 cases of preparing GHZ$_3$-like states with the sequential protocol. (b) $M$-tangling power of the 27 cases of preparing GHZ$_3$-like states with the multi-spin scheme.  The darker blue bars in (a) and (b) correspond to the no-dephasing scenario, whereas the remaining colors include electronic dephasing with dephasing angles $\theta^{-1}=(400,100,50)~\mu$s. For the cases where we consider dephasing, we set $\lambda_0=0.98$. (c) $M$-tangling power for case $\#$ 1 of the sequential scheme for different dephasing angles and $\lambda_0$ parameters. (d) Same as in (c) but for case $\#$ 1 of the multi-spin scheme.}}
    \label{fig:Deph}
\end{figure*}

We begin by considering dephasing errors for the electronic qubit. The dephasing error can be represented via the Kraus operators~\cite{MazziottiPRA2022}:
\begin{equation}
    K_0 =\sqrt{\lambda_0}\begin{pmatrix}
        e^{i\theta t} & 0 \\
        0 & e^{-i\theta t}
    \end{pmatrix}
\end{equation}
\begin{equation}
    K_1=\sqrt{\lambda_1}\begin{pmatrix}
        e^{-i\theta t} & 0 \\
        0 & e^{i\theta t}
    \end{pmatrix}.
\end{equation}

In Appendix~\ref{App:Dephased_M_way_EP}, we prove that the ``dephased'' $M$-tangling power of the target subspace (consisting of the electron and target nuclei) has a simple closed-form expression. In particular, for a CPMG decoupling sequence of unit time, $t$, and of $N$ iterations, it takes the form:
\begin{equation}
    \epsilon_{p,M}(\mathcal{E}_{\text{deph}})=\Big(\frac{2}{3}\Big)^M \frac{1+\mathcal{R}}{2}\prod_{l=2}^{M}(1-G_1^{(l-1)}),
\end{equation}
where $\mathcal{R}$ is given by:
\begin{equation}
\begin{split}
   \mathcal{R}&=\left(\lambda_0^2+\lambda_1^2+2\lambda_0\lambda_1\cos(\theta t)\right)^{2N}\times
   \\&
   ~~~~\left(\lambda_0^2+\lambda_1^2+2\lambda_0\lambda_1\cos(2\theta t)\right)^N.
   \end{split}
\end{equation}
In the case where we compose CPMG sequences of different unit times $t_j$ and iterations $N_j$ (as we do, for example, in the sequential scheme), the function $\mathcal{R}$ takes the form:
\begin{equation}
\begin{split}
    \mathcal{R}&=\prod_j\left(\lambda_0^2+\lambda_1^2+2\lambda_0\lambda_1\cos(\theta t_j)\right)^{2 N_j}~\times
    \\&~~~~~~~~\left(\lambda_0^2+\lambda_1^2+2\lambda_0\lambda_1\cos(2\theta t_j)\right)^{N_j}.
    \end{split}
\end{equation}
We verify that in the limit of $\lambda_0=1$ and $\lambda_1=0$, we recover the ideal $\epsilon_{p,M}(U)$ for the target subspace. We further find that the $M$-tangling power in the presence of electronic dephasing errors is bounded from below by $50\%$. 

We begin by studying the impact of the electronic dephasing for the 39 cases of preparing GHZ$_3$-like states via the sequential scheme. The results are shown in Fig.~\ref{fig:Deph}(a), where we consider $\lambda_0=0.98$, $\lambda_1=1-\lambda_0$, and  dephasing angles ranging from $\theta^{-1}=400~\mu$s to $\theta^{-1}=50~\mu$s. We see that for $\theta^{-1}=400~\mu$s, we preserve high $M$-tangling power over 90$\%$ for most realizations. For a dephasing angle $\theta^{-1}=100~\mu$s, we find several cases with $M$-tangling power around 80-85$\%$. For a dephasing angle $\theta^{-1}=50~\mu$s, the $M$-way correlations can drop to less than $60\%$ across many cases. 

In Fig.~\ref{fig:Deph}(b), we repeat the calculations for the 27 cases of preparing GHZ$_3$-like states using the multi-spin protocol. The $M$-tangling power of the multi-spin protocol remains higher in the presence of electronic dephasing compared to the sequential scheme. We even observe that for the smallest dephasing angle of $\theta^{-1}=50~\mu$s, we have several cases which exceed 80$\%$ $M$-way correlations. This behavior is expected since the multi-spin scheme is, in principle, faster than the sequential protocol, and hence it is less sensitive to electronic dephasing errors.

Figure~\ref{fig:Deph}(c) shows the $M$-tangling power of case ``1'' of the sequential scheme, as a function of $\lambda_0$ and $\theta^{-1}$, whereas Fig.~\ref{fig:Deph}(d) shows the $M$-tangling power of case ``1'' of the multi-spin scheme. In the sequential scheme, we control the nuclei C1 and C18 (with a gate time $\sim 1843.2~\mu$s), whereas, in the multi-spin protocol, we control the nuclei C1 and C2 (with a gate time $\sim 1366.5~\mu$s). Although the results of Fig.~\ref{fig:Deph}(c) and Fig.~\ref{fig:Deph}(d) should not be compared directly due to the different total timings, we note that the multi-spin scheme is more robust to dephasing errors. In Appendix~\ref{App:Seq_vs_Multi_Dephasing}, we provide another example where we compare the performance of the multi-spin scheme with the sequential, but now for case ``16''. Although case ``7'' of the multi-spin scheme has a total sequence time longer than the sequence time of the sequential scheme, we find again that the multi-spin scheme is more robust to dephasing. We attribute this feature to the more complicated dephasing channel of the sequential scheme. The expression of $\epsilon_{p,M}(\mathcal{E}_{\text{deph}})$ for the sequential scheme is described by different unit times $t_j$ and iterations $N_j$ since we are composing CPMG sequences to control each nucleus. We believe that this feature can potentially lead to $M$-way correlations that are more sensitive to electronic dephasing errors.

\subsection{Target subspace \texorpdfstring{$M$}{M}-way entanglement under pulse errors \label{Sec:Additional_Errors_Pulse_Errors}}

\begin{figure*}[!htbp]
    \centering
    \includegraphics[scale=0.8]{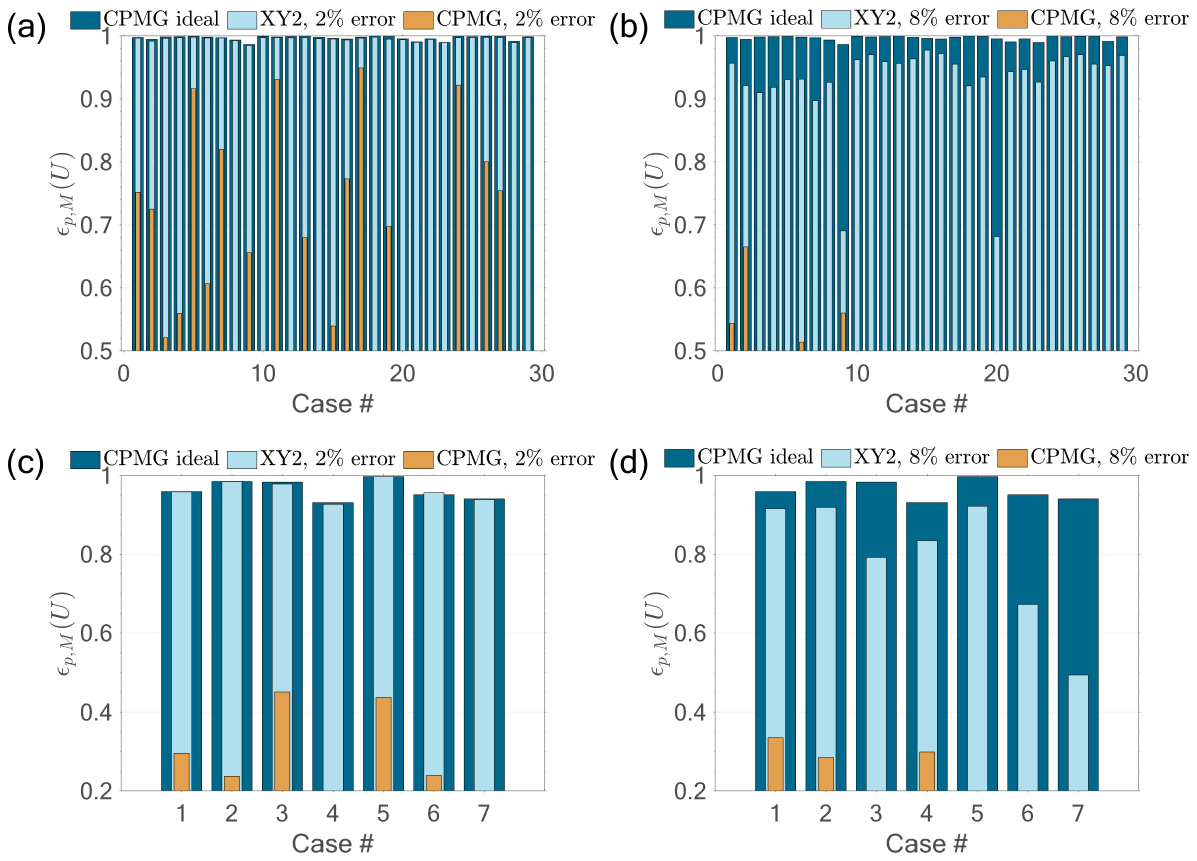}
    \caption{{\color{black}Generation of GHZ$_4$-like states in the presence of systematic over-rotation pulse errors. (a), (b) $M$-tangling power using the 29 cases we found for preparing GHZ$_4$-like states via the sequential scheme, assuming a systematic error of $2\%$ in (a), and of $8\%$ in (b).  (c), (d) $M$-tangling power for the 7 cases of preparing GHZ$_4$-like states using the multi-spin scheme assuming a $2\%$ error in (c), and an $8\%$ error in (d). In all panels, the bars with the highest value of $\epsilon_{p,M}(U)$ correspond to the error-free case. The orange bars show the CPMG including pulse errors, whereas the light blue bars show the XY2 including pulse errors. All bars are scaled by the maximal value of $(2/3)^4$.}}
    \label{fig:SystematicPulseErrors}
\end{figure*}

We now consider the effect of pulse errors during the control of the electronic spin. We assume that the pulse error results in an over- or under-rotation along the $\hat{\textbf{x}}$-axis of the electron. We model such errors by modifying the perfect $R_x(\pi)$ acting on the electron into $R_x(\pi+\epsilon)=e^{-i\pi/2(1+\epsilon) \sigma_x}.$ Such rotation angle errors yield an evolution operator which is no longer block-diagonal. Consequently, we cannot use the $M$-tangling power we found in Eq.~(\ref{Eq:MwayEP}). Additionally, we cannot study the $M$-tangling power for an odd-sized system since, as we mentioned in Sec.~\ref{Sec:MwayEP}, the expression of Eq.~(\ref{Eq:MwayEp_of_any_U}) is only applicable to CR-type evolution operators. This difficulty in describing the $M$-way correlations on the evolution operator level for odd $M$ arises from the fact that  averaging over all initial states is not straightforward due to the expression of the odd $M$-tangle we start with. Nevertheless, by modeling the rotation angle error in this way, we still preserve unitary dynamics, and we can find the impact of such errors on the $M$-tangling power of even-sized systems numerically by making use of Eq.~(\ref{Eq:MwayEp_of_any_U}).

First, we consider systematic errors and study their impact on the cases of preparing GHZ$_4$-like states via the sequential or multi-spin protocols. In Fig.~\ref{fig:SystematicPulseErrors}(a) and Fig.~\ref{fig:SystematicPulseErrors}(b), we show the $M$-tangling power for the 29 cases of preparing GHZ$_4$-states, assuming a systematic over-rotation of $\epsilon=2\%$ and $\epsilon=8\%$ respectively. The dark blue bars depict the error-free case. The light blue bars correspond to the XY2 sequence, whereas the orange bars to the CPMG sequence, including the over-rotation errors. In the case of $2\%$ error, we see that CPMG can provide high values of $M$-way correlations only for a few cases (e.g., case ``5'', case ``11'', case ``17'', case ``24''). However, if we consider the XY2 sequence whose one unit is $t/4 - (\pi)_X - t/2 -(\pi)_Y-t/4$, we find that the impact of the error on the quality of the GHZ states is negligible. For an over-rotation error of $8\%$, the $M$-way correlations accumulated via the CPMG sequence are lower than $50\%$, whereas the XY2 sequence still provides high entanglement across almost all cases (all cases besides case ``9'' and case ``20''). 

We observe a similar behavior of systematic over-rotation errors for the multi-spin scheme. We depict the results for the multi-spin protocol in Fig.~\ref{fig:SystematicPulseErrors}(c) and Fig.~\ref{fig:SystematicPulseErrors}(d), where we assume a systematic over-rotation error of $2\%$ and $8\%$, respectively. For $2\%$ error, we find that $\epsilon_{p,M}(U)$ obtained via the XY2 scheme is extremely robust, although the pulse iterations per case are in general high: $N=128$ for case ``1'', $N=76$ for case ``2'', $N=124$ for case ``3'', $N=136$ for case ``4'', $N=78$ for case ``5'', $N=223$ for case ``6'', and $N=243$ for case ``7''. For an $8\%$ error shown in Fig.~\ref{fig:SystematicPulseErrors}(d), case ``6'' and case ``7'' display the most prominent reduction in $M$-way correlations across all cases, which is consistent with the large number of decoupling unit iterations.

\begin{figure*}[!htbp]
    \centering
    \includegraphics[scale=0.66]{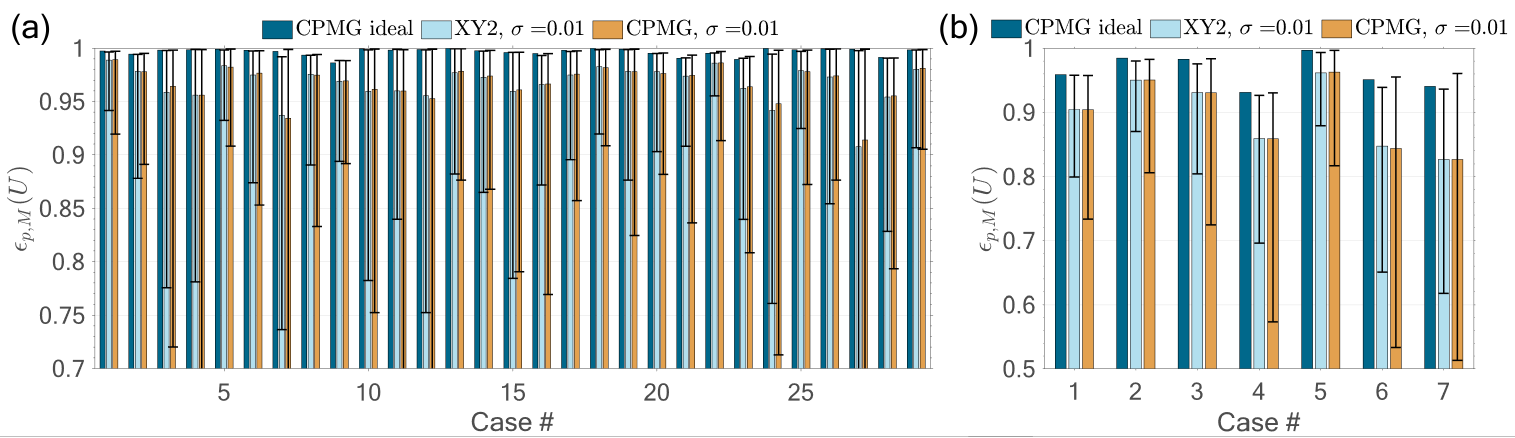}
    \caption{{\color{black}Generation of GHZ$_4$-like states in the presence of random over-/under-rotation pulse errors. (a) Mean $M$-tangling power using the 29 cases we found for preparing GHZ$_4$-like states via the sequential scheme, assuming rotation angle errors sampled from a normal distribution with standard deviation $\sigma=0.01$.  (b) Mean $M$-tangling power for the 7 cases of preparing GHZ$_4$-like states via the multi-spin scheme assuming the same standard deviation. In both (a) and (b), the bars with the highest value of $\epsilon_{p,M}(U)$ correspond to the error-free case. The orange bars correspond to the CPMG case including pulse errors, whereas the light blue bars show the XY2 case including pulse errors. All bars are scaled by the maximal value of $(2/3)^4$. Each error bar for CPMG and XY2 in (a) captures the range of $\epsilon_{p,M}(U)$ where we run 500 trials per case, with different random samplings per sequence iteration. In (b), we perform 1000 random trials per case to display the error bars.}}
    \label{fig:RandomErrors}
\end{figure*}

The robustness of XY-sequences to pulse errors has already been reported extensively in the literature~\cite{Gullion_Journal_Mag_Res1969,DobrovitskiPRB2012,VanDerSarThesis2012,DeLangeThesis2012}. In Ref.~\cite{DeLangeThesis2012}, a quantum process tomography protocol called boostrap was used to determine control pulse errors for the $\pi$ and $\pi/2$ pulses. It was mentioned therein that the $\pi/2$ pulses show twice as much variation than $\pi$-pulse errors indicating that the pulse edges have a larger impact on shorter pulses. The pulse errors were taken as constant during each run and the same for different runs. The reported error values for $\pi$-pulses were $\epsilon_x=\epsilon_y=-0.02$ (rotation errors for $\pi_X$ and $\pi_Y$ pulses), $q_x=0.005$ ($\hat{\textbf{x}}$-axis component of $\pi_Y$ rotation), $q_y=0$ ($\hat{\textbf{y}}$-axis component of $\pi_X$ rotation) and $q_z=\pm 0.05$ ($\hat{\textbf{z}}$-axis component of $\pi_X$ and $\pi_Y$ rotations). More importantly Ramsey, spin-echo, XY4, CPMG, and UDD simulations with those parameters were in strong agreement with experimental data. The estimated pulse error was $1\%$. In Ref.~\cite{VanDerSarThesis2012} XY-sequences were used to cancel out the systematic pulse errors, suppress decoherence of the electron, as well as suppress artificially injected magnetic noise, which was introduced in the same stripline that was used for the control pulses. The fidelity of a single $\pi$-pulse on the electron was again estimated to be 99$\%$ using calibration techniques from Ref.~\cite{deLangeSci10,DobrovitskiPRB2012}. Therefore, since the $\pi$-pulse microwave control can typically exceed 99$\%$ for defect platforms, we expect that the $M$-tangling power we evaluate for the $2\%$ systematic rotation error will be close to the experimentally observed one. In Appendix~\ref{App:angle_and_axes_errors}, we  perform another simulation with the aforementioned error parameters and verify that the $M$-tangling power we obtain for the ideal CPMG sequence is close to the one obtained via the XY2 sequence in the presence of pulse imperfections. This reveals that combining our protocols with the XY2 decoupling sequence is sufficient to prepare high-fidelity GHZ$_M$ states in the presence of realistic experimental errors. 

We continue the pulse error analysis, assuming random over-/under-rotation pulse errors, which we sample from a normal distribution with standard deviation $\sigma=0.01$. In Fig.~\ref{fig:RandomErrors}(a), we show the mean $M$-tangling power for the XY2 (light blue bars) and for the CPMG (orange bars), using the 29 cases of preparing GHZ$_4$-like states with the sequential scheme. In each case, we run 500 independent trials where we sample anew from the normal distribution (per iteration of the decoupling unit), and we evaluate the mean $\epsilon_{p,M}(U)$ across the 500 trials. For comparison, we also display the ideal $\epsilon_{p,M}(U)$ per case. CPMG and XY2 perform on par when random pulse errors are present. The error bars mark the range of $\epsilon_{p,M}(U)$ for the 500 trials per case. We observe that for the sequential scheme, the $M$-tangling power in the presence of random pulse errors is higher than 0.7 across all cases.

In Fig.~\ref{fig:RandomErrors}(b), we show the mean $M$-tangling power for the 7 cases of preparing GHZ$_4$-like states using the multi-spin scheme. We sample errors from the normal distribution with standard deviation $\sigma=0.01$ and repeat the calculation 500 times per case to collect statistics and evaluate the mean $M$-tangling power. Once again, we observe that the  XY2 and CPMG sequences perform on par in the presence of random over-/under-rotation errors. The error bars per case correspond to the range of $\epsilon_{p,M}(U)$ for a total of 500 random trials. 

Overall, we find that the generation of high-quality GHZ$_M$ states is possible with the current state-of-the-art experimental control and constraints. Our formalism provides a detailed analysis of the entanglement generation in the presence of errors and allows us to address the problem of designing optimal entangling gates that saturate all-way correlations. 

}
\section{Conclusions}

Genuine multipartite entangled states are an essential component of quantum networks and quantum computing.
Exploiting the full potential of nuclear spins in defect platforms for large-scale applications requires precise and fast entanglement generation. We showed under what conditions decoupling sequences produce gates capable of maximizing all-way correlations and quantified the entanglement capability of the gates through the  {\color{black}$M$-tangling} power. Using this formalism, we guided the selection of sequence parameters and nuclear spin candidates to prepare high-quality GHZ$_M$-like states by appropriate driving of the electron spin. We improved the sequential entanglement generation scheme, pushing the gate time to as low as 4~ms for preparing GHZ-like states of up to 10 spins. We also studied the possibility of direct entanglement generation, which performs on par with the sequential scheme in decoupling capabilities, with the extra advantage that the latter approach drastically reduces the gate count and speeds up the entanglement generation. Further, we studied the entanglement of mixed states, revealing that the $M$-way correlations of a target subspace are sensitive to residual entanglement with the {\color{black}unwanted} nuclei. We introduced a non-unitary $M$-way entanglement metric, which additionally captures correlations between the target subspace and the unwanted nuclei and showed that it is upper-bounded by the unitary  {\color{black}$M$-tangling} power. {\color{black} We derived a non-unitary $M$-tangling power for the target subspace in the presence of electronic dephasing errors, revealing that the multi-spin protocol can have in principle superior performance compared to conventional schemes due to its shorter implementation time. Finally, we studied the impact of pulse errors on the target subspace $M$-tangling power and showed that our protocols combined with XY decoupling sequences can provide high-fidelity preparation of GHZ states in the presence of realistic experimental errors.} Our results pave the way for the systematic and efficient creation of multi-partite entanglement in spin defect systems for quantum information applications.  

\begin{acknowledgments}
The authors would like to thank Chenxu Liu, Yanzhu Chen, Filippos Tzimkas-Dakis and Isabela Gnasso for useful discussions. This work was supported by the NSF (awards 1741656, 2137953, and 2137645) and the Commonwealth Cyber Initiative
(CCI), an investment in the advancement of cyber R\&D, innovation, and workforce development~\cite{cci}.
\end{acknowledgments}

\appendix

\section{Mathematical treatment of \texorpdfstring{$M$}{M}-way entanglement}

\subsection{Formulas of \texorpdfstring{$M$}{M}-tangles expressed in terms of the state vector\label{App:MTangles}}

For any even $M\geq 4$ the $M$-tangle is given by~\cite{Coffman2000,WongPRA2001}:
\begin{equation}
\begin{split}
    \tau_M(|\psi\rangle) :&=|\langle \psi|\sigma_y^{\otimes M}|\psi^*\rangle|^2 \\&= \langle \psi| \sigma_y^{\otimes M}|\psi^*\rangle \langle \psi^*|\sigma_y^{\otimes M}|\psi\rangle \\&=
    \text{Tr}[|\psi\rangle \langle \psi|\sigma_y^{\otimes M}|\psi^*\rangle \langle \psi^*| \sigma_{y}^{\otimes M}]\\&=\text{Tr}[\rho \sigma_{y}^{\otimes M}\rho^* \sigma_{y}^{\otimes M}]\\&=
    2^M\text{Tr}[\rho^{\otimes 2}\prod_{i=1}^M P^{(-)}_{i,i+M}],
    \end{split}
\end{equation}
where in the last line, we have linearized the equation of the $M$-tangle by introducing a second copy and projecting onto the antisymmetric space of subsystems $i$ and $i+M$, defined as $P^{(-)}_{i,i+M}=1/2(\mathds{1}_{2^{2M}\times 2^{2M}}-\text{SWAP}_{i,i+M})$, with $\text{SWAP}_{i,i+M}=
\sum_{\alpha,\beta\in\{0,1\}}|\alpha\rangle_i|\beta\rangle_{i+M}\langle \beta|_i \langle \alpha|_{i+M}$. For $M=3$, the three-tangle can be expressed in the following form~\cite{Coffman2000}:
\begin{equation}
\begin{split}
    \tau_3(|\psi\rangle):&=\tau_{BC|A}-\tau_{A|B}-\tau_{A|C},
    \end{split}
\end{equation}
where $\tau_{BC|A}=2(1-\text{Tr}[(\text{Tr}_{BC}[\rho])^2])$ is the one-tangle of the electron ($A$ is the system of the electron partitioned from the space of the two nuclei represented by systems $B$ and $C$), and $\tau_{A|B}=\text{Tr}[\rho_{AB}\tilde{\rho}_{AB}]-\sqrt{2}\sqrt{(\text{Tr}[\rho_{AB}\tilde{\rho}_{AB}])^2-\text{Tr}[(\rho_{AB}\tilde{\rho}_{AB})^2]}$  (similar definition holds for $\tau_{A|C}$), with $\tilde{\rho}_{AB}=\sigma_y^{\otimes 2}\rho_{AB}^*\sigma_y^{\otimes 2}$. $\tau_{A|B}$ (or $\tau_{A|C}$) is also known as the entanglement of formation and is alternatively given by $\tau_{A|B}=\text{max}[0,\lambda_1-\lambda_2-\lambda_3-\lambda_4]$, where $\lambda_j$ are the square roots of eigenvalues (in decreasing order) of $\rho_{AB}\tilde{\rho}_{AB}$, with $\rho_{AB}=\text{Tr}_C(\rho)$.

\begin{widetext}
An alternative way to write the three-tangle (as in Ref.~\cite{OsterlohPRA2009}) is:

\begin{equation}
\begin{split}
    \tau_3(|\psi\rangle):&=\Big|-(\langle \psi|\mathds{1}\otimes \sigma_y \otimes \sigma_y|\psi^*\rangle)^2
    +(\langle \psi |\sigma_x \otimes \sigma_y\otimes \sigma_y |\psi^*\rangle)^2 + (\langle \psi |\sigma_z\otimes \sigma_y\otimes \sigma_y|\psi^*\rangle)^2\Big|.
    \end{split}
\end{equation}
This is a complicated expression and averaging this quantity to find the  {\color{black}$M$-tangling} power of an arbitrary evolution operator for $M=3$ is a difficult task. However, for CR-type evolution operators we find that it holds $\langle \psi|\mathds{1} \otimes \sigma_y\otimes \sigma_y |\psi^*\rangle=0=\langle \psi|\sigma_z \otimes \sigma_y\otimes \sigma_y |\psi^*\rangle$. Considering  $\langle \psi|\mathds{1} \otimes \sigma_y\otimes \sigma_y |\psi^*\rangle$ we can show that this vanishes for CR-type evolutions:

\begin{equation}
\begin{split}
\langle \psi |\mathds{1}\otimes \sigma_y \otimes \sigma_y |\psi^*\rangle&=\langle \psi_0 |(\text{CR})^\dagger\mathds{1}\otimes \sigma_y \otimes \sigma_y (\text{CR})^* |\psi_0^*\rangle 
\\&=
\langle \psi_0|\sigma_{jj}\otimes_l [R_{\textbf{n}_j}^{(l)}]^\dagger(\mathds{1}\otimes \sigma_y \otimes \sigma_y)\sigma_{kk}\otimes_{l}(R_{\textbf{n}_k}^{(l)})^*|\psi_0^*\rangle\\&=
\langle \psi_0 |\sigma_{jj}\otimes_{l} [R_{\textbf{n}_j}^{(l)}]^\dagger\sigma_y^{(l)}(R_{\textbf{n}_j}^{(l)})^*|\psi_0^*\rangle\propto \langle \psi_{0,B}|\sigma_y|\psi_{0,B}^*\rangle\langle \psi_{0,C}|\sigma_y|\psi_{0,C}^*\rangle=0,
\end{split}
\end{equation}
\end{widetext}
where we have used the fact that $R_{\textbf{n}_j}^\dagger\sigma_y(R_{\textbf{n}_j})^*=\sigma_y$ as well as the property $\langle \phi |A|\phi\rangle =\langle\phi| \sigma_y C|\phi\rangle=\langle \phi|\sigma_y|\phi^*\rangle=0$, where $A$ is an anti-linear operator whose expectation value vanishes for all states $|\phi\rangle \in\mathcal{H}$ and $C$ denotes complex conjugation. Analogously, one can prove that $\langle \psi|\sigma_z\otimes \sigma_y\otimes \sigma_y|\psi^*\rangle=0$. Thus, these results simplify the expression of the three-tangle for states generated by CR-type evolutions, meaning that we can express it as:
\begin{equation}
\begin{split}\label{Eq:three_tangle}
    \tau_3(|\psi\rangle):&=|\langle \psi|\sigma_x\otimes \sigma_y\otimes \sigma_y|\psi^*\rangle|^2
    \\&=
    \text{Tr}[\rho (\sigma_x \otimes \sigma_y\otimes \sigma_y)\rho^*(\sigma_x \otimes \sigma_y\otimes \sigma_y)].
    \end{split}
\end{equation}
We can linearize the above formula by noticing that we can write the three-tangle by extending the Hilbert space into a 6 qubit system:
\begin{equation}\label{Eq:three_tangle_final}
    \tau_3(|\psi\rangle)=2^3\text{Tr}[\rho^{\otimes 2}P^{(+)}_{14}\prod_{i=2}^3 P_{i,i+3}^{(-)}].
\end{equation}
The above expression can be understood as a ``vectorized'' form of Eq.~(\ref{Eq:three_tangle}), since we note that $|\sigma_y\rangle=\text{vec}(\sigma_y)=-i(|01\rangle-|10\rangle)$, which means $P^{(-)}=1/2(\mathds{1}-\text{SWAP})$ between the sectors $i$ and $i+M$ can be represented as $P^{(-)}=\frac{1}{2}|\sigma_y\rangle \langle \sigma_y|$. We also found  that Eq.~(\ref{Eq:three_tangle_final}) can be generalized to odd $M>3$, for arbitrary initial (pure) product states evolved under a CR-type evolution:
\begin{equation}\label{Eq:odd_tangle}
    \tau_M(|\psi\rangle)=2^M\text{Tr}[\rho^{\otimes 2}P_{1,1+M}^{(+)}\prod_{i=2}^M P_{i,i+M}^{(-)}].
\end{equation}
The most general expression for the odd $M$-tangle for states that undergo arbitrary unitary evolution can be found in Ref.~\cite{DafaQInf2012}. We also verified by numerical inspection that Eq.~(\ref{Eq:odd_tangle}) agrees with the definition of the odd tangle of Ref.~\cite{DafaQInf2012} when the evolution is CR-type.

\subsection{Entangling power of multipartite gates using \texorpdfstring{$M$}{M}-way entanglement metrics \label{App:MwayEP}}

We define the {\color{black}$M$-tangling} power as the average of the $M$-tangle $\epsilon_{p,M}=\langle \tau_M(U\rho_0 U^\dagger)\rangle$, where the average is taken over all initial product states $|\psi_0\rangle$ of $M$ qubits, with $\rho_0=\ket{\psi_0}\bra{\psi_0}$. To combine both the even $M$ and odd $M$ cases, we define $\tilde{P}$ as:
\begin{equation}
    \tilde{P}=\begin{cases}
    \prod_{i=1}^M P^{(-)}_{i,i+M}~\text{even $M$} \\
    P_{1,1+M}^{(+)}\prod_{i=2}^M P^{(-)}_{i,i+M}~\text{ odd $M$}.
    \end{cases}
\end{equation}
In order to perform the average we consider the uniform distribution of product states $|\psi_0\rangle$ of $M$ qubits. For example, for a single-qubit state $\cos (\theta/2)|0\rangle +e^{i\phi}\sin(\theta/2)|1\rangle$, the uniform distribution is $P(\theta,\phi)=1/(4\pi)=(\int \sin\theta d\theta d\phi )^{-1}$. 

Averaging over the uniform distribution in $\mathcal{H}_{M^2}$, we find that the {\color{black}$M$-tangling} power reads:
\begin{equation}
\begin{split}
    \epsilon_{p,M}(U)&=
    2^M\langle\text{Tr}[U^{\otimes 2}\rho_0^{\otimes 2}(U^\dagger)^{\otimes 2}\tilde{P}] \rangle_{|\psi_0\rangle^{\otimes 2}}\\&=2^M\text{Tr}[U^{\otimes 2} \langle \rho_0^{\otimes 2} \rangle_{|\psi_0\rangle^{\otimes 2}} (U^\dagger)^{\otimes 2}\tilde{P}]\\&=
    2^M\text{Tr}\Big[U^{\otimes 2}\int d\mu (|\psi_0\rangle \langle \psi_0|)^{\otimes 2}(U^\dagger)^{\otimes 2}\tilde{P}\Big]\\&=
    2^M \text{Tr}[U^{\otimes 2}\Omega_{p0}(U^\dagger)^{\otimes 2}\tilde{P}],
    \end{split}
\end{equation}
where $\Omega_{p0}=(d+1)^{-M}\prod_{i=1}^M P^{(+)}_{i,i+M}$, with $d=2$ and $P^{(+)}_{i,i+M}=1/2(\mathds{1}_{2M\times 2M}+\text{SWAP}_{i,i+M})$. To prove the expression for $\Omega_{p0}$, it suffices to note that the uniform distribution of product states factorizes, and hence we can consider the total average as averages of the sectors $i$ and $i+M$, $\forall i$. Thus, we have $\Omega_{p0}=\prod_{i}^M \omega_{i,i+M}$. Since $\Omega_{p0}$ is symmetric under the exchange of $i$ and $i+M$, we then have $\omega_{i,i+M}=cP^{(+)}_{i,i+M}$, where $c$ is a constant equal to $c=(d+1)^{-1}$ (see also Ref.~\cite{Zanardi2000}). Alternatively, the combined integral of the sectors $i$ and $i+M$ can be expressed in terms of an integral over the unitary group where the initial state is fixed and we vary the unitary acting on $i$ and $i+M$ systems:
\begin{equation}
\begin{split}
    &\int d\mu(U') (U'\otimes U') \rho_0^{\otimes 2} (U'^\dagger\otimes U'^\dagger)\\&=\frac{1}{d(d-1)}(\text{Tr}[\rho_0]-\text{Tr}[\rho_0 \text{SWAP}_{i,i+M}])P^{(-)}_{i,i+M}
    \\&+
    \frac{1}{d(d+1)}(\text{Tr}[\rho_0]+\text{Tr}[\rho_0 \text{SWAP}_{i,i+M}])P^{(+)}_{i,i+M}
    \\&=
    \frac{2}{d(d+1)}P^{(+)}_{i,i+M}=\frac{1}{d+1}P^{(+)}_{i,i+M},
    \end{split}
\end{equation}
where we have used the fact that $\text{Tr}[\rho_0]=1$ and $\text{Tr}[\rho_0 \text{SWAP}_{i,i+M}]=1$.

Note that for even $M$, the  {\color{black}$M$-tangling} power holds for arbitrary gates $U$, whereas for odd $M$ it holds only for controlled evolutions of the form $U=\sum_{j\in{0,1}}\sigma_{jj}\otimes_{l=1}^{M-1} R_{\textbf{n}_j}^{(l)}$.

\subsection{Proof of \texorpdfstring{$M$}{M}-tangling power for CR-evolution\label{App:MwayEP_CRevol}}
Before going into the proof, it will be useful to find the action of products of symmetric/antisymmetric projectors (which project onto the sectors $i$ and $i+M$) on a trial product state. Suppose we have a collection of states $\{|\phi_l\rangle,|\phi_{l+M}\rangle\}_{l=1}^M$. Suppose further that we have the ordered ket $|\phi_1\rangle_1 \dots |\phi_M\rangle_{M} |\phi_{1+M}\rangle_{1+M} \dots |\phi_{2M}\rangle_{2M}:=\prod_{l=1}^M|\phi_l\rangle_l|\phi_{l+M}\rangle_{l+M}$.  If we act with the projectors $\tilde{P}^{(\pm)}=\prod_{i=1}^M P^{(\pm)}_{i,i+M}$, on the above state we find:
\begin{equation}
\begin{split}
    &2^M\tilde{P}^{(\pm)}\prod_{l=1}^M|\phi_l\rangle_l |\phi_{l+M}\rangle_{l+M}=\\&\prod_{l=1}^M \big[|\phi_l\rangle_l |\phi_{l+M}\rangle_{l+M} \pm |\phi_{l+M}\rangle_l |\phi_{l}\rangle_{l+M}\big].
\end{split}
\end{equation}
Note that for the above equality to hold it is important that both the kets and the symbols inside the kets are labeled by an index; otherwise if the kets are not labeled the above expression holds only up to re-ordering of the unlabeled kets.

To derive a closed-form expression for the  {\color{black}$M$-tangling} power of CR-type gates in defect systems, we start by considering odd $M$. For odd $M$, we start from the general expression:
\begin{equation}
    \epsilon_{p,M}(U)=2^M \text{Tr}[U^{\otimes 2}\Omega_{p0}(U^\dagger)^{\otimes 2} \tilde{P}],
\end{equation}
and we define the vector $|\textbf{m};\textbf{m}{+}M\rangle = |m_1\rangle_1 \dots |m_M\rangle_M |m_{1+M}\rangle_{1+M}\dots |m_{2M}\rangle_{2M}$. The action of the projectors $P^{(+)}_{1,1+M}\prod_{l=2}^M P^{(-)}_{l,l+M}$ on the above vector yields:
\begin{widetext}
\begin{equation}\label{Eq:Pminus}
\begin{split}
    \tilde{P}|\textbf{m};\textbf{m}+M\rangle&=\frac{1}{2^{M}} (|m_1\rangle_1 |m_{1+M}\rangle_{1+M}+|m_{1+M}\rangle_1 |m_1\rangle_{1+M})\prod_{l=2}^M \big[|m_{l}\rangle_l |m_{l+M}\rangle_{l+M}-|m_{l+M}\rangle_l|m_l\rangle_{l+M}\big].
    \end{split}
\end{equation}
Next, we define:
\begin{equation}\label{Eq:Udag}
\begin{split}
(U^\dagger)^{\otimes 2} &= \sigma_{ii}\otimes_{r=2}^{M} (R_{\textbf{n}_i}^{(r-1)})^\dagger\otimes \sigma_{kk}\otimes _{r'=2}^{M} (R_{\textbf{n}_k}^{(r'+M-1)})^\dagger
=[\otimes_{l} (Q^{(l)}_i)]^\dagger    [\otimes_{l} (Q^{(l)}_k)]^\dagger
\end{split}
\end{equation}
and
\begin{equation}
    U^{\otimes 2}=\sigma_{jj}\otimes_{s=2}^{M} R_{\textbf{n}_j}^{(s-1)}\otimes \sigma_{pp}\otimes_{s'=2}^{M-1}R_{\textbf{n}_p}^{(s'-1+M)}=[\otimes_l Q_j^{(l)}][\otimes_l Q_p^{(l)}],
\end{equation}
where we have suppressed the symbol of summation. In the above expressions we have defined $Q_g^{(l)}$ to be $Q_g^{(1)}=\sigma_{gg}$ for $l=1$, or  $Q_g^{(l)}=R_{\textbf{n}_g}^{(l-1)}$, for $l\neq 1$. Using Eq.~(\ref{Eq:Udag}) and Eq.~(\ref{Eq:Pminus}), we evaluate $2^{M}(U^\dagger)^{\otimes 2}\tilde{P}|\textbf{m};\textbf{m}+M\rangle$:
\begin{equation}
\begin{split}
    &2^{M}(U^\dagger)^{\otimes 2}\tilde{P}|\textbf{m};\textbf{m}+M\rangle =
    \Big((Q^{(1)}_i)^{\dagger}|m_1\rangle_1 (Q_k^{(1)})^{\dagger}|m_{1+M}\rangle_{1+M}+(Q^{(1)}_i)^\dagger|m_{1+M}\rangle_1 (Q^{(1)}_k)^\dagger|m_1\rangle_{1+M}
    \Big)\times
    \\&
    \prod_{l=2}^M (Q^{(l)}_i)^\dagger|m_l\rangle_l (Q^{(l)}_k)^\dagger|m_{l+M}\rangle_{l+M}-(Q^{(l)}_i)^\dagger|m_{l+M}\rangle_l (Q^{(l)}_k)^\dagger|m_{l}\rangle_{l+M}
    \end{split}
\end{equation}
We then find the action of $[2(d+1)]^M U^{\otimes 2}\Omega_{p0}$ on the bra $\langle \textbf{m};\textbf{m}+M|$:
\begin{equation}\label{Eq:Left_Even_Odd}
\begin{split}
    \prod_{l=1}^M{}_l\langle m_l|Q_{j}^{(l)}~{}_{l+M}\langle m_{l+M}|Q_{p}^{(l)} (d+1)^M \Omega_{p0}&= \Big({}_1\langle m_1 |Q_j^{(1)} ~{}_{1+M} \langle m_{1+M}|Q_p^{(1)} + {}_1 \langle m_{1+M}|Q_p^{(1)} ~ {}_{1+M}\langle m_1 |Q_j^{(1)}\Big) \times \\&\prod_{l=2}^M{}_l \langle m_l |Q_j^{(l)} ~{}_{l+M} \langle m_{l+M}|Q_p^{(l)} + {}_l \langle m_{l+M}|Q_p^{(l)} ~ {}_{l+M}\langle m_l |Q_j^{(l)}
    \end{split}.
\end{equation}
Thus, we find that $[2(d+1)]^M 2^{M}\text{Tr}[U^{\otimes 2}\Omega_{p0}(U^\dagger)^{\otimes 2}\tilde{P}]$ is equal to:
\begin{equation}
\begin{split}
    &\sum_{i,k,p,j}\sum_{\{m_l\},\{m_{l+M}\}}\Big[(\delta^{m_1}_j \delta^j_i \delta^i_{m_1})(\delta^{m_{1+M}}_p\delta^p_k \delta^k_{m_{1+M}})+(\delta^{m_{1+M}}_p \delta^p_i \delta^i_{m_1})(\delta^{m_1}_j \delta^j_k \delta^k_{m_{1+M}})
    \\&~~~~~~~~~~~~~~~~~~~~~+
    (\delta^{m_1}_j\delta^j_i\delta^i_{m_{1+M}})(\delta^{m_{1+M}}_p\delta^p_k\delta^k_{m_1})+(\delta^{m_{1+M}}_p\delta^p_i \delta^i_{m_{1+M}})(\delta^{m_1}_j\delta^j_k \delta^k_{m_1})
    \Big]\times \\&
    \prod_{l=2}^M \Big\{
    \langle m_l |Q_j^{(l)}(Q^{(l)}_i)^\dagger|m_l\rangle \langle m_{l+M}|Q_p^{(l)}(Q^{(l)}_k)^\dagger|m_{l+M}\rangle
    -\langle m_l |Q_j^{(l)}(Q^{(l)}_i)^\dagger|m_{l+M}\rangle \langle m_{l+M}|Q_p^{(l)}(Q^{(l)}_k)^\dagger|m_l\rangle
    +\\&
    ~~~~~~\langle m_{l+M}|Q_p^{(l)}(Q^{(l)}_i)^\dagger|m_l\rangle \langle m_l |Q_j^{(l)}(Q^{(l)}_k)^\dagger|m_{l+M}\rangle -\langle m_{l+M}|Q_p^{(l)}(Q^{(l)}_i)^\dagger|m_{l+M}\rangle \langle m_l |Q_j^{(l)}(Q^{(l)}_k)^\dagger|m_{l}\rangle
    \Big\}=
    \\&
    \sum_{i,k,p,j}\sum_{\{m_1\},\{m_{1+M}\}}\Big[(\delta^{m_1}_j \delta^j_i \delta^i_{m_1})(\delta^{m_{1+M}}_p\delta^p_k \delta^k_{m_{1+M}})+(\delta^{m_{1+M}}_p \delta^p_i \delta^i_{m_1})(\delta^{m_1}_j \delta^j_k \delta^k_{m_{1+M}})
    \\&~~~~~~~~~~~~~~~~~~~~~+
    (\delta^{m_1}_j\delta^j_i\delta^i_{m_{1+M}})(\delta^{m_{1+M}}_p\delta^p_k\delta^k_{m_1})+(\delta^{m_{1+M}}_p\delta^p_i \delta^i_{m_{1+M}})(\delta^{m_1}_j\delta^j_k \delta^k_{m_1})
    \Big]\times \\&
    \prod_{l=2}^M\Big\{
    \text{Tr}[Q_j^{(l)}(Q^{(l)}_i)^\dagger]\text{Tr}[Q_p^{(l)}(Q^{(l)}_k)^\dagger]-\text{Tr}[Q_j^{(l)}(Q_i^{(l)})^\dagger Q_p^{(l)}(Q^{(l)}_k)^\dagger]+\text{Tr}[Q_p^{(l)}(Q^{(l)}_i)^\dagger Q_j^{(l)}(Q^{(l)}_k)^\dagger]
    \\&
    ~~~-\text{Tr}[Q_p^{(l)}(Q^{(l)}_i)^\dagger]\text{Tr}[Q_j^{(l)}(Q^{(l)}_k)^\dagger]
    \Big\}
    \end{split}
\end{equation}
where we have used the property $\sum_{\{m_i\},\{m_{i+M}\}}\prod_{i}=\prod_i \sum_{\{m_i\},\{m_{i+M}\}}$. We now work with the first expression of the Kronecker delta's which gives:
\begin{equation}
\begin{split}
    &\sum_{p,j}\sum_{m_1,m_{1+M}} (\delta^{m_1}_j  \delta^j_{m_1})(\delta^{m_{1+M}}_p \delta^p_{m_{1+M}})\times 
    \\&
    \prod_{l=2}^M\Big\{
    \text{Tr}[Q_j^{(l)}(Q^{(l)}_j)^\dagger]\text{Tr}[Q_p^{(l)}(Q^{(l)}_p)^\dagger]-\text{Tr}[Q_j^{(l)}(Q^{(l)}_j)^\dagger Q_p^{(l)}(Q^{(l)}_p)^\dagger]+\text{Tr}[Q_p^{(l)}(Q^{(l)}_j)^\dagger Q_j^{(l)}(Q^{(l)}_p)^\dagger]
    \\&
    ~~~-\text{Tr}[Q_p^{(l)}(Q^{(l)}_j)^\dagger]\text{Tr}[Q_j^{(l)}(Q^{(l)}_p)^\dagger]
    \Big\}=\\&
    \sum_{p,j}\sum_{m_1,m_{1+M}} (\delta^{m_1}_j  \delta^j_{m_1})(\delta^{m_{1+M}}_p \delta^p_{m_{1+M}})\times \prod_{l=2}^M\Big\{
    4-\text{Tr}[Q_p^{(l)}(Q^{(l)}_j)^\dagger]\text{Tr}[Q_j^{(l)}(Q^{(l)}_p)^\dagger]
    \Big\}=\\&
    \sum_{p,j} \prod_{l=2}^M\Big\{
    4-\text{Tr}[Q_p^{(l)}(Q^{(l)}_j)^\dagger]\text{Tr}[Q_j^{(l)}(Q^{(l)}_p)^\dagger]\Big\}=\\&
    2\prod_{l=2}^M \Big\{4-(\text{Tr}[Q_0^{(l)}(Q^{(l)}_1)^\dagger])^2 \Big\}=
    \\& 2\prod_{l=2}^M(4-4G_1^{(l-1)})=2 \times 4^{M-1}\prod_{l=2}^M (1-G_1^{(l-1)}),
    \end{split}
\end{equation}
where we have used the fact that $2^{-1}\text{Tr}[Q_0^{(l)}(Q^{(l)}_1)^\dagger]=\cos \frac{\phi_0^{(l-1)}}{2}\cos \frac{\phi_1^{(l-1)}}{2}+(\textbf{n}_0\cdot \textbf{n}_1)^{(l-1)}\sin \frac{\phi_0^{(l-1)}}{2}\sin \frac{\phi_1^{(l-1)}}{2}\equiv \sqrt{G_1^{(l-1)}}$, as well as $4-(\text{Tr}[Q_0^{(l)}(Q^{(l)}_0)^\dagger])^2=4-(\text{Tr}[Q_1^{(l)}(Q^{(l)}_1)^\dagger])^2=0$. The second term with the Kronecker deltas reads:
\begin{equation}
    \begin{split}
    &\sum_{p,j}\sum_{m_1,m_{1+M}} (\delta^{m_{1+M}}_p  \delta^p_{m_1})(\delta^{m_1}_j  \delta^j_{m_{1+M}})\times 
    \\&
        \prod_{l=2}^M\Big\{
    \text{Tr}[Q_j^{(l)}(Q^{(l)}_p)^\dagger]\text{Tr}[Q_p^{(l)}(Q^{(l)}_j)^\dagger]-\text{Tr}[Q_j^{(l)}(Q^{(l)}_p)^\dagger Q_p^{(l)}(Q^{(l)}_j)^\dagger]+\text{Tr}[Q_p^{(l)}(Q^{(l)}_p)^\dagger Q_j^{(l)}(Q^{(l)}_j)^\dagger]
    \\&
    ~~~-\text{Tr}[Q_p^{(l)}(Q^{(l)}_p)^\dagger]\text{Tr}[Q_j^{(l)}(Q^{(l)}_j)^\dagger]
    \Big\}=\\&
    \sum_{p,j}\sum_{m_1,m_{1+M}} (\delta^{m_{1+M}}_p  \delta^p_{m_1})(\delta^{m_1}_j  \delta^j_{m_{1+M}})\times \prod_{l=2}^M\Big\{
    \text{Tr}[Q_j^{(l)}(Q^{(l)}_p)^\dagger]\text{Tr}[Q_p^{(l)}(Q^{(l)}_j)^\dagger]-4
    \Big\}=\\&
    \sum_{p,j} \delta^p_j \delta^j_p \times \prod_{l=2}^M\Big\{
    \text{Tr}[Q_j^{(l)}(Q^{(l)}_p)^\dagger]\text{Tr}[Q_p^{(l)}(Q^{(l)}_j)^\dagger]-4
    \Big\}=0.
    \end{split}
\end{equation}
For similar reasons the third term of Kronecker deltas also vanishes. Working with the last term of Kronecker deltas we find:
\begin{equation}
\begin{split}
&\sum_{p,j}\sum_{\{m_l\},\{m_{l+M}\}}\Big[(\delta^{m_{1+M}}_p \delta^p_{m_{1+M}})(\delta^{m_1}_j\ \delta^j_{m_1})
    \Big]\times 
    \prod_{l=2}^M\Big\{
    \text{Tr}[Q_j^{(l)}(Q^{(l)}_p)^\dagger]\text{Tr}[Q_p^{(l)}(Q^{(l)}_j)^\dagger]-\text{Tr}[Q_p^{(l)}(Q^{(l)}_p)^\dagger]\text{Tr}[Q_j^{(l)}(Q^{(l)}_j)^\dagger]\Big\}
    \\&=
    \sum_{p,j}\delta^p_p \delta^j_j \prod_{l=2}^M\Big\{ \Big(\text{Tr}[Q_j^{(l)}(Q^{(l)}_p)^\dagger]\Big)^2-4\Big\}=2\times 4^{M-1}(-1)^{M-1}\prod_{l=2}^M(1-G_1^{(l-1)})=2\times 4^{M-1}\prod_{l=2}^M(1-G_1^{(l-1)}),
\end{split}    
\end{equation}
where we made use of $(-1)^{M-1}=1$ since $M-1$ is even. Therefore, our final expression for odd $M$ is:
\begin{equation}
    2^M\text{Tr}[U^{\otimes 2}\Omega_{p0}(U^\dagger)^{\otimes 2}\tilde{P}]=2^M \frac{1}{(d+1)^M}\frac{1}{2^M}\frac{1}{2^{M}} 4 \times 4^{M-1}\prod_{l=2}^M(1-G_1^{(l-1)})=\Big(\frac{d}{d+1}\Big)^M \prod_{l=2}^M (1-G_1^{(l-1)}),
\end{equation}
which concludes our proof.

For even $M$, note that $\tilde{P}=\prod_{i=1}^M P^{(-)}_{i,i+M}$. Again, we follow a similar procedure and start by calculating the action of $2^M(U^{\dagger})^{\otimes 2}\tilde{P}$ from the left on the ket $|\textbf{m};\textbf{m}+M\rangle$, which gives:
\begin{equation}\label{Eq:Ket_Even}
\begin{split}
    2^M(U^\dagger)^{\otimes 2}\tilde{P}|\textbf{m};\textbf{m}+M\rangle=&\Big[
    (Q^{(1)}_i)^\dagger|m_1\rangle_1 (Q^{(1)}_k)^\dagger|m_{1+M}\rangle_{1+M}-(Q^{(1)}_i)^\dagger|m_{1+M}\rangle_1 (Q^{(1)}_k)^\dagger|m_{1}\rangle_{1+M}\Big]\times \\&\prod_{l=2}^M (Q^{(l)}_i)^\dagger|m_l\rangle_l (Q^{(l)}_k)^\dagger|m_{l+M}\rangle_{l+M}-(Q^{(l)}_i)^\dagger|m_{l+M}\rangle_l (Q^{(l)}_k)^\dagger|m_{l}\rangle_{l+M}.
    \end{split}
\end{equation}
Now, we combine Eq.~(\ref{Eq:Left_Even_Odd}) with Eq.~(\ref{Eq:Ket_Even}), to obtain $[2(d+1)]^M2^M\text{Tr}[U^{\otimes 2}\Omega_{p0}(U^\dagger)^{\otimes 2}\tilde{P}^{(-)}]$:
\begin{equation}\label{Eq:Trace_Even}
    \begin{split}
        &\sum_{i,k,p,j}\sum_{\substack{\{m_l\} \\ \{m_{l+M}\}}}\Big[\langle m_1|Q_j^{(1)}\langle m_{1+M}|~Q_p^{(1)}+\langle m_{1+M}|Q_p^{(1)}\langle m_1|Q_j^{(1)}\Big]\times
        \\&
        ~~~~~~~~~~~~~~~~\Big[(Q_i^{(1)})^\dagger|m_1\rangle(Q^{(1)}_k)^\dagger|m_{1+M}\rangle-(Q^{(1)}_i)^\dagger|m_{1+M}\rangle (Q^{(1)}_k)^\dagger|m_{1}\rangle\Big] \times 
        \\&
        \prod_{l=2}^M \Big[\langle m_l|Q_j^{(l)}\langle m_{l+M}|~Q_p^{(l)}+\langle m_{l+M}|Q_p^{(l)}\langle m_l|Q_j^{(l)} \Big]
        \Big[(Q^{(l)}_i)^\dagger|m_l\rangle(Q^{(l)}_k)^\dagger|m_{l+M}\rangle-(Q^{(l)}_i)^\dagger|m_{l+M}\rangle (Q^{(l)}_k)^\dagger|m_{l}\rangle\Big]=
        \\&
        \sum_{i,k,p,j}\sum_{m_1,m_{1+M}}\Big[ 
        (\delta^{m_1}_j \delta^j_i \delta^i_{m_1})(\delta^{m_{1+M}}_p \delta^p_k \delta^k_{m_{1+M}})
        -(\delta^{m_1}_j\delta^j_i \delta^i_{m_{1+M}})(\delta^{m_{1+M}}_p\delta^p_k \delta^k_{m_1})
        \\&~~~~~~~~~~~~~~~~~
        +(\delta^{m_{1+M}}_p\delta^p_i \delta^i_{m_1})(\delta^{m_1}_j \delta^j_k \delta^k_{m_{1+M}})-(\delta^{m_{1+M}}_p \delta^p_i \delta^i_{m_{1+M}})(\delta^{m_1}_j \delta^j_k \delta^k_{m_{1}})\Big]\times
        \\&
        \prod_{l=2}^M \Big\{ \text{Tr}[Q_j^{(l)} (Q^{(l)}_i)^\dagger]\text{Tr}[Q_p^{(l)}(Q^{(l)}_k)^\dagger]-\text{Tr}[Q_j^{(l)}(Q^{(l)}_i)^\dagger Q_p^{(l)}(Q^{(l)}_k)^\dagger]+\text{Tr}[Q_p^{(l)}(Q^{(l)}_i)^\dagger Q_j^{(l)}(Q^{(l)}_k)^\dagger]
        \\&
        ~~~-\text{Tr}[Q_p^{(l)}(Q^{(l)}_i)^\dagger]\text{Tr}[Q_j^{(l)}(Q^{(l)}_k)^\dagger] \Big\}.
    \end{split}
\end{equation}
First, we focus on the first term of Kronecker deltas, which gives:
\begin{equation}
    \begin{split}
        &\sum_{i,k,p,j}\sum_{m_1,m_{1+M}}
        (\delta^{m_1}_j \delta^j_i \delta^i_{m_1})(\delta^{m_{1+M}}_p \delta^p_k \delta^k_{m_{1+M}})
        \times
        \\&
        \prod_{l=2}^M \Big\{ \text{Tr}[Q_j^{(l)} (Q_i^{(l)})^\dagger]\text{Tr}[Q_p^{(l)}(Q^{(l)}_k)^\dagger]-\text{Tr}[Q_j^{(l)}(Q^{(l)}_i)^\dagger Q_p^{(l)}(Q^{(l)}_k)^\dagger]+\text{Tr}[Q_p^{(l)}(Q^{(l)}_i)^\dagger Q_j^{(l)}(Q^{(l)}_k)^\dagger]
        \\&
        ~~~-\text{Tr}[Q_p^{(l)}(Q^{(l)}_i)^\dagger]\text{Tr}[Q_j^{(l)}(Q^{(l)}_k)^\dagger] \Big\}=
        \\&
        \sum_{p,j}\sum_{m_1,m_{1+M}}
        (\delta^{m_1}_j  \delta^j_{m_1})(\delta^{m_{1+M}}_p  \delta^p_{m_{1+M}})
        \\&
        \prod_{l=2}^M \Big\{ \text{Tr}[Q_j^{(l)} (Q^{(l)}_j)^\dagger]\text{Tr}[Q_p^{(l)}(Q^{(l)}_p)^\dagger]-\text{Tr}[Q_j^{(l)}(Q^{(l)}_j)^\dagger Q_p^{(l)}(Q^{(l)}_p)^\dagger]+\text{Tr}[Q_p^{(l)}(Q^{(l)}_j)^\dagger Q_j^{(l)}(Q^{(l)}_p)^\dagger]
        \\&
        ~~~-\text{Tr}[Q_p^{(l)}(Q^{(l)}_j)^\dagger]\text{Tr}[Q_j^{(l)}(Q^{(l)}_p)^\dagger] \Big\}=
        \\&
        \sum_{p,j}\sum_{m_1,m_{1+M}}
        (\delta^{m_1}_j  \delta^j_{m_1})(\delta^{m_{1+M}}_p  \delta^p_{m_{1+M}})
        \prod_{l=2}^M \Big\{ 4-\text{Tr}[Q_p^{(l)}(Q^{(l)}_j)^\dagger]\text{Tr}[Q_j^{(l)}(Q^{(l)}_p)^\dagger] \Big\}=2\times 4^{M-1}\prod_{l=2}^{M}(1-G_1^{(l-1)}).
    \end{split}
\end{equation}
It is further easy to show that the second and third terms of Kronecker deltas in Eq.~(\ref{Eq:Trace_Even}) evaluate to 0, for similar reasons as in the odd case. The fourth term of Kronecker deltas produces:
\begin{equation}
    \begin{split}
        &\sum_{i,k,p,j}\sum_{m_1,m_{1+M}}\Big[-(\delta^{m_{1+M}}_p \delta^p_i \delta^i_{m_{1+M}})(\delta^{m_1}_j \delta^j_k \delta^k_{m_{1}})\Big]\times
        \\&
        \prod_{l=2}^M \Big\{ \text{Tr}[Q_j^{(l)} (Q^{(l)}_i)^\dagger]\text{Tr}[Q_p^{(l)}(Q^{(l)}_k)^\dagger]-\text{Tr}[Q_j^{(l)}(Q^{(l)}_i)^\dagger Q_p^{(l)}(Q^{(l)}_k)^\dagger]+\text{Tr}[Q_p^{(l)}(Q^{(l)}_i)^\dagger Q_j^{(l)}(Q^{(l)}_k)^\dagger]
        \\&
        ~~~-\text{Tr}[Q_p^{(l)}(Q^{(l)}_i)^\dagger]\text{Tr}[Q_j^{(l)}(Q^{(l)}_k)^\dagger] \Big\}=
        \\&
        \sum_{p,j}\sum_{m_1,m_{1+M}}\Big[-(\delta^{m_{1+M}}_p  \delta^p_{m_{1+M}})(\delta^{m_1}_j  \delta^j_{m_{1}})\Big]\times
        \\&
        \prod_{l=2}^M \Big\{ \text{Tr}[Q_j^{(l)} (Q^{(l)}_p)^\dagger]\text{Tr}[Q_p^{(l)}(Q^{(l)}_j)^\dagger]-\text{Tr}[Q_j^{(l)}(Q^{(l)}_p)^\dagger Q_p^{(l)}(Q^{(l)}_j)^\dagger]+\text{Tr}[Q_p^{(l)}(Q^{(l)}_p)^\dagger Q_j^{(l)}(Q^{(l)}_j)^\dagger]
        \\&
        ~~~-\text{Tr}[Q_p^{(l)}(Q^{(l)}_p)^\dagger]\text{Tr}[Q_j^{(l)}(Q^{(l)}_j)^\dagger] \Big\}=
        \\&
        \sum_{p,j}\sum_{m_1,m_{1+M}}\Big[-(\delta^{m_{1+M}}_p  \delta^p_{m_{1+M}})(\delta^{m_1}_j  \delta^j_{m_{1}})\Big]\prod_{l=2}^M \Big\{ \text{Tr}[Q_j^{(l)} (Q^{(l)}_p)^\dagger]\text{Tr}[Q_p^{(l)}(Q^{(l)}_j)^\dagger]-4 \Big\}=2\times 4^{M-1}\prod_{l=2}^{M}(1-G_1^{(l-1)}).
    \end{split}
\end{equation}
Therefore, the expression $2^M\text{Tr}[U^{\otimes 2}\Omega_{p0}(U^\dagger)^{\otimes 2}\tilde{P}]$ for even $M$ reads:
\begin{equation}
    \epsilon_{p,M}(U)=2^M\text{Tr}[U^{\otimes 2}\Omega_{p0}(U^\dagger)^{\otimes 2}\tilde{P}]=\frac{1}{(d+1)^M}\frac{1}{2^M}\frac{1}{2^M}2^M \times 4 \times 4^{M-1}\prod_{l=2}^M(1-G_1^{(l-1)})=\Big(\frac{d}{d+1}\Big)^{M}\prod_{l=2}^M(1-G_1^{(l-1)}).
\end{equation}

\end{widetext}

\subsection{\texorpdfstring{$M$}{M}-tangling power of non-unitary quantum evolution \label{App:Non_Uni_Ep}}

In this section we consider a setup where we have $L$ nuclear spins with $K$ of them being the target ones, and $L-K$ being the unwanted spins. Thus, the target subspace has $M=K+1$ qubits, consisting of the electron and $K$ target nuclear spins. In Ref.~\cite{EconomouPRX2023} we showed that the Kraus operators associated with the partial trace operation of the $L-K$ unwanted spins have the closed form expression:

\begin{equation}
    E_i = \sum_{j\in \{0,1\}} f_j^{(i)}\sigma_{jj}\otimes_{l=1}^K R_{\textbf{n}_j}^{(k)},
\end{equation}
with $f_j^{(i)}$ given by:
\begin{equation}
    f_j^{(i)} = \prod_{m_i}\langle 0| R_{\textbf{n}_j}^{(m_i)}|0\rangle \prod_{m'_i}\langle 1| R_{\textbf{n}_j}^{(m'_i)}|0\rangle,
\end{equation}
where for the $i$-th Kraus operator, the first product is taken over the unwanted spins comprising the environment that are in $|0\rangle$, and the second product is taken over the unwanted spins that are in $|1\rangle$. If all spins are in $|0\rangle$, then the second product is 1, whereas if all spins are in $|1\rangle$ the first product is 1. Further, we have $\langle 0|R_{\textbf{n}_j}|0 \rangle=\cos\frac{\phi_j}{2}-in_{j,z}\sin\frac{\phi_j}{2}$ and $\langle 1|R_{\textbf{n}_j}|0 \rangle=-i\sin\frac{\phi_j}{2}(n_{x,j}+i n_{y,j})$.  

The evolution of the target subspace is thus described by the Kraus operators via the quantum channel $\mathcal{E}(\rho)=\sum_k E_k \rho_0E_k^\dag$. To include the impact of the unwanted subspace into the $M$-way entanglement that is generated by the quantum evolution, we define the  {\color{black}$M$-tangling} power of the non-unitary quantum channel:

\begin{equation}
\begin{split}
    &\epsilon_{p,M}(\mathcal{E})=2^M \langle\text{Tr}[ \rho_{\mathcal{E}}^{\otimes 2}\tilde{P}]\rangle_{|\psi_0\rangle^{\otimes 2}}\\&=
    2^M \sum_{r,s=0}^{2^{L-K}-1}\text{Tr}[(E_r\otimes E_s)\langle \rho_0^{\otimes 2}\rangle_{|\psi_0\rangle^{\otimes 2}}(E_r\otimes E_s)^\dagger\tilde{P}]\\&=2^M \sum_{r,s=0}^{2^{L-K}-1}\text{Tr}[(E_r\otimes E_s) \Omega_{p0}(E_r\otimes E_s)^\dagger\tilde{P}].
    \end{split}
\end{equation}
Clearly, in the limit of a unitary channel i.e., $E_r=E_s=U$ (the summation is over the unique Kraus operator) we recover the  {\color{black}$M$-tangling} power of a unitary.
\begin{widetext}

To derive a closed-form expression, we follow a similar procedure as in the unitary case. We begin by defining
\begin{equation}
    E_r^\dagger \otimes E_s^{\dagger} = (f^{(r)}_i)^{*}(f^{(s)}_k)^{*}\sigma_{ii}\otimes_{x=2}^{M} (R_{\textbf{n}_i}^{(x)})^\dagger\otimes \sigma_{kk}\otimes_{x'=2}^M (R_{\textbf{n}_k}^{(x'+M-1)})^\dagger=(f^{(r)}_i)^{*}(f^{(s)}_k)^{*}[\otimes_l (Q^{(l)}_i)]^\dagger[\otimes_l (Q^{(l)}_k)]^\dagger
\end{equation}
and
\begin{equation}
    E_r \otimes E_s = f_j^{(r)}f_p^{(s)}\sigma_{jj}\otimes_{x=2}^{M} R_{\textbf{n}_j}^{(x)}\otimes \sigma_{pp}\otimes_{x'=2}^M R_{\textbf{n}_p}^{(x'+M-1)}=f_j^{(r)}f_p^{(s)}[\otimes_l Q_j^{(l)}][\otimes_l Q_p^{(l)}].
\end{equation}
Starting with the odd $M$ case, we find that $[2(d+1)]^M 2^{M}\sum_{r,s}\text{Tr}[(E_r\otimes E_s)\Omega_{p0}(E_r\otimes E_s)^\dagger\tilde{P}]$ reads:
\begin{equation}
\begin{split}
    &\sum_{r,s}\sum_{ik,pj}\sum_{m_1,m_{1+M}}f_j^{(r)}f_p^{(s)}(f_i^{(r)})^*(f_k^{(s)})^*\Big\{(\delta^{m_1}_j \delta^j_i \delta^i_{m_1})(\delta^{m_{1+M}}_p\delta^p_k \delta^k_{m_{1+M}})+(\delta^{m_{1+M}}_p\delta^p_i \delta^i_{m_1})(\delta^{m_1}_j\delta^j_k \delta^k_{m_{1+M}})
    \\&~~~~~~~~~~~~~~~~~~~~~~~~~~~~~~~~~~~~~~~~~~~~~~~~+
    (\delta^{m_1}_j\delta^j_{i}\delta^i_{m_{1+M}})(\delta^{m_{1+M}}_p\delta^p_k\delta^k_{m_1})+(\delta^{m_{1+M}}_p\delta^p_i\delta^i_{m_{1+M}})(\delta^{m_1}_j\delta^j_k\delta^k_{m_1})
    \Big\}\times 
    \\&
    \prod_{l=2}^M \Big\{\text{Tr}[Q_j^{(l)}(Q^{(l)}_i)^\dagger]\text{Tr}[Q_p^{(l)}(Q^{(l)}_k)^\dagger]-\text{Tr}[Q_j^{(l)}(Q^{(l)}_i)^\dagger Q_p^{(l)}(Q^{(l)}_k)^\dagger]+\text{Tr}[Q_p^{(l)}(Q^{(l)}_i)^\dagger Q_j^{(l)}(Q^{(l)}_k)^\dagger
    \\&
    ~~~-\text{Tr}[Q_p^{(l)}(Q^{(l)}_i)^\dagger]\text{Tr}[Q_j^{(l)}(Q^{(l)}_k)^\dagger]\Big\}
    \\&=
    \sum_{r,s}\sum_{p,j} \Big(|f_j^{(r)}|^2|f_p^{(s)}|^2+f_{j}^{(r)}f_p^{(s)}(f_p^{(r)})^*(f_j^{(s)})^*\Big) \times 
    \prod_{l=2}^M (4-\text{Tr}[Q_p^{(l)}(Q^{(l)}_j)^\dagger]^2)
    \\&=4^{M-1}\sum_{r,s}(|f_1^{(r)}|^2|f_0^{(s)}|^2+|f_0^{(r)}|^2|f_1^{(s)}|^2+f_0^{(r)}f_1^{(s)}(f_1^{(r)})^*(f_0^{(s)})^*+f_1^{(r)}f_0^{(s)}(f_0^{(r)})^*(f_1^{(s)})^*)\prod_{l=2}(1-G_1^{(l-1)})
    \\&=2\times 4^{M-1}\Big(1+\sum_{r,s}\Re[(f_0^{(r)})^*(f_1^{(s)})^*f_1^{(r)}f_0^{(s)}]\Big)\prod_{l=2}(1-G_1^{(l-1)}),
    \end{split}
\end{equation}
where we have made use of the fact that only the first and last terms of products of Kronecker deltas survive. Therefore we find that $\epsilon_{p,M}(\mathcal{E})$ for the odd case is:
\begin{equation}
\begin{split}
   \epsilon_{p,M}(\mathcal{E})&=\frac{1}{2^{2M}}\frac{1}{(d+1)^M}2^M\times 2\times 4^{M-1}\Big(1+\sum_{r,s=0}^{2^{L-K}-1}\Re[(f_0^{(r)})^*(f_1^{(s)})^*f_1^{(r)}f_0^{(s)}]\Big)\prod_{l=2}(1-G_1^{(l-1)})
   \\&=\Big(\frac{d}{d+1}\Big)^M \frac{1}{2}\Big(1+\sum_{r,s=0}^{2^{L-K}-1}\Re[(f_0^{(r)})^*(f_1^{(s)})^*f_1^{(r)}f_0^{(s)}]\Big)\prod_{l=2}(1-G_1^{(l-1)}).
   \end{split}
\end{equation}
whereas for the even case following similar steps we obtain:
\begin{equation}
\resizebox{0.98\hsize}{!}{
    $\begin{split}
        \epsilon_{p,M}(\mathcal{E})&=\Big(\frac{1}{d+1}\Big)^M \frac{1}{2^{2M}}2^M 2^{2M-2} \sum_{r,s=0}^{2^{L-K}-1}\Big(|f_0^{(r)}|^2|f_1^{(s)}|^2+|f_1^{(r)}|^2|f_0^{(s)}|^2+2\Re[(f_0^{(r)})^*(f_1^{(s)})^*f_1^{(r)}f_0^{(s)}]\Big) \prod_{l=2}^M (1-G_1^{(l-1)})\\&=
        \Big(\frac{d}{d+1}\Big)^M \frac{1}{4}\sum_{r,s=0}^{2^{L-K}-1}\Big(|f_0^{(r)}|^2|f_1^{(s)}|^2+|f_1^{(r)}|^2|f_0^{(s)}|^2+2\Re[(f_0^{(r)})^*(f_1^{(s)})^*f_1^{(r)}f_0^{(s)}]\Big) \prod_{l=2}^M (1-G_1^{(l-1)})\\&=
        \Big(\frac{d}{d+1}\Big)^M\frac{1}{2}\Big( 1+ \sum_{r,s=0}^{2^{L-K}-1}\Re[(f_0^{(r)})^*(f_1^{(s)})^*f_1^{(r)}f_0^{(s)}]\Big) \prod_{l=2}^M (1-G_1^{(l-1)}).
    \end{split}$
    }
\end{equation}
In the above expressions we have made use of the completeness relation i.e.:

\begin{equation}
\begin{split}
    &\sum_{r,s=0}^{2^{L-K}-1}E_r^\dagger E_s = \mathds{1}\delta^r_s \Rightarrow
    \sum_{r,s}\sum_{i,j}(f_i^{(r)})^*f_j^{(s)} [\sigma_{ii}\otimes_l (R_{\textbf{n}_i}^{(l)})^\dagger] [\sigma_{jj}\otimes_l R_{\textbf{n}_j}^{(l)}] = \mathds{1}\delta^r_s 
    \\&
    \Rightarrow
    \sum_{r,s}\sum_i (f_i^{(r)})^*f_i^{(s)}\sigma_{ii} \otimes_l \mathds{1}^{(l)}=\mathds{1}\delta^r_s \Rightarrow \\&
    \sum_i \sigma_{ii}[\sum_{r,s} (f_i^{(r)})^*f_i^{(s)}] \otimes_l \mathds{1}^{(l)}=\delta^r_s \mathds{1}\otimes_l \mathds{1}^{(l)} \Rightarrow \begin{pmatrix}
    \sum_{r,s}(f_0^{(r)})^*f_0^{(s)} & 0 \\
    0 & \sum_{r,s}(f_1^{(r)})^*f_1^{(s)} \end{pmatrix}= \begin{pmatrix}
    \delta^r_s & 0 \\
    0 & \delta^r_s
    \end{pmatrix},
    \end{split}
\end{equation}
which means that $\sum_r |f_0^{(r)}|^2=1=\sum_r |f_1^{(r)}|^2$, and hence we can write:
\begin{equation}
    \sum_r |f_0^{(r)}|^2 \times \sum_s |f_1^{(s)}|^2  +\sum_s |f_0^{(s)}|^2 \times \sum_r |f_1^{(r)}|^2=2 \Rightarrow \sum_{r,s} |f_0^{(r)}|^2|f_1^{(s)}|^2+|f_1^{(r)}|^2|f_0^{(s)}|^2 =2.
\end{equation}

\end{widetext}
We note that the expressions for even/odd number of qubits in the target subspace are identical. Further, the non-unitary  {\color{black}$M$-tangling} power is upper bounded by the unitary  {\color{black}$M$-tangling} power.

{\color{black}
\begin{widetext}
Let us now simplify the expression that appears due to the summation over Kraus operators. We first define the set $S=\{1,2,\dots,L-K\}$ of unwanted spin indices. Let $P(S)$ be the power set of $S$, i.e.:
\begin{equation}
\begin{split}
    \mathcal{P}(S):&=\Big\{\{\},\{1\},\{2\},\dots,\{L-K\},\dots
    \\&
    ~~~~~~\{1,2\},\{1,3\},\dots,\{1,L-K\},\dots
    \\&
    ~~~~~~\{1,2,\dots,L-K\}\Big\}.
    \end{split}
\end{equation}

This power set is useful in order to write down which unwanted spins are in $|0\rangle$ or $|1\rangle$ in the Kraus operators. Let $p\in \mathcal{P}(S)$ be an element of the power set and $l\in p$ be an unwanted spin index within the subset $p$. We now start from:

\begin{equation}
\begin{split}
    &\Re [\sum_{r,s=0}^{2^{L-K}-1}(f_0^{(r)})^*(f_1^{(s)})^*f_1^{(r)}f_0^{(s)}]=\Bigg|\sum_{r=0}^{2^{L-K}-1} (f_0^{(r)})^*f_1^{(r)}\Bigg|^2=
    \\&
    \Bigg|\sum_{p\in \mathcal{P}(S)} \Big\{ \Big(\prod_{l\in p} \langle 0|(R_{\textbf{n}_0}^{(l)})^*|0\rangle \prod_{l'\in S\backslash p}\langle 1|(R_{\textbf{n}_0}^{(l')})^*|0\rangle\Big)\Big(\prod_{l\in p}\langle 0|R_{\textbf{n}_1}^{(l)}|0\rangle \prod_{l'\in S\backslash p}\langle 1|R_{\textbf{n}_1}^{(l')}|0\rangle\Big) \Big\}\Bigg|^2=
    \\&
    \Bigg|\sum_{p\in \mathcal{P}(S)} \Big\{ \prod_{l\in p} \langle 0|(R_{\textbf{n}_0}^{(l)})^*|0\rangle \langle 0|R_{\textbf{n}_1}^{(l)}|0\rangle \prod_{l'\in S\backslash p}\langle 1|(R_{\textbf{n}_0}^{(l')})^*|0\rangle \langle 1|R_{\textbf{n}_1}^{(l')}|0\rangle\Big\}\Bigg|^2=
    \\&
    \Bigg|\sum_{p\in \mathcal{P}(S)}\prod_{l\in p}\alpha_l \prod_{l'\in S\backslash p}\beta_{l'}\Bigg|^2.
    \end{split}
\end{equation}
In the last line we have defined $\alpha_l=\langle 0|(R_{\textbf{n}_0}^{(l)})^*|0\rangle\langle 0| R_{\textbf{n}_1}^{(l)}|0\rangle$, and $\beta_l=\langle 1|(R_{\textbf{n}_0}^{(l)})^*|0\rangle\langle 1| R_{\textbf{n}_1}^{(l)}|0\rangle$. To evaluate the above expression, let us look into some examples. If we have one unwanted nuclear spin, then $\mathcal{P}(S)=\{\{\},\{1\}\}$, and so it holds:
\begin{equation}
    \sum_{p\in \mathcal{P}(S)}\prod_{l\in p}\alpha_l \prod_{l'\in S\backslash p}\beta_{l'}=\alpha_1\cdot 1+1\cdot \beta_1 =\alpha_1+\beta_1.
\end{equation}
If we have 2 unwanted nuclear spins, then we find:
\begin{equation}
        \sum_{p\in \mathcal{P}(S)}\prod_{l\in p}\alpha_l \prod_{l'\in S\backslash p}\beta_{l'}=\alpha_1\alpha_2+\alpha_1\beta_2+\alpha_2\beta_1+\beta_1\beta_2=(\alpha_1+\beta_1)(\alpha_2+\beta_2).
\end{equation}
Similarly, if we have 3 unwanted nuclear spins, the summation over the power set evaluates to:
\begin{equation}
\begin{split}
     \sum_{p\in \mathcal{P}(S)}\prod_{l\in p}\alpha_l \prod_{l'\in S\backslash p}\beta_{l'}&=\alpha_1\alpha_2\alpha_3+\alpha_1\beta_2\beta_3+\alpha_2\beta_1\beta_3+\alpha_3\beta_1\beta_2 
     +
     \alpha_1\alpha_2\beta_3+\alpha_1\alpha_3\beta_2+\alpha_2\alpha_3\beta_1+\beta_1\beta_2\beta_3
     \\&
     =\alpha_1(\alpha_2\alpha_3+\beta_2\beta_3)+\beta_1(\alpha_2\beta_3+\alpha_3\beta_2)+\alpha_1(\alpha_2\beta_3+\alpha_3\beta_2)+\beta_1(\alpha_2\alpha_3+\beta_2\beta_3)
     \\&
     =(\alpha_1+\beta_1)(\alpha_2\alpha_3+\beta_2\beta_3)+(\alpha_1+\beta_1)(\alpha_2\beta_3+\alpha_3\beta_2)
     \\&
     =(\alpha_1+\beta_1)(\alpha_2\alpha_3+\beta_2\beta_3+\alpha_2\beta_3+\alpha_3\beta_2)
     \\&
     =(\alpha_1+\beta_1)(\alpha_2(\alpha_3+\beta_3)+\beta_2(\alpha_3+\beta_3))
     \\&=
     (\alpha_1+\beta_1)(\alpha_2+\beta_2)(\alpha_3+\beta_3).
     \end{split}
\end{equation}
Therefore, we can now generalize this result to an arbitrary number of $L-K$ unwanted nuclear spins as:
\begin{equation}
\begin{split}
&\Bigg|\sum_{p\in \mathcal{P}(S)}\prod_{l\in p}\alpha_l \prod_{l'\in S\backslash p}\beta_{l'}\Bigg|^2=
\Bigg|\prod_{j=1}^{L-K}(\alpha_j+\beta_j)\Bigg|^2=
 \prod_{j=1}^{L-K}|\alpha_j+\beta_j|^2=\prod_{j=1}^{L-K}|(\alpha_j+\beta_j)_R+i(\alpha_j+\beta_j)_I|^2=
 \\&
 \prod_{j=1}^{L-K}\Bigg\{\left(\sqrt{G_1^{(j)}}\right)^2+\Big(n_{z,0}^{(j)}\cos\frac{\phi_1^{(j)}}{2}\sin\frac{\phi_0^{(j)}}{2}-n_{z,1}^{(j)}\cos\frac{\phi_0^{(j)}}{2}\sin\frac{\phi_1^{(j)}}{2}-(n_{x,1}^{(j)}n_{y,0}^{(j)}-n_{x,0}^{(j)}n_{y,1}^{(j)})\sin\frac{\phi_0^{(j)}}{2}\sin\frac{\phi_1^{(j)}}{2}\Big)^2\Bigg\}=
 \\&
 \prod_{j=1}^{L-K}\Bigg\{G_1^{(j)}+\Big(n_{z,0}^{(j)}\cos\frac{\phi_1^{(j)}}{2}\sin\frac{\phi_0^{(j)}}{2}-n_{z,1}^{(j)}\cos\frac{\phi_0^{(j)}}{2}\sin\frac{\phi_1^{(j)}}{2}-(n_{x,1}^{(j)}n_{y,0}^{(j)}-n_{x,0}^{(j)}n_{y,1}^{(j)})\sin\frac{\phi_0^{(j)}}{2}\sin\frac{\phi_1^{(j)}}{2}\Big)^2\Bigg\}.
\end{split}
\end{equation}
We see that again $G_1$ quantities related to unwanted nuclear spins appear in the expression. Finally, the full expression of the non-unitary entangling power reads:
\begin{equation}
\begin{split}
    &\epsilon_{p,M}(\mathcal{E})=\frac{\epsilon_{p,M}(U)}{2} \Bigg(1
    +
    \prod_{  \substack{j\in \\ \text{unwanted nuc.}}}\Bigg\{G_1^{(j)}
    +
    \\&~~~~~~~~~~~~~~~~~
    \Big(n_{z,0}^{(j)}\cos\frac{\phi_1^{(j)}}{2}\sin\frac{\phi_0^{(j)}}{2}-n_{z,1}^{(j)}\cos\frac{\phi_0^{(j)}}{2}\sin\frac{\phi_1^{(j)}}{2}-(n_{x,1}^{(j)}n_{y,0}^{(j)}-n_{x,0}^{(j)}n_{y,1}^{(j)})\sin\frac{\phi_0^{(j)}}{2}\sin\frac{\phi_1^{(j)}}{2}\Big)^2\Bigg\}\Bigg).
    \end{split}
\end{equation}
\end{widetext}
}

{\color{black}

\section{\texorpdfstring{$M$}{M}-tangling power of target subspace under dephasing channel \label{App:Dephased_M_way_EP}}

In this section, we extend our formalism to include dephasing errors that can occur on the electron during the entanglement generation. The dephasing on the electron can be expressed via the following Kraus operators~\cite{MazziottiPRA2022}:

\begin{eqnarray}
K_0(\tilde{t})&=&\sqrt{\lambda_0}\begin{pmatrix}
        e^{i\theta \tilde{t}} & 0 \\
        0 & e^{-i\theta \tilde{t}}
    \end{pmatrix} \nonumber \\
    K_1(\tilde{t})&=&\sqrt{\lambda_1}\begin{pmatrix}
        e^{-i\theta \tilde{t}} & 0 \\
        0 & e^{i\theta \tilde{t}}
    \end{pmatrix},
\end{eqnarray}
where $\lambda_0+\lambda_1=1$. Assuming that dephasing occurs during the free evolution periods, the electron-nuclear spin register evolves under the new Kraus operators:
\begin{equation}
\begin{split}
\mathcal{D}_{\text{deph},m}^{\text{free}}(t)&=[K_m(t) \otimes_{l=1}^K \mathds{1}^{(l)}] U_\text{f}(t) 
\\&= \sum_{j,q}\kappa_{qq}^{(m)}(t)\sigma_{qq}\sigma_{jj}\otimes_{l=1}^K \tilde{R}_{\textbf{n}_j}^{(l)}
\\&=
\sum_{j,q}\kappa_{qq}^{(m)}(t)\sigma_{qj}\delta^q_j \otimes_{l=1}^K \tilde{R}_{\textbf{n}_j}^{(l)}
\\&=
\sum_{j}\kappa_{jj}^{(m)}(t)\sigma_{jj} \otimes_{l=1}^K \tilde{R}_{\textbf{n}_j}^{(l)}.
\end{split}
\end{equation}
Note that we also use the notation $\tilde{R}_{\textbf{n}_j}$ to indicate that this nuclear rotation is the one that happens in a single free evolution unit and should not be confused with  $R_{\textbf{n}_j}$, which is the total rotation induced by a single unit of a decoupling sequence (e.g., a CPMG or UDD$_n$ sequence).

\begin{widetext}
Without loss of generality, we assume the CPMG unit, during which we have the total Kraus operators (neglecting global phases):
\begin{equation}
\begin{split}
\mathcal{D}_{\text{deph},\textbf{m}}^{\text{CPMG}}&=
\mathcal{D}_{\text{deph},m_3}^{\text{free}}(t/4)[R_x(\pi)\otimes_{l}\mathds{1}^{(l)}]\mathcal{D}_{\text{deph},m_2}^{\text{free}}(t/2)[R_x(\pi)\otimes_{l}\mathds{1}^{(l)}]\mathcal{D}_{\text{deph},m_1}^{\text{free}}(t/4)
\\&=\sum_{j,r,q}{\kappa}^{(m_1)}_{jj}(t/4){\kappa}^{(m_2)}_{rr}(t/2){\kappa}^{(m_3)}_{qq}(t/4) \sigma_{jj}R_{x}(\pi)\sigma_{rr}R_{x}(\pi)\sigma_{qq}\otimes_{l=1}^K 
\tilde{R}_{\textbf{n}_j}^{(l)}\tilde{R}^{(l)}_{\textbf{n}_r}\tilde{R}^{(l)}_{\textbf{n}_q}
\\&=
\sum_{j,r,q}{\kappa}^{(m_1)}_{jj}(t/4){\kappa}^{(m_2)}_{rr}(t/2){\kappa}^{(m_3)}_{qq}(t/4)\sigma_{jr}R_x(\pi)\sigma_{qq}[\delta^j_0\delta^1_r+\delta^j_1\delta^0_r]\otimes_{l=1}^K 
\tilde{R}_{\textbf{n}_j}^{(l)}\tilde{R}^{(l)}_{\textbf{n}_r}\tilde{R}^{(l)}_{\textbf{n}_q}
\\&=
\sum_{j,r,q}{\kappa}^{(m_1)}_{jj}(t/4){\kappa}^{(m_2)}_{rr}(t/2){\kappa}^{(m_3)}_{qq}(t/4)\sigma_{jq}[\delta^r_0\delta^1_q+\delta^r_1\delta^0_q][\delta^j_0\delta^1_r+\delta^j_1\delta^0_r]\otimes_{l=1}^K 
\tilde{R}_{\textbf{n}_j}^{(l)}\tilde{R}^{(l)}_{\textbf{n}_r}\tilde{R}^{(l)}_{\textbf{n}_q}
\\&=
\sum_{j,r,q}{\kappa}^{(m_1)}_{jj}(t/4){\kappa}^{(m_2)}_{rr}(t/2){\kappa}^{(m_3)}_{qq}(t/4)\sigma_{jq}[\delta^r_0\delta^1_q\delta^j_0\delta^1_r+\delta^r_1\delta^0_q\delta^j_0\delta^1_r+\delta^r_0\delta^1_q\delta^j_1\delta^0_r+\delta^r_1\delta^0_q\delta^j_1\delta^0_r]\otimes_{l=1}^K 
\tilde{R}_{\textbf{n}_j}^{(l)}\tilde{R}^{(l)}_{\textbf{n}_r}\tilde{R}^{(l)}_{\textbf{n}_q}
\\&=
\sum_{j,r,q}{\kappa}^{(m_1)}_{jj}(t/4){\kappa}^{(m_2)}_{rr}(t/2){\kappa}^{(m_3)}_{qq}(t/4)\sigma_{jq}[\delta^r_1\delta^0_q\delta^j_0\delta^1_r+\delta^r_0\delta^1_q\delta^j_1\delta^0_r]\otimes_{l=1}^K 
\tilde{R}_{\textbf{n}_j}^{(l)}\tilde{R}^{(l)}_{\textbf{n}_r}\tilde{R}^{(l)}_{\textbf{n}_q}.
\end{split}
\end{equation}
We label the Kraus operators of one unit with the bitstring $\textbf{m}=(m_1,m_2,m_3)$, where each $m_j$ can take the value of 0 or 1. The only non-zero terms occur for $(q,j,r)=(0,0,1)$ and $(q,j,r)=(1,1,0)$, respectively. Thus, the Kraus operator reads:
\begin{equation}
\begin{split}
\mathcal{D}_{\text{deph},\textbf{m}}^{\text{CPMG}}&=\sum_{j}{\kappa}^{(m_1)}_{jj}(t/4){\kappa}^{(m_3)}_{jj}(t/4){\kappa}^{(m_2)}_{j\oplus 1 j\oplus 1}(t/2)\sigma_{jj}\otimes_l\tilde{R}_{\textbf{n}_j}^{(l)}\tilde{R}_{\textbf{n}_{j\oplus 1}}^{(l)}\tilde{R}_{\textbf{n}_j}^{(l)}
\\&={\color{black}\sum_j \sqrt{\lambda_{m_1}}e^{i \theta t/4 (-1)^{j\oplus m_1}}\sqrt{\lambda_{m_3}}e^{i \theta t/4 (-1)^{j\oplus m_3}}\sqrt{\lambda_{m_2}}e^{i \theta t/2 (-1)^{j\oplus m_2\oplus 1}}\sigma_{jj}\otimes_l\tilde{R}_{\textbf{n}_j}^{(l)}\tilde{R}_{\textbf{n}_{j\oplus 1}}^{(l)}\tilde{R}_{\textbf{n}_j}^{(l)}}
\\&={\color{black}\sum_j \sqrt{\lambda_{m_1}\lambda_{m_2}\lambda_{m_3}}e^{i\theta t/4[(-1)^{j\oplus m_1}+(-1)^{j\oplus m_3}]}e^{i \theta t/2 (-1)^{j\oplus m_2\oplus 1}}\sigma_{jj}\otimes_l\tilde{R}_{\textbf{n}_j}^{(l)}\tilde{R}_{\textbf{n}_{j\oplus 1}}^{(l)}\tilde{R}_{\textbf{n}_j}^{(l)}}
\\&={\color{black}\sum_j \sqrt{\lambda_{m_1}\lambda_{m_2}\lambda_{m_3}} e^{i\theta t/4[(-1)^{j\oplus m_1}+(-1)^{j\oplus m_3}+2(-1)^{j\oplus m_2\oplus 1}]}\sigma_{jj}\otimes_l\tilde{R}_{\textbf{n}_j}^{(l)}\tilde{R}_{\textbf{n}_{j\oplus 1}}^{(l)}\tilde{R}_{\textbf{n}_j}^{(l)}}
\\&
={\color{black}\sum_j g_j(m_1,m_2,m_3,\theta,t)\sigma_{jj}\otimes_l R_{\textbf{n}_j}^{(l)}},
\end{split}
\end{equation}
where we have defined:
\begin{equation}g_j(m_1,m_2,m_3,\theta,t)=\sqrt{\lambda_{m_1}\lambda_{m_2 }\lambda_{m_3}}e^{i\theta t/4[(-1)^{m_1\oplus j}+(-1)^{m_3\oplus j}+2(-1)^{m_2\oplus j\oplus 1}]}.
\end{equation}
Also, we define $\oplus$ as addition modulo 2. It is easy to verify that the Kraus operators for one decoupling unit satisfy $\sum_{m_1,m_2,m_3} [\mathcal{D}_{\text{deph},\textbf{m}}^{\text{CPMG}}]^\dagger \mathcal{D}_{\text{deph},\textbf{m}}^{\text{CPMG}} = \mathds{1}$ as follows:
\begin{equation}
    \begin{split}
        \sum_{m_1,m_2,m_3} [\mathcal{D}_{\text{deph},\textbf{m}}^{\text{CPMG}}]^\dagger \mathcal{D}_{\text{deph},\textbf{m}}^{\text{CPMG}} &=\sum_{m_1,m_2,m_3,j,k} \lambda_{m_1}\lambda_{m_2}\lambda_{m_3}\sigma_{jj}\sigma_{kk}\otimes_{l}\left(R_{\textbf{n}_j}^{(l)}\right)^\dagger R_{\textbf{n}_k} ^{(l)}
        \\&=\sum_{m_1,m_2,m_3}\lambda_{m_1}\lambda_{m_2}\lambda_{m_3}\sum_j\sigma_{jj} 
 \otimes_l \mathds{1}^{(l)}
        \\&=\sum_{m_1,m_2}\lambda_{m_1}\lambda_{m_2}(\lambda_{0}+\lambda_1)\mathds{1}
        \\&=\sum_{m_1}\lambda_{m_1}(\lambda_{0}+\lambda_1)^2\mathds{1}
        \\&=(\lambda_{0}+\lambda_1)^3\mathds{1}
        =\mathds{1},
    \end{split}
\end{equation}
\end{widetext}
where we have used the fact that $\lambda_0+\lambda_1=1$. For multiple decoupling units, the Kraus operator is simply $\mathcal{D}_{\text{deph},\textbf{m}}^{\text{CPMG}}=\prod_{k=1}^N\mathcal{D}_{\text{deph},\textbf{m}^{(k)}}^{\text{CPMG}}$, where $N$ is the number of iterations of a single unit. In this notation, each $\textbf{m}^{(k)}$ consists of indices $(m_{1}^{(k)},m_{2}^{(k)},m_{3}^{(k)})$, and each one of them takes the value 0 or 1. Each Kraus operator is distinguished by a different ordering of 0s and 1s in the total vector $\textbf{m}$. For $N$ iterations, we have $2^{3N}$ such Kraus operators. For brevity of notation, we will label the composite Kraus operators that result after $N$ CPMG iterations as $\mathcal{D}_{N,q}$, where $q=1,...,2^{3N}$ labels which Kraus operator we are referring to out of all Kraus operators.

Let us now look into the $M$-tangling power, where we focus on the subspace of the electron and the $K$ target nuclear spins, and we ignore contributions from unwanted nuclei. 
This target subsystem now evolves under the non-unitary channel $\mathcal{E}_{\text{deph}}(\rho)=\sum_{r} \mathcal{D}_{N,r}\rho[\mathcal{D}_{N,r}]^\dagger$. We first start by defining $\mathcal{D}_{N,r}=\sum_j \tilde{g}_{j}^{(r)}\sigma_{jj}\otimes_l R_{\textbf{n}_j}^{(l)}$. {\color{black}Note, that here $R_{\textbf{n}_j}^{(l)}$ is the rotation after $N$ iterations, acting on the $l$-th nuclear spin.} Thus, we can express the non-unitary $M$-tangling power of the channel as:
\begin{widetext}
\begin{equation}
\begin{split}
\epsilon_{p,M}(\mathcal{E}_{\text{deph}})&=2^M\sum_{r,s=0}^{2^{3N}-1}\text{Tr}[\mathcal{D}_{N,r}\otimes\mathcal{D}_{N,s}\langle \rho_0^{\otimes 2}\rangle (\mathcal{D}_{N,r}\otimes \mathcal{D}_{N,s})^\dagger \tilde{P} ]
\\&=
\left(\frac{d}{d+1}\right)^M \frac{1}{2}\left(1+\sum_{r,s}\Re[(\tilde{g}_0^{(r)})^*(\tilde{g}_1^{(s)})^*\tilde{g}_1^{(r)}(\tilde{g}_0^{(s)})]\right)\prod_{l=2}^M(1-G_1^{(l-1)})
\\&=
\left(\frac{d}{d+1}\right)^M \frac{1}{2}\left(1+\mathcal{R}\right)\prod_{l=2}^M(1-G_1^{(l-1)}),
\end{split}
\end{equation}
\end{widetext}
where we have made use of the results we found in the previous section. We have also defined $\mathcal{R}=\sum_{r,s}\Re[(\tilde{g}_0^{(r)})^*(\tilde{g}_1^{(s)})^*\tilde{g}_1^{(r)}(\tilde{g}_0^{(s)})]$. Note that in this case, the entangling power expresses only the evolution in the target subspace. In other words, we do not include the correlations that arise from the unwanted spins. In the limit of no dephasing, i.e., $\lambda_0=1$, $\lambda_1=0$ and $\theta=0$, the only Kraus operator that survives corresponds to the bitstring of $\textbf{m}=[0,\dots,0]^T$, and we recover the unitary channel. We can verify this from the above equation, since we will have $r=s=0$ and $\tilde{g}_0^{(r)}=\tilde{g}_1^{(r)}=\sqrt{\lambda}_0=1$.

The expression for $\tilde{g}_j^{(r)}$ is defined as follows:

\begin{widetext}
\begin{equation}
    \begin{split}
        \tilde{g}_j^{(r)} &= \prod_{k=1}^N \sqrt{\lambda_{m_1^{(k)}}\lambda_{m_2^{(k)}}\lambda_{m_3^{(k)}}}\text{exp}\left[\frac{i\theta t}{4}((-1)^{j\oplus m_1^{(k)}} +(-1)^{j\oplus m_3^{(k)}} +2(-1)^{j\oplus m_2^{(k)}\oplus 1} )\right]
        \\&=\sqrt{\prod_{k=1}^N \lambda_{m_1^{(k)}}\lambda_{m_2^{(k)}}\lambda_{m_3^{(k)}}}\text{exp}\left[\frac{i\theta t}{4}\sum_{k=1}^N\left((-1)^{j\oplus m_1^{(k)}} +(-1)^{j\oplus m_3^{(k)}} +2(-1)^{j\oplus m_2^{(k)}\oplus 1} \right)\right].
    \end{split}
\end{equation}
Using the above definitions and considering only 1 CPMG iteration, we find the following expression for the sum:
\begin{equation}
    \begin{split}
       \mathcal{R}= 
(\lambda_0^2+\lambda_1^2+2\lambda_0\lambda_1\cos(\theta t))^2(\lambda_0^2+\lambda_1^2+2\lambda_0\lambda_1\cos(2\theta t)).
    \end{split}
\end{equation}
For $N$ CPMG iterations we can show that $\mathcal{R}$ is given by:
\begin{equation}
    \begin{split}
        \mathcal{R}&=\Re\Big[\sum_{r,s}f^{(r)}(f^{(s)})^*\Big]
        \\&=\Big|\sum_r f^{(r)}\Big|^2
        \\&=\Big|\prod_{k=1}^N f^{(m_1^{(k)})}f^{(m_2^{(k)})}f^{(m_3^{(k)})}\Big|^2
        \\&=\prod_{k=1}^N \Big|f^{(m_1^{(k)})}f^{(m_2^{(k)})}f^{(m_3^{(k)})}\Big|^2
        \\&=\Big((\lambda_0^2+\lambda_1^2+2\lambda_0\lambda_1\cos(\theta t))^{2}(\lambda_0^2+\lambda_1^2+2\lambda_0\lambda_1\cos(2\theta t))\Big)^N.
    \end{split}
\end{equation}
In the first line we defined $f^{(r)}=(\tilde{g}_0^{(r)})^* \tilde{g}_1^{(r)}$.
Using the above result, we find that the expression of the $M$-tangling power of the target subspace in the presence of dephasing errors is:
\begin{equation}
    \epsilon_{p,M}(\mathcal{E}_{\text{deph}})=\Big(\frac{d}{d+1}\Big)^M\frac{1+(\lambda_0^2+\lambda_1^2+2\lambda_0\lambda_1\cos(\theta t))^{2N}(\lambda_0^2+\lambda_1^2+2\lambda_0\lambda_1\cos(2\theta t))^{N}}{2}\prod_{l=2}^M(1-G_1^{(l-1)}).
\end{equation}
It is easy to verify that in the limit of   $\lambda_0=1$, ($\lambda_0=0$) $\lambda_1=0$ ($\lambda_1=1$) we get the unitary $M$-tangling power as expected. 

If we are composing $p$ CPMG sequences of different unit times $t_j$ and iterations $N_j$ then we will have:
\begin{equation} 
    \mathcal{R}=
    \prod_{j=1}^p \Big(\lambda_0^2+\lambda_1^2+2\lambda_0\lambda_1 \cos(\theta t_j)\Big)^{2N_j} \Big(\lambda_0^2+\lambda_1^2+2\lambda_0\lambda_1 \cos(2\theta t_j)\Big)^{N_j}
\end{equation}
\end{widetext}
Based on the above expression we verify that in the limit where all $p$ sequences have the same time $t=t_j$, $\forall j\in[1,p]$, we recover the previous expression for $\mathcal{R}$. In particular, the above expression is the one we use for the sequential protocol when the electron undergoes dephasing.

\section{Comparison of sequential with multi-spin schemes in the presence of electronic dephasing \label{App:Seq_vs_Multi_Dephasing}}

In this section, we compare the performance of the multi-spin with the sequential scheme in the presence of electronic dephasing errors. In Fig.~\ref{fig:Deph_Case3}(a) we show $\epsilon_{p,M}(\mathcal{E}_{\text{deph}})$ of case ``16'' of preparing GHZ$_3$-like states for the sequential scheme. In Fig.~\ref{fig:Deph_Case3}(b) we show again $\epsilon_{p,M}(\mathcal{E}_{\text{deph}})$ of case ``7'' of preparing GHZ$_3$-like states for the multi-spin scheme. The nuclei we control with the sequential scheme are C5 and C13, and with the multi-spin scheme, C4 and C5.  The total sequence time for the sequential scheme is $1357.9~\mu$s, and for the multi-spin is $1916.6~\mu$s. Although the multi-spin scheme requires slightly longer time than the sequential one, it is again more robust to dephasing errors. This feature could be attributed to the different unit times $t_j$  and sequence iterations $N_j$ that enter the expression of $\epsilon_{p,M}(\mathcal{E}_{\text{deph}})$ for the sequential scheme since we are composing CPMG sequences of different timings and iterations in the sequential protocol.

\begin{figure*}[!htbp]
    \centering
    \includegraphics[scale=0.72]{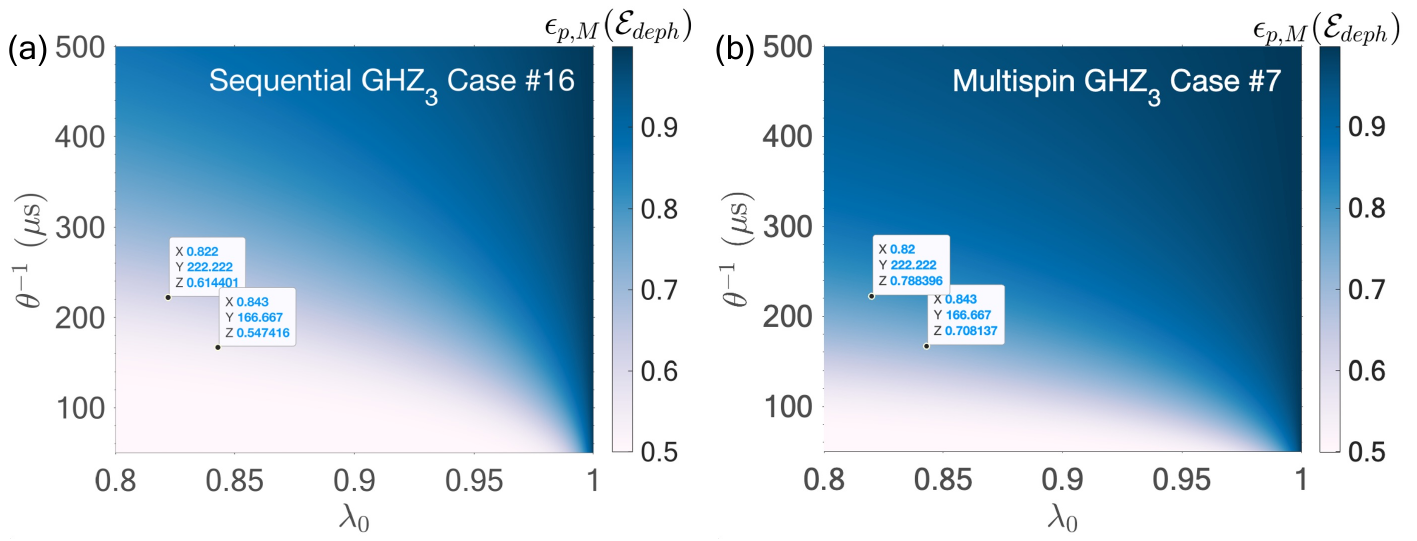}
    \caption{{\color{black}$M$-tangling power of the target subspace when the electron undergoes dephasing. (a) $\epsilon_{p,M}(\mathcal{E}_{\text{deph}})$ for case $\#~16$ for preparing GHZ$_3$-like states via the sequential scheme as a function of the dephasing angle $\theta$ and $\lambda_0$. (b) Same as in (a) for case $\#~7$ of the multi-spin scheme.}}
    \label{fig:Deph_Case3}
\end{figure*}

}

{\color{black}
\section{Comparison of XY2 and CPMG performance to pulse errors\label{App:Comparison_XY2_vs_CPMG}}

\begin{figure*}[!htbp]
    \centering
    \includegraphics[scale=0.67]{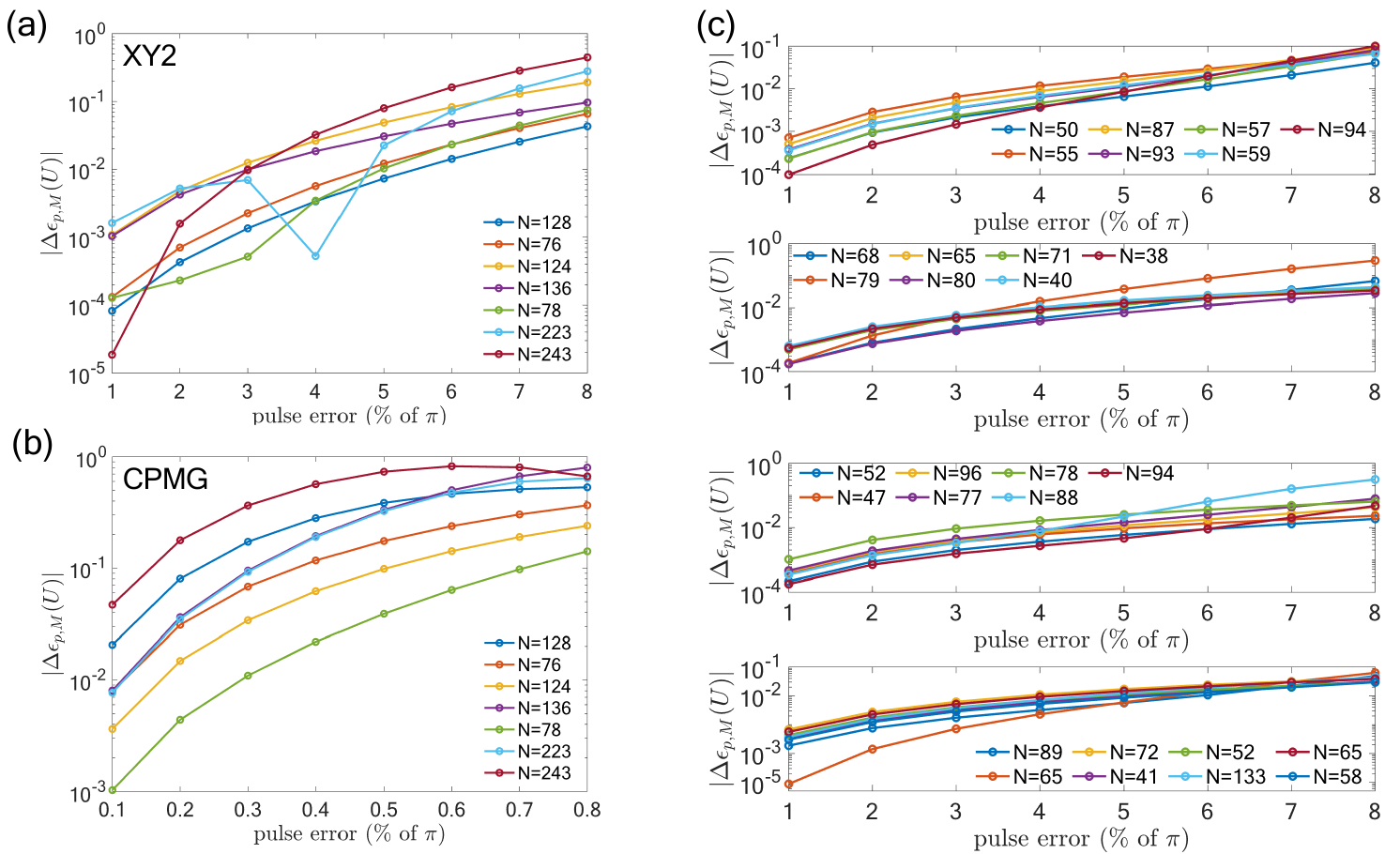}
    \caption{{\color{black}(a) $\Delta \epsilon_{p,M}(U):=|\epsilon_{p,M}(U)^{\epsilon=0}-\epsilon_{p,M}(U)^{\epsilon \neq 0}|$ as a function of systematic over-rotation pulse error for the 7 cases of preparing GHZ$_4$-like states via the multi-spin scheme with the XY2 sequence. (b) Same as (a), but using CPMG. (c) Same as (a), but for the 29 cases of preparing GHZ$_4$-like states via the sequential scheme, using XY2.}}
    \label{fig:Pulse_Errors_CPMG_VS_XY2}
\end{figure*}

In this section, we study over-/under-rotation errors for the electronic control pulses. We consider the optimal cases we found in the main text for preparing GHZ$_4$-like states via the sequential or multi-spin scheme. The systematic pulse errors are simulated as $R_{x/y}(\pi+\epsilon)=e^{-i\pi/2(1+\epsilon)\sigma_{x/y}}$ for the CPMG and XY2 sequences. We calculate the deviation between the $M$-tangling power of the target subspace in the absence of errors, $\epsilon_{p,M}(U)^{\epsilon=0}$, with the one in the presence of errors, $\epsilon_{p,M}(U)^{\epsilon \neq 0}$. We define this deviation as $\Delta \epsilon_{p,M}(U)=|\epsilon_{p,M}(U)^{\epsilon=0}-\epsilon_{p,M}(U)^{\epsilon \neq 0}|$.

In Fig.~\ref{fig:Pulse_Errors_CPMG_VS_XY2}(a), we show $\Delta \epsilon_{p,M}(U)$ for the 7 cases of preparing GHZ$_4$-like cases via the multi-spin scheme, assuming the XY2 sequence. The different colored lines correspond to each one of the 7 cases, and the labels correspond to the number of total sequence iterations. We see that for small pulse errors ($\leq 3-4 \%$), the deviation in estimating this quantity is on the order $10^{-3}-10^{-2}$. Additionally, we expect that as the number of pulses increases, the deviation is, in principle, enhanced [e.g., see dark red line versus green line]. This feature, however, does not always hold since the dynamics are case-dependent and tend to corrupt the GHZ preparation differently. In Fig.~\ref{fig:Pulse_Errors_CPMG_VS_XY2}(b), we repeat the same calculation assuming the CPMG sequence for a range of systematic pulse errors from $0.1-0.8\%$. As expected, the ability of the CPMG sequence to prepare high-quality GHZ$_4$-like states is severely impacted by the pulse errors.

In Fig.~\ref{fig:Pulse_Errors_CPMG_VS_XY2}(c), we display in four panels the deviation in $M$-tangling power for the 29 cases of preparing GHZ$_4$-like states using the sequential protocol. The labels correspond to the total number of iterations, $\sum_{j=1}^3 N_j$. In this scenario, we are assuming the XY2 sequence. We notice a similar performance as for the multi-spin scheme, and the results again reveal the robustness of XY2 under moderate pulse errors ($\leq 3-4\%$).

}

{\color{black}

\section{Rotation angle and rotation axis errors\label{App:angle_and_axes_errors}}

In this section we consider the effect of rotation angle and rotation axis errors on the $M$-tangling power. We consider as an example the 7 cases of preparing GHZ$_4$-like states via the multi-spin scheme. To showcase the robustness of the XY2 decoupling sequence, we consider both rotation axis and rotation angle errors, assuming the estimated parameters from Ref.~\cite{DeLangeThesis2012}, which we also mentioned in Sec.~\ref{Sec:Additional_Errors_Pulse_Errors}. In particular, we model the erroneous $\pi_X$ and $\pi_Y$ pulses as:
\begin{equation}
    R_x(\pi+\epsilon_x)=e^{-i(\pi+\epsilon_x)/2(\sqrt{1-q_z^2}\sigma_x+q_z\sigma_z)},
\end{equation}
\begin{equation}
    R_y(\pi+\epsilon_y)=e^{-i(\pi+\epsilon_y)/2(q_x\sigma_x+\sqrt{1-q_x^2-q_z^2}\sigma_y+q_z\sigma_z)}.
\end{equation}

In Fig.~\ref{fig:GHZ4_Multispin_Error_Rates_Delft} we show the ideal $M$-tangling power (dark blue bar) assuming a perfect CPMG decoupling sequence. The light blue bars correspond to the XY2 decoupling sequence which includes the errors $\epsilon_x=\epsilon_y=-0.02$, $q_z=0.05$, $q_x=0.005$. We observe that the deviation in the expected $M$-way correlations in the presence of both rotation axis and rotation angle errors is sufficiently small, and hence the XY2 decoupling sequence can be reliably used in experiments to prepare high-fidelity GHZ states. 

\begin{figure}[!htbp]
    \centering
    \includegraphics[scale=0.7]{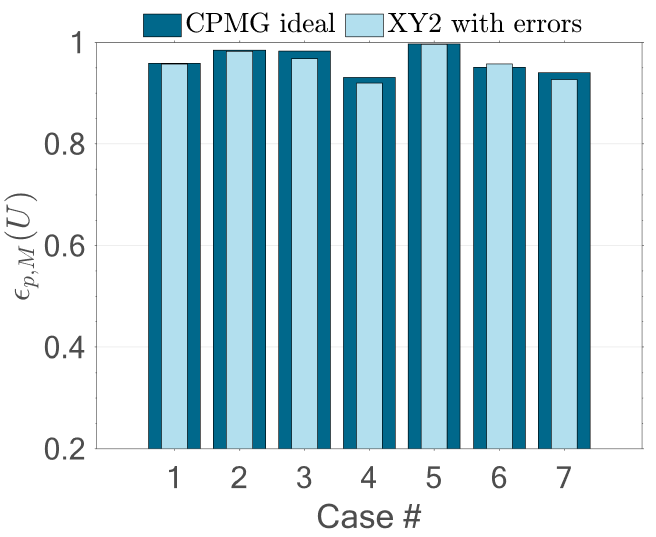}
    \caption{{\color{black}Robustness of XY2 decoupling sequence in the presence of pulse and rotation axis errors. The dark blue bars show the  $\epsilon_{p,M}(U)$ we find via the ideal CPMG sequence for the 7 cases of preparing GHZ$_4$-like states via the multi-spin scheme. The light blue bars show the $\epsilon_{p,M}(U)$ obtained via the XY2 decoupling sequence assuming both rotation axis and rotation angle errors.}}
    \label{fig:GHZ4_Multispin_Error_Rates_Delft}
\end{figure}

\section{Uncertainty in HF parameters}
In this section, we consider the effect of errors due to uncertainty in the experimentally measured HF parameters of the nuclear spins. We use the same optimal sequence parameters we found for the HF parameters we considered in the main text and additionally shift the HF parameters $A$ and $B$ of the target nuclear spins by 0.01 or 0.05 kHz. In Fig.~\ref{fig:HFdeviation_0.01}, we consider a 0.01 kHz shift and plot the deviation $\Delta \epsilon_{p,M}(U):=\epsilon_{p,M}(U)-\epsilon_{p,M}^\text{uncertain}(U)$ for the cases of preparing GHZ states up to 6 qubits for the sequential scheme in Fig.~\ref{fig:HFdeviation_0.01}(a) and for the multi-spin scheme in Fig.~\ref{fig:HFdeviation_0.01}(b). We observe that the presence of a 0.01 kHz uncertainty in the HF parameters only produces a negligible uncertainty on the order of $10^{-3}$ relative to our ideal calculations for $\epsilon_{p,M}(U)$. In Fig.~\ref{fig:HFdeviation_0.05}, we repeat the same calculations for the sequential scheme in Fig.~\ref{fig:HFdeviation_0.05}(a) and for the multi-spin scheme in Fig.~\ref{fig:HFdeviation_0.05}(b), assuming a shift of the HF parameters by 0.05 kHz. We note that the error in the estimation of $\epsilon_{p,M}(U)$ is on the order of $10^{-2}-10^{-3}$. Thus, based on the experimentally achievable accuracy in the measured HF parameters~\cite{TaminiauNat2019}, we can estimate the $M$-tangling power to a relatively narrow confidence interval. Optimization within a small range of the optimal parameters $t^*$ and $N^*$ that we find for definite values of HF parameters can guide the experimental calibration of the timing of the pulses and optimal sequence iterations.

\begin{figure*}[!htbp]
    \centering
    \includegraphics[scale=0.72]{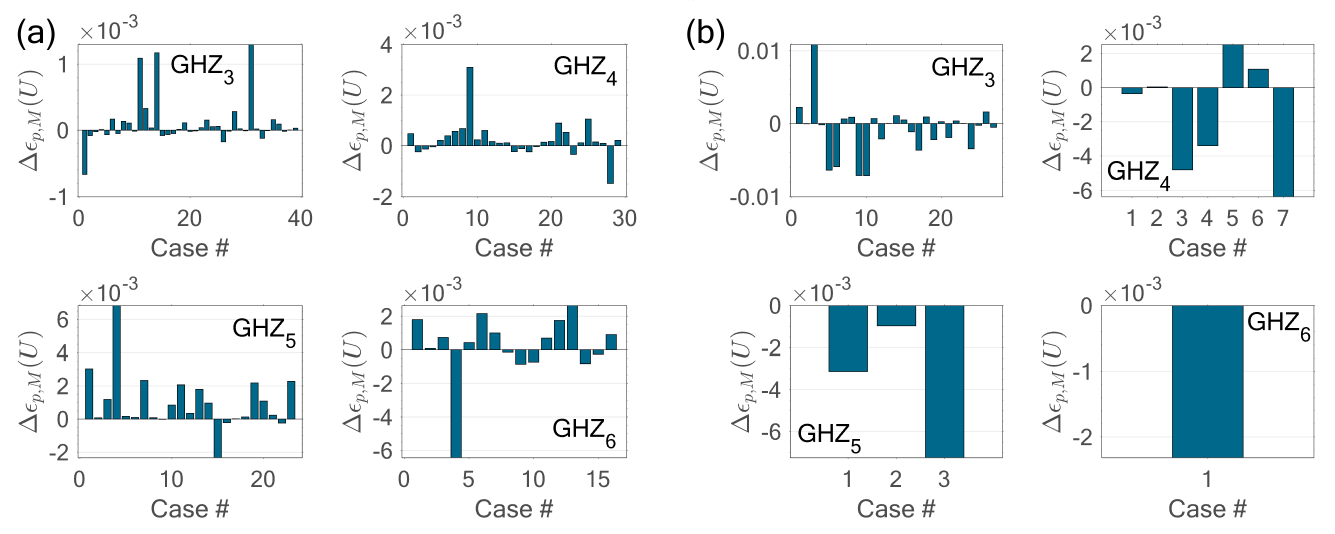}
    \caption{{\color{black}(a) $\Delta \epsilon_{p,M}(U):=\epsilon_{p,M}(U)-\epsilon_{p,M}^{\text{uncertain}}(U)$ for the sequential scheme. $\epsilon_{p,M}^{\text{uncertain}}(U)$ is the error value if we introduce a $0.01$ kHz shift in the HF parameters of the target spins. (b) $\Delta \epsilon_{p,M}(U)$ for the multi-spin scheme. $\epsilon_{p,M}^{\text{uncertain}}(U)$ is the error value if we introduce a $0.01$ kHz shift in the HF parameters of the target spins. The various panels show the deviation in the $M$-tangling power for GHZ$_3$-like, GHZ$_4$-like, GHZ$_5$-like, and GHZ$_6$-like states.}}
    \label{fig:HFdeviation_0.01}
\end{figure*}

\begin{figure*}[!htbp]
    \centering
    \includegraphics[scale=0.68]{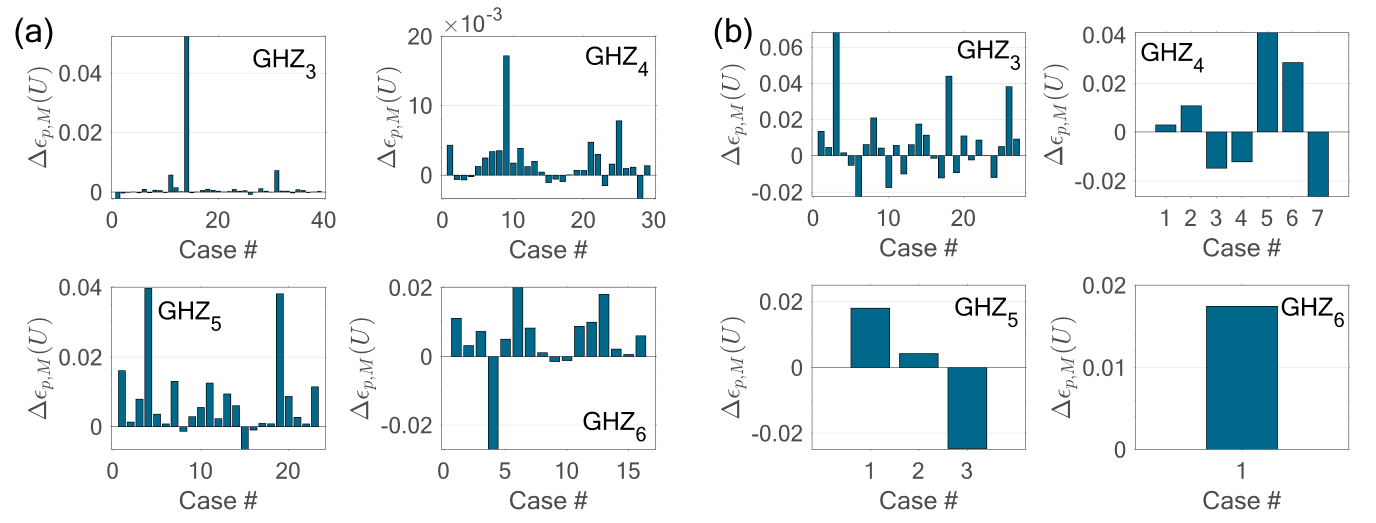}
    \caption{{\color{black}(a) $\Delta \epsilon_{p,M}(U):=\epsilon_{p,M}(U)-\epsilon_{p,M}^{\text{uncertain}}(U)$ for the sequential scheme. $\epsilon_{p,M}^{\text{uncertain}}(U)$ is the error value if we introduce a $0.05$ kHz shift in the HF parameters of the target spins. (b) $\Delta \epsilon_{p,M}(U)$ for the multi-spin scheme. $\epsilon_{p,M}^{\text{uncertain}}(U)$ is the error value if we introduce a $0.05$ kHz shift in the HF parameters of the target spins. The various panels show the deviation in the $M$-tangling power for GHZ$_3$-like, GHZ$_4$-like, GHZ$_5$-like, and GHZ$_6$-like states.}}
    \label{fig:HFdeviation_0.05}
\end{figure*}
}

\section{Optimization process for GHZ\texorpdfstring{$_M$}{M}-like generation \label{App:Optimization_Procedure}}

Here we highlight the optimization procedure for generating GHZ$_M$-like states using the sequential or multi-spin protocols. In Fig.~\ref{fig:FlowChart} of the main text, we show a diagram of the optimization procedure for the sequential protocol. First, we need to provide as inputs:
\begin{itemize}
    \item the size of the GHZ-like state we want to prepare ($|\text{GHZ}|$),
    \item the maximum resonance number ($k_{\text{max}}$) to search for each nuclear spin,
    \item the maximum time for the total sequence and of individual gates ($T_{max}$),
    \item the tolerances of unwanted/target nuclear one-tangles,
    \item the gate error tolerance.
\end{itemize}

\begin{table*}[!htbp]
\begin{center}
\resizebox{\textwidth}{!}{
\begin{tabular}{ |c|c|c|c|c|c|c|c|c|} 
 \hline
 GHZ  & Total gate time & Individual gate time  & Gate error  & Target   & Unwanted  & {\color{black}$k_{\text{max}}$} & {\color{black} $\delta t $} & {\color{black} time step}\\ 
 size & tol. ($\mu$s) & tol. ($\mu$s) & tol. & one-tangle tol. & one-tangle tol. & &{\color{black} ($\mu$s)} &{\color{black} ($\mu$s)}\\
 \hline
 3  & 2000 & {\color{black}2000}  & 0.1  & 0.99 & 0.1 & {\color{black}18} & {\color{black}0.2} & {\color{black} $5\times 10^{-3}$}\\
 4  & 2000 & {\color{black}2000} & 0.1  & 0.99 & 0.1 & {\color{black} 18 } & {\color{black}0.2} & {\color{black} $5\times 10^{-3}$}\\
 5  & 2300 & {\color{black}2300} & 0.1  & 0.9  & 0.1 & {\color{black} 18 } & {\color{black}0.2} & {\color{black} $5\times 10^{-3}$}\\
 6  & 2500 & {\color{black}2500} & 0.11 & 0.9  & 0.12 & {\color{black} 18} & {\color{black}0.2} & {\color{black} $5\times 10^{-3}$}\\
 7  & 3300  & {\color{black}3300} & 0.12 & 0.9  & 0.12 & {\color{black} 18} & {\color{black}0.2} & {\color{black} $5\times 10^{-3}$}\\
 8  & 3700 & {\color{black}2000} & 0.13 & 0.9  & 0.12 &  {\color{black} 18}& {\color{black}0.2} & {\color{black} $5\times 10^{-3}$}\\
 9  & 4000 & {\color{black}1400} & 0.13 & 0.85 & 0.15 & {\color{black} 18} & {\color{black}0.2} & {\color{black} $5\times 10^{-3}$}\\
 10 & 4000 & {\color{black}1400} & 0.19 & 0.87 & 0.22 & {\color{black}18} & {\color{black}0.2} & {\color{black} $5\times 10^{-3}$}\\
 \hline
\end{tabular}
}
\end{center}
\caption{Tolerances (tol.) {\color{black} and relevant parameters} for the optimization of the generation of GHZ$_M$-like states using the sequential protocol.}
\label{Tab:1}
\end{table*}

Then, for each nuclear spin, we search over all resonances by varying the unit time around each resonance {\color{black}by $\delta t=\pm 0.2~\mu$s, and with a time step of $5\times 10^{-3}~\mu$s and select as optimal the unit time $t$ where $\textbf{n}_0\cdot \textbf{n}_1$ is as close as possible to the value $-1$. Having fixed the unit time, we then use the analytical expressions for the minima of $G_1$ which give us the number of iterations that maximize the nuclear one-tangle [see Ref.~\cite{EconomouPRX2023} for the expressions of the minima].} Since $G_1$ is periodic, there are multiple numbers of iterations that can minimize $G_1$, and so we usually truncate to about {\color{black}15} maxima {\color{black}of the one-tangles}, {\color{black}which we also post-select such that $N\cdot t\leq T_{\text{max}}$}. {\color{black}Note that here $T_{\text{max}}$ could be the gate time restriction we impose for the gate time of a single decoupling sequence, rather than the total gate time restriction for the composition of all sequential gates}. We then inspect all elements in the sets of $\{t,N\}$ and keep those that give a sequence with gate time smaller than $T_{max}$. After this step, we calculate the remaining nuclear one-tangles and check if they are smaller than the unwanted one-tangle tolerance. If this is not satisfied, we reject that particular $(t,N)$ case, whereas if this is satisfied, we store the unit time and number of iterations. We repeat this process for each nuclear spin, such that we have all the possible unit times and iterations that can give maximal entanglement of the target spins with the electron while keeping the unwanted one-tangles minimal. At this stage, we have multiple unit times and iterations that satisfy these requirements.  

\begin{table*}[!htbp]
\begin{center}
\begin{tabular}{ |c|c|c|c|c|c|c|c|} 
 \hline
 GHZ  & Gate time  & Gate error  & Target   & Unwanted & {\color{black}$k_{\text{max}}$} & {\color{black}$\delta t$} & {\color{black} time values}\\ 
 size & tol. ($\mu$s) & tol. & one-tangle tol. & one-tangle tol. & & {\color{black}($\mu$s)} &  \\
 \hline
 3 & 2000 &  {\color{black}0.11}   & 0.9    & 0.1  & {\color{black}10} & {\color{black} 0.25} & {\color{black} 500} \\
 4 & 2000 & {\color{black}0.101}   & 0.9    & 0.1 & {\color{black}10} & {\color{black} 0.25} &  {\color{black}500}\\
 5 & 2300 & 0.1   & 0.84   & {\color{black}0.125} & {\color{black}10} & {\color{black} 0.25} & {\color{black}400}\\
 6 & 2500 & 0.13  & 0.88   & 0.12 & {\color{black}10} & {\color{black} 0.25} & {\color{black} 200}\\
 7 & 2800 & 0.13  & 0.85   & 0.15 & {\color{black}10} & {\color{black} 0.25} & {\color{black} 200}\\
 8 & 3000 & 0.15  & 0.85   & 0.15 & {\color{black}10} & {\color{black} 0.25} & {\color{black} 200}\\
 9 & 3000 & 0.15  & 0.82   & 0.15 & {\color{black}10} & {\color{black} 0.25} & {\color{black} 200}\\
 \hline
\end{tabular}
\end{center}
\caption{Tolerances (tol.) {\color{black} and relevant parameters} for the optimization of the generation of GHZ$_M$-like states using the multi-spin protocol.}
\label{Tab:2}
\end{table*}

We then combine the unit times and iterations corresponding to sets associated with the selection of any $|\text{GHZ}|-1$ nuclei out of the entire nuclear register, such that the total decoupling sequence does not exceed the maximum time, $T_{max}$. For each such nuclear spin combination, we evolve all nuclei individually under the composite evolution and obtain their nuclear one-tangles. At this step, each nuclear spin combination is associated with multiple unit times and iterations that we could consider, so we need to choose which $t$ and $N$ we keep for each distinct nuclear spin combination. We choose those $t$ and $N$ which give rise to a maximal $M$-tangle at the end of the composite $|\text{GHZ}|$-1 entangling gates. Different choices could be made here, e.g., selecting the $t$ and $N$ that give the shortest gate time or penalizing both gate time and the deviation from the maximal possible $M$-tangle using an appropriate cost function. At this stage, we have narrowed down nuclear spin candidates and associated each nuclear spin combination with a particular sequence composed of times $t_j$ and iterations $N_j$ that gives rise to maximal $M$-way entanglement. For each of these cases, we inspect whether the unwanted nuclear one-tangles of the composite evolution are below the unwanted one-tangle tolerance we imposed. If this is satisfied, we accept this case and proceed with the calculation of the gate error. Finally, we accept this case if the gate error is lower than the gate error tolerance. In Table~\ref{Tab:1}, we provide the tolerances {\color{black} and relevant parameters} we set for each GHZ size.

Regarding the multi-spin protocol, we are not composing gates since we are interested in a single-shot operation that generates direct entanglement of the electron with multiple nuclei. We explain here the procedure we follow according to the flowchart of Fig.~\ref{fig:FlowChart}. We start with one nuclear spin selected from the entire register and find its resonance time for some particular resonance number $k$. We vary the unit time $t$ within $\pm 0.25~\mu$s and define an upper bound for the maximum number of iterations that respects the gate time restriction for the particular unit time. Then, for all unit times and $N\in[1,N_{max}]$ we obtain all the nuclear one-tangles using the knowledge of rotation angles and the dot product of the nuclear axes for each nucleus given one iteration. Then, for all possible times and iterations, we check how many one-tangles are above the target one-tangle tolerance and how many are below the unwanted one-tangle tolerance. If we have $|\text{GHZ}|-1$ one-tangles above the target one-tangle tolerance, and all other nuclear one-tangles are below the unwanted one-tangle tolerance, then we accept this case. We repeat this process by choosing a different resonance number $k$, up to some $k_{\text{max}}$. Then, we select a different nuclear spin from the register as our starting point and repeat all the aforementioned steps.   

After completing the above stage, we have multiple times and iterations for which we maximize $|\text{GHZ}|-1$ target one-tangles simultaneously. We then rearrange all these cases corresponding to different unit times and iterations in terms of unique spin combinations. For each spin combination, we select the time and iterations of the single-shot operation, which give the maximal $M$-tangle. Similar to the sequential scheme, the choice of narrowing down particular $t$ and $N$ could be based on minimal gate time or a cost function that both minimizes gate time and maximizes the $M$-tangle for the particular spin combination. Finally, for each spin combination we evolve all unwanted nuclear spins individually, such that we pass this information to the calculation of the gate error. If the gate error is below the tolerance we imposed, we then accept this case.

\section{One-tangles of Sequential Scheme for generation of GHZ\texorpdfstring{$_M$}{M}-like states \label{App:OneTangles_Seq_GHZ}}

\begin{figure*}
    \centering
    \includegraphics[scale=0.68]{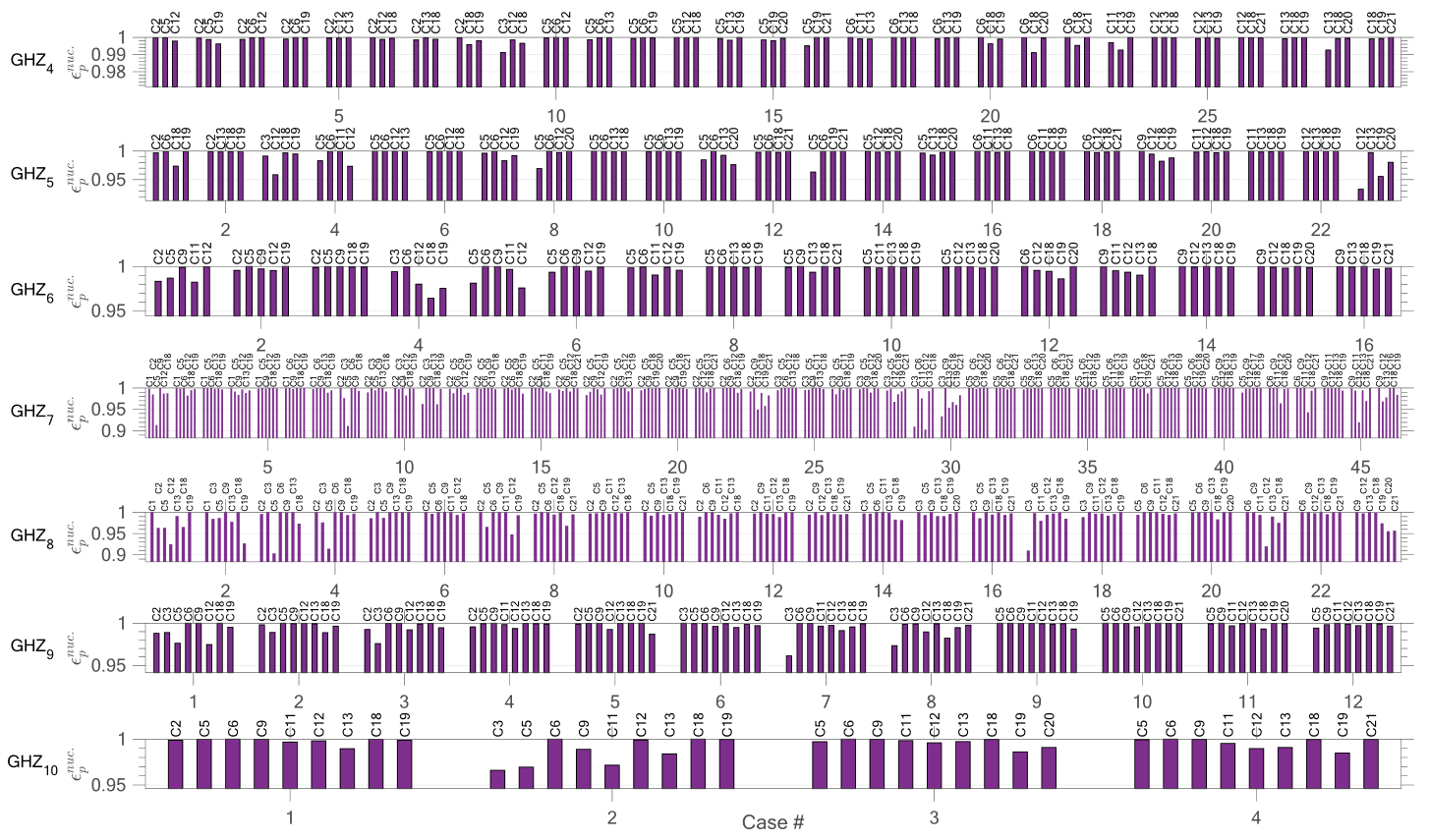}
    \caption{One-tangles of nuclear spins after composing $M-1$ sequential entangling gates to create the cases of Fig.~\ref{fig:GHZM_Sequential}. Each case corresponds to a distinct DD sequence that selects different nuclei from the nuclear spin register. Each bar shows the one-tangle of a particular nuclear spin, and the text above the bars corresponds to the nuclear spin labels. From top to bottom, we show the nuclear one-tangles for the generation of GHZ$_4$-like up to GHZ$_{10}$-like electron-nuclear entangled states.} 
    \label{fig:Nuclear_One_Tangles}
\end{figure*}

Here we present the nuclear one-tangles corresponding to the sequential entanglement scheme of Fig.~\ref{fig:GHZM_Sequential}. The nuclear one-tangles for all different realizations of entangling $M-1$ nuclei with the electron to prepare GHZ$_M$-like states are shown in Fig.~\ref{fig:Nuclear_One_Tangles}. The labels above each bar of each case refer to some $^{13} \text{C}$ nuclear spin of the register labeled as $\text{C}j$, with $j\in[1,27]$. 

\section{Eigendecomposition of mixed state \label{App:MixedState}}
To keep the discussion general, we consider $L$ nuclear spins, with $K$ of them being the target ones and hence, $L-K$ being the unwanted nuclei. To derive the optimal ensemble for the calculation of entanglement of the mixed state, we make the assumption that the initial state of the system is any arbitrary product pure state:
\begin{equation}
    |\psi_0\rangle =|\psi_{\text{el}}\rangle \otimes_{l=1}^L |\psi_{\text{nuc}}^{(l)}\rangle,
\end{equation}
where $|\psi_{\text{el}}\rangle=\alpha|0\rangle+\beta|1\rangle$, with $\alpha,\beta \in \mathbb{C}$ and $|\alpha|^2+|\beta|^2=1$. Under a CR-type evolution the electron-nuclear system evolves into:
\begin{equation}
    |\psi\rangle=U|\psi_0\rangle = \sum_{j\in \{0,1\}}\sigma_{jj}|\psi_{\text{el}}\rangle \otimes_{l=1}^LR^{(l)}_{\textbf{n}_j}|\psi_{\text{nuc}}^{(l)}\rangle.
\end{equation}
Thus, the full density matrix reads:
\begin{equation}
    \rho = \sum_{j,k \in\{0,1\}}\sigma_{jj}\rho_{\text{el}}\sigma_{kk} \otimes_l R_{\textbf{n}_j}^{(l)}\rho_{\text{nuc}}^{(l)} (R_{\textbf{n}_k}^\dagger)^{(l)}.
\end{equation}
Next, suppose that we wish to trace out the last $L-K$ spins from the density operator (assuming that we have ordered the basis such that these appear last in the Kronecker product):
\begin{equation}
\begin{split}
    \rho_{\text{red}}&=\sum_{j,k}[\sigma_{jj}\rho_{\text{el}}\sigma_{kk} \otimes_{l=1}^K R_{\textbf{n}_j}^{(l)}\rho_{\text{nuc}}^{(l)}(R_{\textbf{n}_k}^\dagger)^{(l)}]\\&\times \prod_{l=1}^{L-K} \text{Tr}[R_{\textbf{n}_j}^{(l+K)}\rho_{\text{nuc}}^{(l+K)}(R_{\textbf{n}_k}^\dagger)^{(l+K)}]\\&=\sum_{j,k}f_{jk}\sigma_{jj}\rho_{\text{el}}\sigma_{kk}\otimes_{l=1}^K R_{\textbf{n}_j}^{(l)}\rho_{\text{nuc}}^{(l)}(R_{\textbf{n}_k}^\dagger)^{(l)},
    \end{split}
\end{equation}
where we have defined $f_{jk}=\prod_{l=1}^{L-K} \text{Tr}[R_{\textbf{n}_j}^{(l+K)}\rho_{\text{nuc}}^{(l+K)}(R_{\textbf{n}_k}^\dagger)^{(l+K)}]$, with $f_{00}=1=f_{11}$ and $f_{10}=f_{01}^*$. Next, we are interested in finding the eigenvectors and eigenvalues of the reduced density matrix. To do so, we apply the inverse CR gates on the target subspace of the electron and $K$ target nuclei to find:
\begin{equation}
\begin{split}
    \rho_{\text{red}}'&=\sum_{jk}f_{jk}\sigma_{jj}\rho_{\text{el}}\sigma_{kk}\otimes_{l=1}^K \rho_{\text{nuc}}^{(l)}\\&=
    \begin{pmatrix}
    |\alpha|^2 & \alpha \beta^* f_{01} \\
    \alpha^*\beta f_{10} & |\beta|^2
    \end{pmatrix}\otimes_l \rho_{\text{nuc}}^{(l)}.
    \end{split}
\end{equation}
Thus, we can easily find the eigenvalues of the electron's reduced density matrix which read:
\begin{equation}
    \lambda_{\pm}=\frac{1}{2}\Big(1 \pm \sqrt{(|\alpha|^2-|\beta|^2)^2+4|\alpha|^2|\beta|^2|f_{01}|^2}\Big),
\end{equation}
whereas the eigenvectors are given by:
\begin{equation}
    v_{\pm}=\frac{1}{\sqrt{1+(\frac{|\beta|^2-\lambda_\pm}{c})^2}}\begin{pmatrix}
    -\frac{|\beta|^2-\lambda_\pm}{c} \\ 1
    \end{pmatrix},
\end{equation}
with $c=\alpha^*\beta f_{10}$ (provided $c\neq 0$). Now, to find the eigenvectors of the total reduced density matrix, we need to reapply the controlled gates of the target subspace on the matrix whose columns are the two eigenvectors:
\begin{equation}
    [|v_+'\rangle,|v_-'\rangle]^T=U [|v_+\rangle \otimes_l |\psi_{\text{nuc}}^{(l)}\rangle,|v_-\rangle \otimes_l |\psi_{\text{nuc}}^{(l)}\rangle]^T.
\end{equation}
Thus, we now have the diagonalized density operator:
\begin{equation}
    \rho_{\text{diag}}=\lambda_+|v_+'\rangle\langle v_+'|+\lambda_-|v_-'\rangle\langle v_-'|,
\end{equation}
which means that we can use the trial state:
\begin{equation}
    |\psi_{\text{trial}}\rangle = \sqrt{\lambda_+}|v_+'\rangle -e^{i\chi}\sqrt{1-\lambda_+}|v_-'\rangle,
\end{equation}
to find the entanglement of the mixed reduced density matrix of the electron and the target nuclei. Note that given arbitrary initial states, the highest rank that the reduced density matrix can have (irrespective of the initial pure state of the system, the number of total nuclei, or the number of nuclei we trace out) is 2, due to the form of the controlled evolution operator. 

In the case when the electron starts from a non-superposition state e.g., when $\beta=0$, we find that the two eigenvalues are $\lambda_{+}=1$ and $\lambda_-=0$. The corresponding eigenvectors
are $|v_+\rangle=|0\rangle$ and $|v_-\rangle=|1\rangle$, which after reapplying the controlled gates read:
\begin{equation}
    [|v_+'\rangle , |v'_-\rangle]^T=U[|0\rangle\otimes_{l}|\psi_{\text{nuc}}^{(l)}\rangle,|1\rangle\otimes_{l}|\psi_{\text{nuc}}^{(l)}\rangle]^T.
\end{equation}
In this scenario, since $\lambda_-=0$, the reduced density matrix corresponds to a pure product state.

\section{Concurrence and three-tangle \label{App:C_VS_Tau3}}

Here we compare the entanglement of rank-2 mixed states involving two qubits with the entanglement of rank-2 mixed states involving three qubits. For two-qubit pure states the concurrence is defined as $C(\rho)=\sqrt{2(1-\text{Tr}[\rho_A^2])}=\sqrt{2(1-\text{Tr}[\rho_B^2])}$, where $\rho_{A(B)}$ is the reduced density matrix of system A (B). The entanglement of a mixed two-qubit state can be computed with similar methods as in the main text by minimizing the concurrence of a trial state. We assume that we have a trial state of the form:
\begin{equation}
    |\psi_{\text{trial},2}\rangle = \sqrt{p}|\Phi^+\rangle + \sqrt{1-p}e^{i\chi}|\Phi^-\rangle,
\end{equation}
where $|\Phi^{\pm}\rangle =1/\sqrt{2}(|00\rangle\pm|11\rangle)$ are the two Bell states. We further consider a rank-2 mixed three-qubit state which is the mixture of two $|\text{GHZ}^\pm\rangle=1/\sqrt{2}(|0\rangle^{\otimes 3}\pm |1\rangle^{\otimes 3})$ states. Thus, similarly to the main text, we define the trial state:
\begin{equation}
    |\psi_{\text{trial},3}\rangle = \sqrt{p}|\text{GHZ}^+\rangle+\sqrt{1-p}e^{i\chi}|\text{GHZ}^-\rangle .
\end{equation}

\begin{figure}[!htbp]
    \centering
    \includegraphics[scale=0.7]{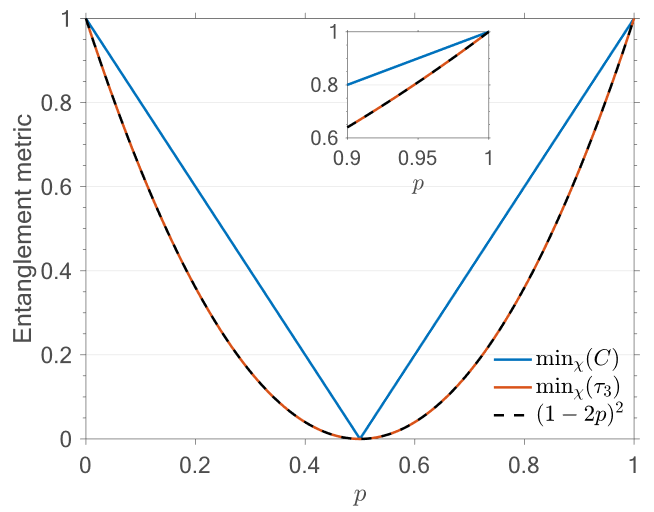}
    \caption{Minimum concurrence of a mixture of Bell states (blue), and minimum three-tangle of a mixture of $|\text{GHZ}^\pm\rangle$ states (red). The black dashed line is the analytical expression of the minimum three-tangle. }
    \label{fig:C_VS_Tau3}
\end{figure}

For each value of $p\in [0,1]$  we vary the relative phase of the two terms, $\chi$, for both $|\psi_{\text{trial},2}\rangle$ and $|\psi_{\text{trial},3}\rangle$, and obtain the minimal concurrence, or three-tangle, respectively. We plot the results as a function of $p$ in Fig.~\ref{fig:C_VS_Tau3}. We observe that the minimum concurrence, as well as the minimum three-tangle are both convex functions. In fact, the minimum concurrence is given by $C=2|p-1/2|$ and the minimum three-tangle by $(1-2p)^2$. We note that for all values of $p$ (except for $p=1/2$ where we have maximal mixed states) the minimum three-tangle is lower than the minimum concurrence, and hence the three-way entanglement is more sensitive to the relative ratio of the superposition terms in the mixed state than the concurrence of two-qubit mixed states.

\subsection{Parameters for 27 nuclear spins \label{App:27Spins}}

The HF parameters for the 27 nuclear spin register we consider in the main text are presented in Table~\ref{tab:3} and can also be found in Refs.~\cite{TaminiauNat2019,BradleyThesis2021}. 
\begin{table}[!htbp]
\centering
\resizebox{\columnwidth}{!}{
\begin{tabular}{c|c|c||c|c|c}
     C atoms & $\frac{A}{2\pi}$ (kHz) & $\frac{B}{2\pi}$ (kHz) & C atoms & $\frac{A}{2\pi}$ (kHz) & $\frac{B}{2\pi}$ (kHz) \\  

C1 & -20.72 & 12 & C14 & -19.815 & 5.3\\
C2 & -23.22 & 13 & C15 & -13.971 & 9\\
C3 & -31.25 & 8 & C16 & -4.66 & 7\\
C4 & -14.07 & 13 & C17 & -5.62 & 5\\
C5 & -11.346 & 59.21 & C18 & -36.308 & 26.62\\
C6 & -48.58 & 9 & C19 & 24.399 & 24.81\\
C7 & -8.32 & 3 & C20 & 2.690 & 11\\
C8 & -9.79 & 5 & C21 & 1.212 & 13\\
C9 & 213.154 & 3 & C22 & 7.683 & 4\\
C10 & 17.643 & 8.6 & C23 & -3.177 & 2\\
C11 & 14.548 & 10 & C24 & -4.225 & 0\\
C12 & 20.569 & 41.51 & C25 & -3.873 & 0\\
C13 & 8.029 & 21.0 & C26 & -3.618 & 0\\
C27 & -4.039 & 0
\\
\end{tabular}
}
    \caption{Hyperfine parameters of the $^{13}$C atoms we considered in the paper.}
    \label{tab:3}
\end{table}


\begin{thebibliography}{10}

\bibitem{BriegelPRL2001}
Robert Raussendorf and Hans~J. Briegel.
\newblock ``A one-way quantum computer''.
\newblock \href{https://dx.doi.org/10.1103/PhysRevLett.86.5188}{Phys. Rev. Lett. {\bf 86}, 5188--5191}~(2001).

\bibitem{BriegelNat2009}
H.~J. Briegel, D.~E. Browne, W.~Dur, R.~Raussendorf, and M.~{Van den Nest}.
\newblock ``Measurement-based quantum computation''.
\newblock \href{https://dx.doi.org/https://doi.org/10.1038/nphys1157}{Nature {\bf 5}, 19--26}~(2009).

\bibitem{Raussendorf2012}
Robert Raussendorf and Tzu-Chieh Wei.
\newblock ``Quantum computation by local measurement''.
\newblock \href{https://dx.doi.org/https://doi.org/10.1146/annurev-conmatphys-020911-125041}{Annual Review of Condensed Matter Physics {\bf 3}, 239--261}~(2012).

\bibitem{FBQCSparrow2021}
Sara Bartolucci, Patrick Birchall, Hector Bombin, Hugo Cable, Chris Dawson, Mercedes Gimeno-Segovia, Eric Johnston, Konrad Kieling, Naomi Nickerson, Mihir Pant, Fernando Pastawski, Terry Rudolph, and Chris Sparrow.
\newblock ``Fusion-based quantum computation''.
\newblock  \href{https://doi.org/10.1038/s41467-023-36493-1}{Nat. Commun. {\bf 14}, 912}~(2023).

\bibitem{HilleryPRA1999}
Mark Hillery, Vladim\'{\i}r Bu\ifmmode~\check{z}\else \v{z}\fi{}ek, and Andr\'e Berthiaume.
\newblock ``Quantum secret sharing''.
\newblock \href{https://dx.doi.org/10.1103/PhysRevA.59.1829}{Phys. Rev. A {\bf 59}, 1829--1834}~(1999).

\bibitem{TittelPRA2001}
W.~Tittel, H.~Zbinden, and N.~Gisin.
\newblock ``Experimental demonstration of quantum secret sharing''.
\newblock \href{https://dx.doi.org/10.1103/PhysRevA.63.042301}{Phys. Rev. A {\bf 63}, 042301}~(2001).

\bibitem{Chen2005}
K.~Chen and H.-K. Lo.
\newblock ``Conference key agreement and quantum sharing of classical secrets with noisy ghz states''.
\newblock In Proceedings. International Symposium on Information Theory, 2005. ISIT 2005.
\newblock \href{https://dx.doi.org/10.1109/ISIT.2005.1523616}{Pages 1607--1611}.
\newblock ~(2005).

\bibitem{ChangQIP2013}
Y.-J. Chang, C.-W. Tsai, and T.~Hwang.
\newblock ``Multi-user private comparison protocol using ghz class states''.
\newblock \href{https://dx.doi.org/https://doi.org/10.1007/s11128-012-0454-z}{Quantum Inf. Process. {\bf 12}, 1077–1088}~(2013).

\bibitem{BellNatCommun2014SecretSharing}
B.~A. Bell, D.~Markham, D.~A. Herrera-Martí, A.~Marin, W.~J. Wadsworth, J.~G. Rarity, and M.~S. Tame.
\newblock ``Experimental demonstration of graph-state quantum secret sharing''.
\newblock \href{https://dx.doi.org/10.1038/ncomms6480}{Nat. Commun. {\bf 5}, 5480}~(2014).

\bibitem{BensonNewJPhys2014}
M.~Leifgen, T.~Schröder, F.~Gädeke, R.~Riemann, V.~Métillon, E.~Neu, C.~Hepp, C.~Arend, C.~Becher, K.~Lauritsen, and O.~Benson.
\newblock ``Evaluation of nitrogen- and silicon-vacancy defect centres as single photon sources in quantum key distribution''.
\newblock \href{https://dx.doi.org/10.1088/1367-2630/16/2/023021}{New. J. Phys. {\bf 16}, 023021}~(2014).

\bibitem{MunroPRA2017}
Nicol\'o Lo~Piparo, Mohsen Razavi, and William~J. Munro.
\newblock ``Memory-assisted quantum key distribution with a single nitrogen-vacancy center''.
\newblock \href{https://dx.doi.org/10.1103/PhysRevA.96.052313}{Phys. Rev. A {\bf 96}, 052313}~(2017).

\bibitem{MonroeSciAvd2017}
Norbert~M. Linke, Mauricio Gutierrez, Kevin~A. Landsman, Caroline Figgatt, Shantanu Debnath, Kenneth~R. Brown, and Christopher Monroe.
\newblock ``Fault-tolerant quantum error detection''.
\newblock \href{https://dx.doi.org/DOI: 10.1126/sciadv.1701074}{Sci. Adv. {\bf 3}, e1701074}~(2017).

\bibitem{MorenoQIP2018}
M.~G.~M. Moreno, A.~Fonseca, and M.~M. Cunha.
\newblock ``Using three-partite ghz states for partial quantum error detection in entanglement-based protocols''.
\newblock \href{https://dx.doi.org/https://doi.org/10.1007/s11128-018-1960-4}{Quantum Inf. Process. {\bf 17}, 191}~(2018).

\bibitem{NickersonNatCommun2013}
N.~H. Nickerson, Y.~Li, and S.~C. Benjamin.
\newblock ``Topological quantum computing with a very noisy network and local error rates approaching one percent''.
\newblock \href{https://dx.doi.org/10.1038/ncomms2773}{Nat. Commun. {\bf 4}, 1756}~(2013).

\bibitem{BellNatCommun2014}
B.~A. Bell, D.~A. Herrera-Martí, M.~S. Tame, D.~Markham, W.~J. Wadsworth, and J.~G. Rarity.
\newblock ``Experimental demonstration of a graph state quantum error-correction code''.
\newblock \href{https://dx.doi.org/10.1038/ncomms4658}{Nat. Commun. {\bf 5}, 3658}~(2014).

\bibitem{WratchtrupNature2014}
G.~Waldherr, Y.~Wang, S.~Zaiser, M.~Jamali, T.~Schulte-Herbrüggen, H.~Abe, T.~Ohshima, J.~Isoya, J.~F. Du, P.~Neumann, and J.~Wrachtrup.
\newblock ``Quantum error correction in a solid-state hybrid spin register''.
\newblock \href{https://dx.doi.org/https://doi.org/10.1038/nature12919}{Nature {\bf 506}, 204--207}~(2014).

\bibitem{TaminiauNatNano2014}
T.~H. Taminiau, J.~Cramer, T.~van~der Sar, V.~V. Dobrovitski, and R.~Hanson.
\newblock ``Universal control and error correction in multi-qubit spin registers in diamond''.
\newblock \href{https://dx.doi.org/https://doi.org/10.1038/nnano.2014.2}{Nat. Nanotechnol. {\bf 9}, 171--176}~(2014).

\bibitem{CramerNatCommun2016}
J.~Cramer, N.~Kalb, M.~A. Rol, B.~Hensen, M.~S. Blok, M.~Markham, D.~J. Twitchen, R.~Hanson, and T.~H. Taminiau.
\newblock ``Repeated quantum error correction on a continuously encoded qubit by real-time feedback''.
\newblock \href{https://dx.doi.org/10.1038/ncomms11526}{Nat. Commun. {\bf 7}, 11526}~(2016).

\bibitem{TaminiauNature2022}
M.~H. Abobeih, Y.~Wang, J.~Randall, S.~J.~H. Loenen, C.~E. Bradley, M.~Markham, D.~J. Twitchen, B.~M. Terhal, and T.~H. Taminiau.
\newblock ``Fault-tolerant operation of a logical qubit in a diamond quantum processor''.
\newblock \href{https://dx.doi.org/https://doi.org/10.5281/zenodo.6461872}{Nature {\bf 606}, 884–889}~(2022).

\bibitem{EldredgePRA2018}
Zachary Eldredge, Michael Foss-Feig, Jonathan~A. Gross, S.~L. Rolston, and Alexey~V. Gorshkov.
\newblock ``Optimal and secure measurement protocols for quantum sensor networks''.
\newblock \href{https://dx.doi.org/10.1103/PhysRevA.97.042337}{Phys. Rev. A {\bf 97}, 042337}~(2018).

\bibitem{KoczorNewJPhys2020}
B.~Koczor, S.~Endo, T.~Jones, Y.~Matsuzaki, and S.~C. Benjamin.
\newblock ``Variational-state quantum metrology''.
\newblock \href{https://dx.doi.org/10.1088/1367-2630/ab965e}{New J. Phys. {\bf 22}, 083038}~(2020).

\bibitem{HansonNature2013}
H.~Bernien, B.~Hensen, W.~Pfaff, G.~Koolstra, M.~S. Blok, L.~Robledo, T.~H. Taminiau, M.~Markham, D.~J. Twitchen, L.~Childress, and R.~Hanson.
\newblock ``Heralded entanglement between solid-state qubits separated by three metres''.
\newblock \href{https://dx.doi.org/https://doi.org/10.1038/nature12016}{Nature {\bf 497}, 86--90}~(2013).

\bibitem{HumphreysNat18}
P.~C. Humphreys, N.~Kalb, J.~P.~J. Morits, R.~N. Schouten, R.~F.~L. Vermeulen, D.~J. Twitchen, M.~Markham, and R.~Hanson.
\newblock ``Deterministic delivery of remote entanglement on a quantum network''.
\newblock \href{https://dx.doi.org/https://doi.org/10.1038/s41586-018-0200-5}{Nature {\bf 558}, 268--273}~(2018).

\bibitem{PompiliSci21}
M.~Pompili, S.~L.~N. Hermans, S.~Baier, H.~K.~C. Beukers, P.~C. Humphreys, R.~N. Schouten, R.~F.~L. Vermeulen, M.~J. Tiggelman, L.~dos Santos~Martins, B.~Dirkse, S.~Wehner, and R.~Hanson.
\newblock ``Realization of a multinode quantum network of remote solid-state qubits''.
\newblock \href{https://dx.doi.org/DOI: 10.1126/science.abg1919}{Sci. {\bf 372}, 259--264}~(2021).

\bibitem{HermansNature2022}
S.~L.~N. Hermans, M.~Pompili, H.~K.~C. Beukers, S.~Baier, J.~Borregaard, and R.~Hanson.
\newblock ``Qubit teleportation between non-neighbouring nodes in a quantum network''.
\newblock \href{https://dx.doi.org/https://doi.org/10.1038/s41586-022-04697-y}{Nature {\bf 605}, 663--668}~(2022).

\bibitem{WrachtrupNatCommun2016}
S.~Zaiser, T.~Rendler, I.~Jakobi, T.~Wolf, S.-Y. Lee, S.~Wagner, V.~Bergholm, T.~Schulte-Herbrüggen, P.~Neumann, and J.~Wrachtrup.
\newblock ``Enhancing quantum sensing sensitivity by a quantum memory''.
\newblock \href{https://dx.doi.org/https://doi.org/10.1038/ncomms12279}{Nat. Commun. {\bf 7}, 12279}~(2016).

\bibitem{CappellaroPRapplied2019}
Alexandre Cooper, Won Kyu~Calvin Sun, Jean-Christophe Jaskula, and Paola Cappellaro.
\newblock ``Environment-assisted quantum-enhanced sensing with electronic spins in diamond''.
\newblock \href{https://dx.doi.org/10.1103/PhysRevApplied.12.044047}{Phys. Rev. Applied {\bf 12}, 044047}~(2019).

\bibitem{WrachtrupNpj2021}
V.~Vorobyov, S.~Zaiser, N.~Abt, J.~Meinel, D.~Dasari, P.~Neumann, and J.~Wrachtrup.
\newblock ``Quantum fourier transform for nanoscale quantum sensing''.
\newblock \href{https://dx.doi.org/https://doi.org/10.1038/s41534-021-00463-6}{Npj Quantum Inf. {\bf 7}, 124}~(2021).

\bibitem{KalbSci2017}
N.~Kalb, A.~A. Reiserer, P.~C. Humphreys, J.~J.~W. Bakermans, S.~J. Kamerling, N.~H. Nickerson, S.~C. Benjamin, D.~J. Twitchen, M.~Markham, and R.~Hanson.
\newblock ``Entanglement distillation between solid-state quantum network nodes''.
\newblock \href{https://dx.doi.org/DOI: 10.1126/science.aan0070}{Sci. {\bf 356}, 928--932}~(2017).

\bibitem{TaminiauPRL2012}
T.~H. Taminiau, J.~J.~T. Wagenaar, T.~van~der Sar, F.~Jelezko, V.~V. Dobrovitski, and R.~Hanson.
\newblock ``Detection and control of individual nuclear spins using a weakly coupled electron spin''.
\newblock \href{https://dx.doi.org/10.1103/PhysRevLett.109.137602}{Phys. Rev. Lett. {\bf 109}, 137602}~(2012).

\bibitem{CiracPRL1997}
S.~F. Huelga, C.~Macchiavello, T.~Pellizzari, A.~K. Ekert, M.~B. Plenio, and J.~I. Cirac.
\newblock ``Improvement of frequency standards with quantum entanglement''.
\newblock \href{https://dx.doi.org/10.1103/PhysRevLett.79.3865}{Phys. Rev. Lett. {\bf 79}, 3865--3868}~(1997).

\bibitem{CarvalhoPRL2004}
Andr\'e R.~R. Carvalho, Florian Mintert, and Andreas Buchleitner.
\newblock ``Decoherence and multipartite entanglement''.
\newblock \href{https://dx.doi.org/10.1103/PhysRevLett.93.230501}{Phys. Rev. Lett. {\bf 93}, 230501}~(2004).

\bibitem{BradleyPRX19}
C.~E. Bradley, J.~Randall, M.~H. Abobeih, R.~C. Berrevoets, M.~J. Degen, M.~A. Bakker, M.~Markham, D.~J. Twitchen, and T.~H. Taminiau.
\newblock ``A ten-qubit solid-state spin register with quantum memory up to one minute''.
\newblock \href{https://dx.doi.org/10.1103/PhysRevX.9.031045}{Phys. Rev. X {\bf 9}, 031045}~(2019).

\bibitem{LukinPRL2019}
C.~T. Nguyen, D.~D. Sukachev, M.~K. Bhaskar, B.~Machielse, D.~S. Levonian, E.~N. Knall, P.~Stroganov, R.~Riedinger, H.~Park, M.~Lon\ifmmode~\check{c}\else \v{c}\fi{}ar, and M.~D. Lukin.
\newblock ``Quantum network nodes based on diamond qubits with an efficient nanophotonic interface''.
\newblock \href{https://dx.doi.org/10.1103/PhysRevLett.123.183602}{Phys. Rev. Lett. {\bf 123}, 183602}~(2019).

\bibitem{LukinPRB2019}
C.~T. Nguyen, D.~D. Sukachev, M.~K. Bhaskar, B.~Machielse, D.~S. Levonian, E.~N. Knall, P.~Stroganov, C.~Chia, M.~J. Burek, R.~Riedinger, H.~Park, M.~Lon\ifmmode~\check{c}\else \v{c}\fi{}ar, and M.~D. Lukin.
\newblock ``An integrated nanophotonic quantum register based on silicon-vacancy spins in diamond''.
\newblock \href{https://dx.doi.org/10.1103/PhysRevB.100.165428}{Phys. Rev. B {\bf 100}, 165428}~(2019).

\bibitem{BourassaNatMater2020}
A.~Bourassa, C.r~P. Anderson, K.~C. Miao, M.~Onizhuk, H.~Ma, A.~L. Crook, H.~Abe, J.~Ul-Hassan, T.~Ohshima, N.~T. Son, G.~Galli, and D.~D. Awschalom.
\newblock ``Entanglement and control of single nuclear spins in isotopically engineered silicon carbide''.
\newblock \href{https://dx.doi.org/https://doi.org/10.1038/s41563-020-00802-6}{Nat. Mater. {\bf 19}, 1319–1325}~(2020).

\bibitem{TaminiauNat2019}
M.~H. Abobeih, J.~Randall, C.~E. Bradley, H.~P. Bartling, M.~A. Bakker, M.~J. Degen, M.~Markham, D.~J. Twitchen, and T.~H. Taminiau.
\newblock ``Atomic-scale imaging of a 27-nuclear-spin cluster using a quantum sensor''.
\newblock \href{https://dx.doi.org/https://doi.org/10.1038/s41586-019-1834-7}{Nature {\bf 576}, 411--415}~(2019).

\bibitem{EconomouPRX2023}
Evangelia Takou, Edwin Barnes, and Sophia~E. Economou.
\newblock ``Precise control of entanglement in multinuclear spin registers coupled to defects''.
\newblock \href{https://dx.doi.org/10.1103/PhysRevX.13.011004}{Phys. Rev. X {\bf 13}, 011004}~(2023).

\bibitem{CarrPurcellPhysRev54}
H.~Y. Carr and E.~M. Purcell.
\newblock ``Effects of diffusion on free precession in nuclear magnetic resonance experiments''.
\newblock \href{https://dx.doi.org/10.1103/PhysRev.94.630}{Phys. Rev. {\bf 94}, 630--638}~(1954).

\bibitem{MeiboomGill58}
S.~Meiboom and D.~Gill.
\newblock ``Modified spin‐echo method for measuring nuclear relaxation times''.
\newblock \href{https://dx.doi.org/https://doi.org/10.1063/1.1716296}{Rev. Sci. Instrum. {\bf 29}, 688--691}~(1958).

\bibitem{deLangeSci10}
G.~de~Lange, Z.~H. Wang, D.~Ristè, V.~V. Dobrovitski, and R.~Hanson.
\newblock ``Universal dynamical decoupling of a single solid-state spin from a spin bath''.
\newblock \href{https://dx.doi.org/DOI: 10.1126/science.1192739}{Sci. {\bf 330}, 60--63}~(2010).

\bibitem{Gullion_Journal_Mag_Res1969}
Terry Gullion, David~B Baker, and Mark~S Conradi.
\newblock ``New, compensated carr-purcell sequences''.
\newblock \href{https://dx.doi.org/https://doi.org/10.1016/0022-2364(90)90331-3}{Journal of Magnetic Resonance (1969) {\bf 89}, 479--484}~(1990).

\bibitem{UhrigNewJPhys08}
G.~S. Uhrig.
\newblock ``Exact results on dynamical decoupling by $\pi$ pulses in quantum information processes''.
\newblock \href{https://dx.doi.org/10.1088/1367-2630/10/8/083024}{New J. Phys. {\bf 10}, 083024}~(2008).

\bibitem{UhrigPRL07}
G\"otz~S. Uhrig.
\newblock ``Keeping a quantum bit alive by optimized $\ensuremath{\pi}$-pulse sequences''.
\newblock \href{https://dx.doi.org/10.1103/PhysRevLett.98.100504}{Phys. Rev. Lett. {\bf 98}, 100504}~(2007).

\bibitem{RBLiuNatNanotechnol2011}
N.~Zhao, J.-L. Hu, S.-W. Ho, J.~T.~K. Wan, and R.~B. Liu.
\newblock ``Atomic-scale magnetometry of distant nuclear spin clusters via nitrogen-vacancy spin in diamond''.
\newblock \href{https://dx.doi.org/https://doi.org/10.1038/nnano.2011.22}{Nat. Nanotechnol {\bf 6}, 242–246}~(2011).

\bibitem{HansonPRB2012}
Zhi-Hui Wang, G.~de~Lange, D.~Rist\`e, R.~Hanson, and V.~V. Dobrovitski.
\newblock ``Comparison of dynamical decoupling protocols for a nitrogen-vacancy center in diamond''.
\newblock \href{https://dx.doi.org/10.1103/PhysRevB.85.155204}{Phys. Rev. B {\bf 85}, 155204}~(2012).

\bibitem{Dong2020}
W.~Dong, F.~A. Calderon-Vargas, and S.~E Economou.
\newblock ``Precise high-fidelity electron–nuclear spin entangling gates in nv centers via hybrid dynamical decoupling sequences''.
\newblock \href{https://dx.doi.org/10.1088/1367-2630/ab9bc0}{New J. Phys. {\bf 22}, 073059}~(2020).

\bibitem{PfaffNatPhys2013}
W.~Pfaff, T.~H. Taminiau, L.~Robledo, Bernien H, M.~Markham, D.~J. Twitchen, and R.~Hanson.
\newblock ``Demonstration of entanglement-by-measurement of solid-state qubits''.
\newblock \href{https://dx.doi.org/https://doi.org/10.1038/nphys2444}{Nat. Phys. {\bf 9}, 29--33}~(2013).

\bibitem{AbobeihThesis2021}
M.~Abobeih.
\newblock ``From atomic-scale imaging to quantum fault-tolerance with spins in diamond''.
\newblock \href{https://dx.doi.org/https://doi.org/10.4233/uuid:cce8dbcb-cfc2-4fa2-b78b-99c803dee02d}{PhD thesis}.
\newblock Delft University of Technology.
\newblock ~(2021).

\bibitem{GitHubCode}
Evangelia Takou.
\newblock ````{Code} to simulate the {GHZ} states generation''''.
\newblock \url{https://github.com/eva-takou/GHZ_States_Public}~(2023).

\bibitem{Chruscinski2014}
D.~Chruscinski and G.~Sarbicki.
\newblock ``Entanglement witnesses: construction, analysis and classification''.
\newblock \href{https://dx.doi.org/10.1088/1751-8113/47/48/483001}{J. Phys. A: Math. Theor. {\bf 47}, 483001}~(2014).

\bibitem{CarvachoSciRep2017}
G.~Carvacho, F.~Graffitti, V.~D'Ambrosio, B.~C. Hiesmayr, and F.~Sciarrino.
\newblock ``Experimental investigation on the geometry of ghz states''.
\newblock \href{https://dx.doi.org/https://doi.org/10.1038/s41598-017-13124-6}{Sci Rep. {\bf 7}, 13265}~(2017).

\bibitem{ZhaoPRA2019}
Qi~Zhao, Gerui Wang, Xiao Yuan, and Xiongfeng Ma.
\newblock ``Efficient and robust detection of multipartite greenberger-horne-zeilinger-like states''.
\newblock \href{https://dx.doi.org/10.1103/PhysRevA.99.052349}{Phys. Rev. A {\bf 99}, 052349}~(2019).

\bibitem{CerezoPRL2021}
Jacob~L. Beckey, N.~Gigena, Patrick~J. Coles, and M.~Cerezo.
\newblock ``Computable and operationally meaningful multipartite entanglement measures''.
\newblock \href{https://dx.doi.org/10.1103/PhysRevLett.127.140501}{Phys. Rev. Lett. {\bf 127}, 140501}~(2021).

\bibitem{Coffman2000}
Valerie Coffman, Joydip Kundu, and William~K. Wootters.
\newblock ``Distributed entanglement''.
\newblock \href{https://dx.doi.org/10.1103/PhysRevA.61.052306}{Phys. Rev. A {\bf 61}, 052306}~(2000).

\bibitem{WongPRA2001}
Alexander Wong and Nelson Christensen.
\newblock ``Potential multiparticle entanglement measure''.
\newblock \href{https://dx.doi.org/10.1103/PhysRevA.63.044301}{Phys. Rev. A {\bf 63}, 044301}~(2001).

\bibitem{DafaQInf2012}
Dafa Li.
\newblock ``The n-tangle of odd n qubits''.
\newblock \href{https://dx.doi.org/https://doi.org/10.1007/s11128-011-0256-8}{Quantum Inf. Process. {\bf 11}, 481–492}~(2012).

\bibitem{Horodecki2009}
Ryszard Horodecki, Pawe\l{} Horodecki, Micha\l{} Horodecki, and Karol Horodecki.
\newblock ``Quantum entanglement''.
\newblock \href{https://dx.doi.org/10.1103/RevModPhys.81.865}{Rev. Mod. Phys. {\bf 81}, 865--942}~(2009).

\bibitem{Makhlin2002}
Yuriy Makhlin.
\newblock ``Nonlocal properties of two-qubit gates and mixed states, and the optimization of quantum computations''.
\newblock \href{https://dx.doi.org/https://doi.org/10.1023/A:1022144002391}{Quantum Inf. Process. {\bf 1}, 243--252}~(2002).


\bibitem{Liarxiv2010}
X.~Li and D.~Li.
\newblock ``Relationship between the n-tangle and the residual entanglement of even n qubits''.
\newblock  \href{https://dl.acm.org/doi/abs/10.5555/2011451.2011462}{Quantum Info. Comput. {\bf 10}, 1018-1028}~(2010).




\bibitem{BradleyThesis2021}
C.~E. Bradley.
\newblock ``Order from disorder: Control of multi-qubit spin registers in diamond''.
\newblock \href{https://dx.doi.org/https://doi.org/10.4233/uuid:acafe18b-3345-4692-9c9b-05e970ffbe40}{PhD thesis}.
\newblock Delft University of Technology.
\newblock ~(2021).

\bibitem{OsterlohPRA2008}
Andreas Osterloh, Jens Siewert, and Armin Uhlmann.
\newblock ``Tangles of superpositions and the convex-roof extension''.
\newblock \href{https://dx.doi.org/10.1103/PhysRevA.77.032310}{Phys. Rev. A {\bf 77}, 032310}~(2008).

\bibitem{LohmayerPRL2006}
Robert Lohmayer, Andreas Osterloh, Jens Siewert, and Armin Uhlmann.
\newblock ``Entangled three-qubit states without concurrence and three-tangle''.
\newblock \href{https://dx.doi.org/10.1103/PhysRevLett.97.260502}{Phys. Rev. Lett. {\bf 97}, 260502}~(2006).

\bibitem{nielsen_chuang_2010}
Michael~A. Nielsen and Isaac~L. Chuang.
\newblock ``Quantum computation and quantum information: 10th anniversary edition''.
\newblock \href{https://dx.doi.org/https://doi.org/10.1017/CBO9780511976667}{Cambridge University Press}. ~(2010).

\bibitem{KongPRA2015}
Fan-Zhen Kong, Jun-Long Zhao, Ming Yang, and Zhuo-Liang Cao.
\newblock ``Entangling power and operator entanglement of nonunitary quantum evolutions''.
\newblock \href{https://dx.doi.org/10.1103/PhysRevA.92.012127}{Phys. Rev. A {\bf 92}, 012127}~(2015).

\bibitem{MazziottiPRA2022}
Anthony~W. Schlimgen, Kade Head-Marsden, LeeAnn~M. Sager-Smith, Prineha Narang, and David~A. Mazziotti.
\newblock ``Quantum state preparation and nonunitary evolution with diagonal operators''.
\newblock \href{https://dx.doi.org/10.1103/PhysRevA.106.022414}{Phys. Rev. A {\bf 106}, 022414}~(2022).

\bibitem{DobrovitskiPRB2012}
Zhi-Hui Wang, Wenxian Zhang, A.~M. Tyryshkin, S.~A. Lyon, J.~W. Ager, E.~E. Haller, and V.~V. Dobrovitski.
\newblock ``Effect of pulse error accumulation on dynamical decoupling of the electron spins of phosphorus donors in silicon''.
\newblock \href{https://dx.doi.org/10.1103/PhysRevB.85.085206}{Phys. Rev. B {\bf 85}, 085206}~(2012).

\bibitem{VanDerSarThesis2012}
T.~Van der Sar.
\newblock ``Quantum control of single spins and single photons in diamond''.
\newblock \href{https://dx.doi.org/http://resolver.tudelft.nl/uuid:c8b7eb3d-2e42-4357-ba96-4bef1f352aac}{PhD thesis}.
\newblock Delft University of Technology.
\newblock ~(2012).

\bibitem{DeLangeThesis2012}
G.~De Lange.
\newblock ``Quantum control and coherence of interacting spins in diamond''.
\newblock \href{https://dx.doi.org/https://doi.org/10.4233/uuid:7e730d04-c04c-404f-a2a8-4a8e62a99823}{PhD thesis}.
\newblock Delft University of Technology.
\newblock ~(2012).

\bibitem{cci}
``https://cyberinitiative.org/''.

\bibitem{OsterlohPRA2009}
Christopher Eltschka, Andreas Osterloh, and Jens Siewert.
\newblock ``Possibility of generalized monogamy relations for multipartite entanglement beyond three qubits''.
\newblock \href{https://dx.doi.org/10.1103/PhysRevA.80.032313}{Phys. Rev. A {\bf 80}, 032313}~(2009).

\bibitem{Zanardi2000}
Paolo Zanardi, Christof Zalka, and Lara Faoro.
\newblock ``Entangling power of quantum evolutions''.
\newblock \href{https://dx.doi.org/10.1103/PhysRevA.62.030301}{Phys. Rev. A {\bf 62}, 030301}~(2000).

\end{thebibliography}

\end{document}